\documentclass{aa}

\usepackage{enumerate}

\makeatletter
\g@addto@macro{\UrlBreaks}{\UrlOrds}
\makeatother
\usepackage{CJKutf8}

\newcommand{\hb}{H$\beta$}
\newcommand{\hg}{H$\gamma$}
\newcommand{\hd}{H$\delta$}

\newcommand{\hda}{\hd$_\mathrm{A}$}

\newcommand{\sfr}{SFR$_\mathrm{100Myrs}$}
\newcommand{\ssfr}{sSFR$_\mathrm{100Myrs}$}
\usepackage{import}
\usepackage{amsmath}
\usepackage{amssymb}
\usepackage{natbib}
\bibpunct{(}{)}{;}{a}{}{,}
\usepackage{lscape}
\usepackage{xcolor}
\usepackage[normalem]{ulem}
\usepackage{blindtext}
\usepackage{graphicx}
\usepackage[para]{threeparttable}
\usepackage{txfonts}
\usepackage{placeins}
\usepackage[colorlinks=true,linkcolor=blue,citecolor=blue,urlcolor=blue]{hyperref}
\begin{document} 

   \title{More is better: strong constraints on the stellar properties of LEGA-C $z\sim1$ galaxies with {\tt Prospector}}
   \titlerunning{Strong constraints on the stellar properties of LEGA-C $z\sim1$ galaxies with {\tt Prospector}} 
   \author{Angelos Nersesian\inst{1,2}
          \and
          Arjen van der Wel\inst{2,3}
          \and
          Anna R. Gallazzi\inst{4}
          \and
          Yasha Kaushal\inst{5}
          \and 
          Rachel Bezanson\inst{5}
          \and
          Stefano Zibetti\inst{4}
          \and
          Eric F. Bell\inst{6}
          \and 
          Francesco D'Eugenio\inst{7,8}
          \and
          Joel Leja\inst{9,10,11}
          \and
          Marco Martorano\inst{2}
          \and 
          Po-Feng Wu\inst{12, 13, 14}
          }
    
   \institute{STAR Institute, Universit\'e de Li{\`e}ge, Quartier Agora, All\'ee du six Aout 19c, B-4000 Liege, Belgium\\
        \email{\textcolor{blue}{angelos.nersesian@uliege.be}}
    \and
        Sterrenkundig Observatorium Universiteit Gent, Krijgslaan 281 S9, B-9000 Gent, Belgium
    \and
        Max-Planck Institut f\"{u}r Astronomie K\"{o}nigstuhl, D-69117, Heidelberg, Germany
    \and 
        Osservatorio Astrofisico di Arcetri, Largo Enrico Fermi 5, I-50125 Firenze, Italy
    \and
        Department of Physics and Astronomy and PITT PACC, University of Pittsburgh, Pittsburgh, PA 15260, USA
    \and
        Department of Astronomy, University of Michigan, 1085 South University Avenue, Ann Arbor, MI 48109, USA
    \and
        Kavli Institute for Cosmology, University of Cambridge, Madingley Road, Cambridge, CB3 0HA, UK
    \and 
        Cavendish Laboratory - Astrophysics Group, University of Cambridge, 19 JJ Thomson Avenue, Cambridge, CB3 0HE, UK
    \and
        Department of Astronomy and Astrophysics, 525 Davey Lab, The Pennsylvania State University, University Park, PA 16802, USA
    \and
        Institute for Gravitation and the Cosmos, The Pennsylvania State University, University Park, PA 16802, USA
    \and
        Institute for Computational and Data Sciences, The Pennsylvania State University, University Park, PA 16802, USA
    \and
        Institute of Astrophysics, National Taiwan University, Taipei 10617, Taiwan
    \and
        Department of Physics and Center for Theoretical Physics, National Taiwan University, Taipei 10617, Taiwan
    \and
        Physics Division, National Center for Theoretical Sciences, Taipei 10617, Taiwan
        }

   \date{Received 18 October 2024 / Accepted 05 February 2025}

 
  \abstract
   {}
   {We present the stellar properties of 2908 galaxies (1208 quiescent and 1700 star forming) at $0.6 < z < 1.0$ from the Large Early Galaxy Astrophysics Census (LEGA-C) survey. We emphasize the importance of high signal-to-noise, high spectral resolution spectroscopy in the inference of stellar population properties of galaxies.}
   {We estimated the galaxy properties with the Bayesian spectral energy distribution (SED) framework {\tt Prospector}. We fit spectroscopy and broadband photometry together, drawn from the LEGA-C DR3 and UltraVISTA catalogs, respectively.}
   {We report a positive correlation between light-weighted ages and stellar velocity dispersion ($\sigma_\star$). The trend with $\sigma_\star$ is weaker for the mass-weighted ages and stellar metallicity ($Z_\star$). At fixed $\sigma_\star$, we find a tentative correlation between $Z_\star$ and stellar age. On average, quiescent galaxies are characterized by high $Z_\star$; they are $\sim 1.1$~Gyr older, less dusty, and have steeper dust attenuation slopes (due to a lower optical depth) compared to star-forming galaxies. Conversely, star-forming galaxies are characterized by significantly higher dust optical depths and shallower (grayer) attenuation slopes. Low-mass (high-mass) star-forming galaxies have lower (higher) $Z_\star$, while their stellar populations are on average younger (older). A key pragmatic result of our study is that a linear-space metallicity prior is preferable to a logarithmic-space one when using photometry alone, as the latter biases the posteriors downward.}
   {Spectroscopy greatly improves stellar population measurements and is required to provide meaningful constraints on age, metallicity, and other properties. Pairing spectroscopy with photometry helps to resolve the dust–age–metallicity degeneracy. Spectroscopic data yield more accurate mass- and light-weighted ages, with ages inferred from photometry alone suffering such large uncertainties that their utility is limited. Stellar metallicities are constrained by our spectroscopy, but precise measurements remain challenging (and impossible with photometry alone), particularly in the absence of Mg and Fe lines redward of 5000~\AA~in the observed spectrum.}

   \keywords{galaxies: high-redshift --  
             galaxies: statistics --
             galaxies: evolution --
             galaxies: fundamental parameters 
            }

   \maketitle
%

\section{Introduction}

Determining the properties of the stellar populations in galaxies is essential to describe how galaxies assemble their mass over cosmic time. The simplest method to infer these properties and determine whether the process of star formation is ongoing or not is through the analysis of broadband photometry. A common practice is to combine the light of different optical and near-infrared (NIR) bands (galaxy colors) to determine the stellar population properties \citep[e.g.,][]{Strateva_2001AJ....122.1861S, Baldry_2004ApJ...600..681B, Bell_2004ApJ...608..752B, Labbe_2005ApJ...624L..81L, Faber_2007ApJ...665..265F, Wuyts_2007ApJ...655...51W, Williams_2009ApJ...691.1879W, Salim_2014SerAJ.189....1S}. While optical colors and multiwavelength photometry (both broadband and narrowband) are useful for studying galaxies, they are insufficient on their own to fully constrain a galaxy’s chemical composition and star-formation history (SFH) \citep[e.g.,][]{Chaves_Montero_2020MNRAS.495.2088C, Nersesian_2024A&A...681A..94N, Csizi_2024A&A...689A..37C}.

It is well known that inferences from galaxy photometry suffer strongly from the age-dust-metallicity degeneracies \citep[e.g.,][]{Worthey_1994ApJS...94..687W, Silva_1998ApJ...509..103S, Devriendt_1999A&A...350..381D, Pozzetti_2000MNRAS.317L..17P, Bell_2001ApJ...550..212B, Walcher_2011Ap&SS.331....1W}. One way to mitigate these degeneracies, and rigorously determine the SFHs and chemical composition of galaxies, is by complementing the photometric spectral energy distributions (SEDs) with spectroscopic information \citep[e.g.,][]{Worthey_1994ApJS...94..687W, Bruzual_2003MNRAS.344.1000B, Trager_2000AJ....120..165T, Gallazzi_2009ApJS..185..253G}. In fact, the importance of optical spectroscopy on the study of stellar populations was established early on, with the measurements of the strengths of specific absorption-line features in nearby galaxies \citep[e.g., using the Lick index system,][]{Worthey_1994ApJS...94..687W}. The use of the Lick index system allowed the constraints on the mean stellar ages and metallicities of galaxies \citep[e.g.,][]{Jorgensen_1999MNRAS.306..607J, Trager_2000AJ....119.1645T, Trager_2000AJ....120..165T, Kuntschner_2000MNRAS.315..184K, Kauffmann_2003MNRAS.341...33K, Gallazzi_2006MNRAS.370.1106G, McDermid_2015MNRAS.448.3484M, Wu_2018ApJ...855...85W}. The recent studies by \citet{Johnson_2021ApJS..254...22J} and \citet{Tacchella_2022ApJ...926..134T} also highlight the importance of spectroscopy to constrain the stellar parameters, along with the necessity of both spectroscopy and photometry to alleviate the dust–age–metallicity degeneracies.

Advancements in the development of stellar population synthesis (SPS) models based on high spectral resolution stellar spectra, both empirical and theoretical, allow for the production of highly resolved synthetic galaxy spectra. Some examples of these SPS models that incorporate the latest developments in stellar evolution theory include: {\tt GALAXEV} \citep{Bruzual_2003MNRAS.344.1000B}, \citet{Maraston_2005MNRAS.362..799M, Maraston_2011MNRAS.418.2785M, Maraston_2020MNRAS.496.2962M}, the Flexible Stellar Populations Synthesis ({\tt FSPS}) code \citep{Conroy_2009ApJ...699..486C, Conroy_2010ApJ...712..833C}, and {\tt BPASS} \citep{Eldridge_2017PASA...34...58E, Stanway_2018MNRAS.479...75S, Byrne_2022MNRAS.512.5329B}. By comparing and matching the observed spectra with these highly complex SPS models, it is possible to characterize the relevant stellar properties of galaxies including the metallicity and age (either mass-weighted or light-weighted) of the stellar populations. An accurate estimation of these two properties sets the stage for an accurate description of the chemical enrichment and SFH of galaxies.

The emergence of large photometric and spectroscopic surveys both at a low-redshift (e.g., SDSS, \citealt{York_2000AJ....120.1579Y}; GAMA, \citealt{Driver_2011MNRAS.413..971D}), as well as at intermediate-redshift (e.g., COSMOS, \citealt{Scoville_2007ApJS..172....1S, Weaver_2022ApJS..258...11W}; 3D-HST, \citealt{Brammer_2012ApJS..200...13B, Momcheva_2016ApJS..225...27M}; MOSDEF, \citealt{Kriek_2015ApJS..218...15K}; VIPERS, \citealt{Scodeggio_2018A&A...609A..84S}), in combination with the high-spectral resolution SPS models \citep[see review by][]{Conroy_2013ARA&A..51..393C}, led to the development of full-spectrum fitting tools \citep[see reviews by][]{Walcher_2011Ap&SS.331....1W, Pacifici_2023ApJ...944..141P}. These state-of-the-art SED fitting tools have taken up the role of estimating the integrated properties of galaxies. A few examples among the most recently developed SED fitting frameworks are {\tt pPXF} \citep{Cappellari_2004PASP..116..138C, Cappellari_2017MNRAS.466..798C}, {\tt STARLIGHT} \citep{Cid_Fernandes_2005MNRAS.358..363C}, {\tt STECMAP} \citep{Ocvirk_2006MNRAS.365...46O}, {\tt VESPA} \citep{Tojeiro_2007MNRAS.381.1252T}, {\tt MAGPHYS} \citep{da_Cunha_2008MNRAS.388.1595D}, {\tt FIT 3D} \citep{Sanchez_2016RMxAA..52...21S, Lacerda_2022NewA...9701895L}, {\tt BEAGLE} \citep{Chevallard_2016MNRAS.462.1415C}, {\tt FIREFLY} \citep{Wilkinson_2017MNRAS.472.4297W}, {\tt FADO} \citep{Gomes_2017A&A...603A..63G}, {\tt Dense Basis} \citep{Iyer_2017ApJ...838..127I, Iyer_2019ApJ...879..116I}, {\tt BAGPIPES} \citep{Carnall_2018MNRAS.480.4379C}, {\tt CIGALE} \citep{Boquien_2019A&A...622A.103B}, {\tt PROSPECT}, \citep{Robotham_2020MNRAS.495..905R}, {\tt Prospector} \citep{Leja_2017ApJ...837..170L, Johnson_2021ApJS..254...22J}, and {\tt piXedfit} \citep{Abdurrouf_2021ApJS..254...15A}. Different SED fitting codes offer different capabilities such as being able to (1) analyze multiwavelength photometry from ground-based and space observatories, (2) analyze optical spectra either from slit or integral-field spectroscopic surveys, and (3) combine photometry and spectroscopy with a calibration model. 

\begin{figure*}[t]
    \centering
    \includegraphics[width=\textwidth]{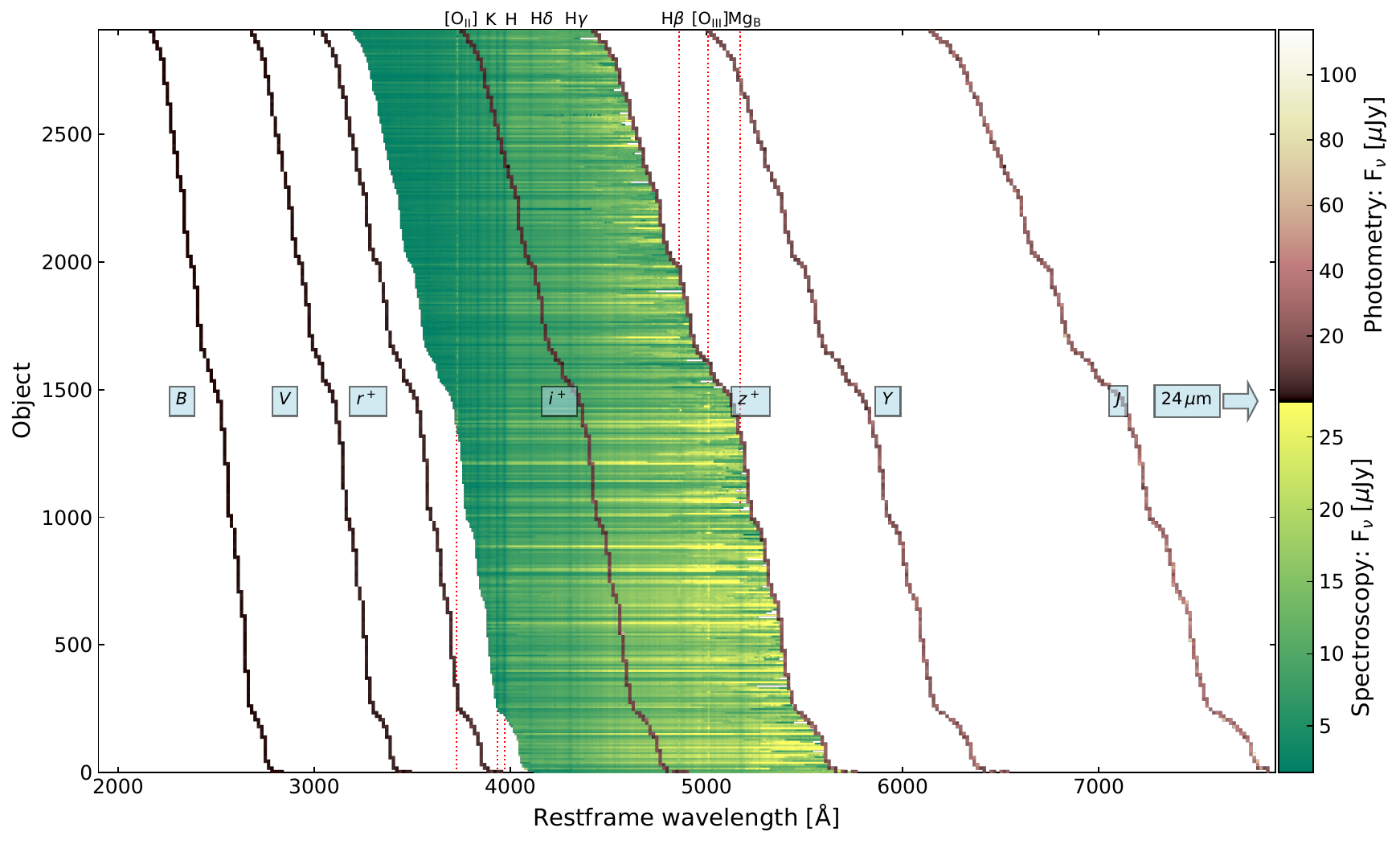}
    \caption{Compilation of 2908 LEGA-C spectrophotometric SEDs. The flux density of the continuum emission of the observed LEGA-C spectra is shown with the spring colorbar. The individual curves represent the UltraVISTA optical and NIR broad bands ($B$, $V$, $r^+$, $i^+$, $z^+$, $Y$, and $J$). The MIPS~24~$\mu$m emission is not shown in the figure due to visualization purposes. The dotted red lines indicate the central wavelength of various prominent spectral features. Galaxies are sorted with increasing spectroscopic redshift.}
    \label{fig:spectrophotometric_seds}
\end{figure*}

In the local Universe, high-quality, and even spatially-resolved, optical spectra are easy to obtain (e.g., ATLAS3D, \citealt{Cappellari_2011MNRAS.413..813C}; CALIFA, \citealt{Sanchez_2013A&A...554A..58S}; SAMI, \citealt{Bryant_2015MNRAS.447.2857B}; MaNGA, \citealt{Bundy_2015ApJ...798....7B}). However, obtaining high-quality spectroscopy for galaxies at higher redshifts is often more challenging, as such surveys must balance between achieving a high signal-to-noise ratio (S/N) and covering a large sample size. In addition, with increasing redshift the contribution of the sky background becomes more dominant, essentially dimming the surface brightness of galaxies \citep{Hogg_1999astro.ph..5116H}. To account for this, many spectroscopic surveys targeted only bright emission lines originating from the ionized gas in galaxies. 

The advent of the Large Early Galaxy Astrophysics Census \citep[LEGA-C;][]{van_der_Wel_2016ApJS..223...29V, van_der_Wel_2021ApJS..256...44V} survey has changed the landscape of intermediate-redshift spectroscopy. The LEGA-C survey is an exceptional dataset that contains about 4,000 high S/N restframe optical spectra at observed redshift $0.6 \le z \le 1$ (or at a lookback time of $\sim 7$~Gyr). The galaxy sample was $K_\mathrm{s}$-band selected from UltraVista targets in the COSMOS field. So far, the analysis of the LEGA-C spectra has produced many interesting results, from the formation timescales and evolution pathways of galaxies at observed redshift $z_\mathrm{spec} \approx 0.8$ \citep{Chauke_2018ApJ...861...13C, Wu_2018ApJ...868...37W, Kaushal_2024ApJ...961..118K}, to the stellar kinematics \citep{Bezanson_2018ApJ...858...60B, Straatman_2018ApJS..239...27S, Cole_2020ApJ...890L..25C, de_Graaff_2021ApJ...913..103D, D_Eugenio_2023MNRAS.525.2765D, D_Eugenio_2023MNRAS.525.2789D, Barrientos_Acevedo_2023MNRAS.524..907B}, and the interstellar medium (ISM) properties \citep{Spilker_2018ApJ...860..103S, Barisic_2020ApJ...903..146B, Lewis_2024ApJ...964...59L}. 

Spectroscopic analyses of the stellar properties of galaxies at a significant redshift have been limited to a small number of targets \citep[e.g.,][]{Choi_2014ApJ...792...95C, Gallazzi_2014ApJ...788...72G, Belli_2015ApJ...799..206B}. The LEGA-C survey, with its large statistical sample, offers a valuable opportunity to gain deeper insights into the stellar properties of galaxies during an epoch ($\sim 7$~Gyr ago), when nearly 50\% of all stars in the Universe had formed since $z \sim 1$ \citep[e.g.,][]{Ilbert_2010ApJ...709..644I, Muzzin_2013ApJ...777...18M}, and galaxies likely hosted different stellar populations.

In this paper, we present the spectrophotometric analysis of 2908 star-forming and quiescent galaxies drawn from the LEGA-C DR3. To be more specific, we combine the high spectral resolution spectroscopy of LEGA-C with the broadband photometry from the COSMOS/UltraVISTA catalog \citep{Muzzin_2013ApJS..206....8M, Muzzin_2013ApJ...777...18M}, in the restframe wavelength range 2000~\AA~to 12~$\mu$m. We applied the SED fitting framework {\tt Prospector}\footnote{\url{https://github.com/bd-j/prospector}} \citep{Johnson_2021ApJS..254...22J} to simultaneously fit the spectroscopy and photometry. One of the main goals of this paper is to highlight the importance of spectroscopy in the inference of the stellar properties of galaxies, through a comparison of the output physical quantities retrieved whether spectroscopy is combined with photometry or not. We perform various quality tests in order to confirm the robustness of our fits to the data. We quantify the differences between the model and the observed spectrophotometric SEDs, first by applying a $\chi^2$ test between the observed and model spectra, and then by measuring the strength of two age- and metallicity-sensitive features, the \hda~and Fe4383 from the model spectra \citep[similar to ][]{Nersesian_2024A&A...681A..94N}.

Another objective of this paper is to introduce a dataset that includes the intrinsic stellar properties of the LEGA-C galaxies as retrieved from the SED fitting with {\tt Prospector}. Within the LEGA-C team, we have used different methodologies to fit the LEGA-C DR3 spectra, with the purpose of determining the stellar population properties of the galaxies, as well as their SFHs. In \citet{Kaushal_2024ApJ...961..118K}, we used the Bayesian SPS code {\tt BAGPIPES} to perform a spectrophotometric analysis, similar to this work. The main differences between the two approaches is the use of a nonparametric SFH and emission line marginalization in {\tt Prospector} compared to {\tt BAGPIPES}. In Gallazzi et al. (in prep.), we adopt a different fitting method, deriving the stellar properties through Lick index plus photometry modeling, utilizing stochastic parametric SFHs with the {\tt BaStA} code \citep{Zibetti_2017MNRAS.468.1902Z, Zibetti_2022MNRAS.512.1415Z}. A comparison of the stellar properties recovered from the three different methodologies, as well as a catalog with the measured values, will be released in Gallazzi et al. (in prep.).

As a first analysis of the rich dataset produced with {\tt Prospector}, we set out to characterize the LEGA-C galaxies at $z \sim 1$ based on their global stellar properties. We perform a statistical analysis on several key galaxy properties, study the stellar population scaling relations, and explore the average underlying trends between stellar age, stellar metallicity and dust attenuation that give rise to strong trends in the $UVJ$ color-color and SFR--$M_\star$ diagrams.

This paper is structured as follows. In Section~\ref{sec:data_and_sample} we describe the datasets we used and the properties of our galaxy sample. In Section~\ref{sec:sed_fitting} we present the physical model that was adopted within the {\tt Prospector} inference framework to describe the spectrophotometric SEDs of the LEGA-C galaxies. In Section~\ref{sec:results} we present the final products of SED fitting with {\tt Prospector}, and discuss the underlying scaling relations between age, metallicity and dust in the $UVJ$ and SFR--$M_\star$ diagrams. Finally, in Section~\ref{sec:conclusions} we summarize our key findings and conclusions.


\section{Data and sample} \label{sec:data_and_sample}

\subsection{Spectroscopy} \label{subsec:spectroscopy}

The LEGA-C \citep{van_der_Wel_2016ApJS..223...29V, van_der_Wel_2021ApJS..256...44V} survey contains 4081 high signal-to-noise (S/N) restframe optical spectra at a lookback time of $\sim 7$~Gyr. Out of these 4081 spectra, 3741 belong to unique galaxy targets, while 340 spectra are duplicate observations. A redshift-dependent limit was applied on the $K_\mathrm{s}$-band magnitude to select the galaxy sample, $K_\mathrm{s, lim} = 20.7 - 7.5 \log\left(\left(1 + z\right) / 1.8\right)$. The $K_\mathrm{s}$-band photometry was taken from the Ultra Deep Survey with the VISTA telescope (UltraVISTA) catalog \citep{Muzzin_2013ApJS..206....8M, Muzzin_2013ApJ...777...18M}. The redshift-dependent selection of the LEGA-C galaxies resulted in a mass-complete sample in DR3 (92\% completeness above $\log (M_\star/\mathrm{M}_\odot) = 10.3$). For more details about the design and the goals of the LEGA-C survey we refer the readers to \citet{van_der_Wel_2016ApJS..223...29V},   \citet{Straatman_2018ApJS..239...27S}, and \citet{van_der_Wel_2021ApJS..256...44V}.

The now retired VIMOS spectrograph \citep{Le_Fevre_2003SPIE.4841.1670L} at the ESO Very Large Telescope (VLT) was used to obtain the spectroscopic observations of LEGA-C over the course of four years. The typical observed wavelength range of LEGA-C is 6300~\AA$~< \lambda <~$8800~\AA~or restframe $\sim$ 3000~\AA~$< \lambda <~$5550~\AA, with an effective spectral resolution of $R\sim3500$. The optical spectrum contains a significant number of absorption line features that show small galaxy-to-galaxy variations in terms of flux density. Many of these absorption features are often too weak to be accurately measured, and their detection may be further compromised by various systematic effects. In particular, a bias can be introduced in the equivalent width measurement of the absorption line indices, if the variance of the spectrum is considered or not. Other systematic effects can be introduced during the steps of sky subtraction, noise modeling, and wavelength calibration. In \citet{van_der_Wel_2021ApJS..256...44V}, a comprehensive description is given of the steps taken to minimize systematic effects and biases, by employing an approximately bias-free methodology. Along with the reduced LEGA-C spectra, a catalog was released with the Lick indices of 20 spectral absorption features, corrected for emission. In Gallazzi et al. (in prep.), an updated catalog will be released, increasing the number of objects with Lick index measurements, and effectively reducing the uncertainties of the measurements by a factor of 1.3.

\subsection{Photometry} \label{subsec:photometry}

We combined the spectroscopic data of LEGA-C with the broadband photometry from the UltraVISTA catalog \citep{Muzzin_2013ApJS..206....8M, Muzzin_2013ApJ...777...18M}. UltraVISTA contains the point spread function (PSF) matched photometry of 30 bands, covering the wavelength range $0.15$-$24~\mu$m. The flux density and the associated error in each band was measured within a $2\arcsec.1$ aperture from the PSF-matched images. The photometry in all bands has already been corrected for Galactic dust extinction using the dust maps from \citep{Schlegel_1998ApJ...500..525S}.

Here we only used a subset of eight photometric bands for the SED fitting. In particular, we used the optical broad-band data observed with Subaru/SuprimeCam \citep[$B$, $V$, $r^+$, $i^+$, $z^+$;][]{Taniguchi_2007ApJS..172....9T}, the NIR $Y$ and $J$ bands taken from the UltraVISTA survey \citep{McCracken_2012A&A...544A.156M}, and the $24~\mu$m channel from Spitzer’s MIPS camera \citep{Sanders_2007ApJS..172...86S}. There are four reasons for this choice: (1) the LEGA-C sample was selected based on the catalog by \citet{Muzzin_2013ApJS..206....8M}, (2) a zero-point correction was applied to the UltraVISTA photometry by \citet{van_der_Wel_2021ApJS..256...44V} to ensure that the flux densities are independent of the SPS models, (3) despite the applied zero-point correction to the photometric bands ($B$, $V$, $r^+$, $i^+$, $z^+$, $Y$, $J$, $H$, $K_\mathrm{s}$), \citet{van_der_Wel_2021ApJS..256...44V} showed a persistent systematic mismatch between a synthesized photometry and the UltraVISTA $H-K_\mathrm{s}$ color, and (4) the LEGA-C spectra were calibrated using the $BVr^+i^+z^+YJ$ filter set, because of the mismatch in the $H-K_\mathrm{s}$ color. 

\subsection{Primary working galaxy sample} \label{subsec:final_sample}

In our analysis, we just use the LEGA-C galaxies with a {\tt PRIMARY} flag from the DR3 catalog. The {\tt PRIMARY} flag indicates whether a galaxy was selected from the $K_\mathrm{s}$-band selected parent sample (1) or as a filler (0) \citep{van_der_Wel_2021ApJS..256...44V}. Our final working sample contains 2908 galaxies in the redshift range $0.6 \le z \le 1$. Hereafter, we refer to this sample of 2908 galaxies as primary galaxy sample. Figure~\ref{fig:spectrophotometric_seds} shows the spectrophotometric SEDs of these 2908 targets sorted with increasing redshift.

The average S/N of the spectra in our primary sample is $\sim17.2$~\AA$^{-1}$. The $16^\mathrm{th}$, $50^\mathrm{th}$, and $84^\mathrm{th}$ percentile of the stellar velocity dispersion ($\sigma_\star$) is 126.7, 166, and 206.6~km~s$^{-1}$, respectively. We separated galaxies into 1208 quiescent and 1700 star-forming based on a $UVJ$ diagram and the definition of \citet{Muzzin_2013ApJ...777...18M}. The restframe $U - V$ and $V - J$ colors were calculated through fitting template spectra to the UltraVISTA photometric SEDs \citep{Straatman_2018ApJS..239...27S}. One final detail to note is that among the 2908 targets, 279 are duplicate observations. Some of these duplicate observations show differences in wavelength coverage and emission-line characteristics \citep{van_der_Wel_2021ApJS..256...44V}. In this work, we treat the duplicate observations as individual galaxies.

\section{SED fitting with {\tt Prospector}} \label{sec:sed_fitting}

In this section, we present the physical model that was adopted to describe the spectrophotometric SEDs of the LEGA-C galaxies. We also perform several quality tests of the fits. The physical model was generated within the {\tt Prospector} inference framework \citep{Johnson_2021ApJS..254...22J}. {\tt Prospector} is an SED fitting code that adopts a Bayesian forward modeling and Monte Carlo sampling of the parameter space. {\tt Prospector} can generate gridless "on-the-fly" SEDs by combining stellar, nebular, and dust models into composite stellar populations. One of the strong capabilities of {\tt Prospector} is the treatment of photometric and spectroscopic data using a flexible spectrophotometric calibration. Another advantage of {\tt Prospector} is the availability of nonparametric SFHs with various prior distributions and parameterization schemes. The benefit of using a nonparametric SFH is that we impose a much weaker prior on the shape of the functional form of the SFH. It has also been shown that a nonparametric SFH can outperform the traditional parametric SFH by vastly improving the stellar mass estimates \citep{Lower_2020ApJ...904...33L}. A summary of the free parameters and range of the prior distributions of our physical model is given in Table~\ref{tab:free_params_and_priors}. Our physical model largely follows the model presented in \citet{Leja_2017ApJ...837..170L, Leja_2018ApJ...854...62L, Leja_2019ApJ...877..140L}. Lastly, we used the nested sampler {\tt dynesty} \citep{Skilling_2004AIPC..735..395S, Skilling_2006AIPC..872..321S, Koposov_2022zndo...7388523K} to simultaneously estimate both the Bayesian evidence and the posterior distributions. {\tt dynesty} also allows the dynamic sampling of the parameter space to maximize a chosen objective function as the fit proceeds.

\begin{table*}
\caption{Free Parameters and their associated prior distribution functions in the {\tt Prospector} physical model.}
\begin{center}
\scalebox{0.88}{
\begin{threeparttable}
\begin{tabular}{llr}
\hline 
\hline\\
Parameter & Description & Prior distribution \\
\hline\\
$z$ & redshift & uniform: LEGA-C spectroscopic redshift $z_\mathrm{spec} \pm 0.005$\\[0.2cm]
$\log\left(M_\star/\mathrm{M}_\odot\right)$ & total stellar mass formed & uniform: $\mathrm{min}=7$, $\mathrm{max}=12$\\[0.2cm]
$\log\left(Z_\star/\mathrm{Z}_\odot\right)$ & stellar metallicity & uniform: $\mathrm{min}=-1.98$, $\mathrm{max}=0.2$\\[0.2cm]
SFR ratios & Ratio of adjacent SFRs in the   & Student’s t-distribution with\\
           & eight-bin nonparametric SFH, & $\sigma=0.3$ and $\nu=2$\\
           & seven free parameters in total & \\

\hline\\
$\hat{\tau}_\mathrm{dust, 2}$ & diffuse dust optical depth & clipped normal: $\mathrm{min}=0$, $\mathrm{max}=4$, $\mathrm{mean}=0.3$, $\sigma=1$\\[0.2cm]
$\hat{\tau}_\mathrm{dust, 1}$ & birth-cloud dust optical depth & clipped normal in ($\hat{\tau}_\mathrm{dust, 1}/\hat{\tau}_\mathrm{dust, 2}$): $\mathrm{min}=0$, $\mathrm{max}=4$, $\mathrm{mean}=0.3$, $\sigma=1$\\[0.2cm]
$n$ & slope of \citet{Kriek_2013ApJ...775L..16K} dust law & uniform: $\mathrm{min}=-1$, $\mathrm{max}=0.4$\\

\hline\\
$U_\mathrm{min}$ & Minimum radiation field intensity & fixed at $1.0$ \\[0.2cm]
$\gamma$ & Mass fraction of dust in high radiation intensity & fixed at $0.01$ \\[0.2cm]
$q_\mathrm{PAH}$ & PAH fraction & uniform: $\mathrm{min}=0.5$, $\mathrm{max}=7$ \\
           
\hline\\
$\log\left(Z_\mathrm{gas}/\mathrm{Z}_\odot\right)$ & gas-phase metallicity & uniform: $\mathrm{min}=-2$, $\mathrm{max}=0.5$\\[0.2cm]
$\log U$ & Ionization parameter & uniform: $\mathrm{min}=-4$, $\mathrm{max}=-1$\\[0.2cm]
$\sigma_\mathrm{eline}$ & Marginalization over emission line amplitudes & uniform: $\mathrm{min}=30$, $\mathrm{max}=300$\\

\hline\\
$cp$ & Photometric calibration & uniform: $\mathrm{min}=10^{-5}$, $\mathrm{max}=0.5$\\[0.2cm]
$cs$ & Spectroscopic calibration & uniform: $\mathrm{min}=10^{-5}$, $\mathrm{max}=0.5$\\[0.2cm]
$\nu$ & Spectral white noise & uniform: $\mathrm{min}=1$, $\mathrm{max}=3$\\
\hline
\end{tabular}
\end{threeparttable}}
\label{tab:free_params_and_priors}
\end{center}
\end{table*}

\subsection{Physical model} \label{subsec:physical_model}

We created a physical model within {\tt Prospector}, allowing 20 free parameters. {\tt Prospector} utilizes the {\tt FSPS} code \citep{Conroy_2009ApJ...699..486C} to derive the spectra of the stellar populations. The {\tt FSPS} framework offers many options regarding the stellar spectral libraries and stellar isochrones. In this work, we adopted the default SPS parameters in {\tt FSPS}, that is the MILES stellar library \citep{Sanchez_Blazquez_2006MNRAS.371..703S} and the MIST isochrones \citep{Choi_2016ApJ...823..102C}. MILES is an empirical stellar library with a restframe wavelength range from 3525 to 7400~\AA, and high spectral resolution (FWHM~$=2.5$~\AA). The MIST models are based on the open-source stellar evolution package MESA \citep{Paxton_2011ApJS..192....3P, Paxton_2013ApJS..208....4P, Paxton_2015ApJS..220...15P, Paxton_2018ApJS..234...34P}. A 5\% error floor is added to the flux uncertainties to account for the uncertainties in the underlying stellar models. Finally, throughout this paper we adopted a \citet{Chabrier_2003PASP..115..763C} initial mass function (IMF). 

Regarding the SFH in our model, we chose the `continuity' SFH with a Student's-t prior distribution described thoroughly in \citet{Leja_2019ApJ...876....3L}. Based on the regularization schemes by \citet{Ocvirk_2006MNRAS.365...46O} and \citet{Tojeiro_2007MNRAS.381.1252T}, the `continuity' prior favors a piecewise constant SFH without sharp transitions in SFR(t). We use eight time elements to describe the nonparametric SFH, specified in lookback time. To capture any recent variations in the SFH of a particular galaxy, we fixed the first two time bins at 0--30~Myr and 30--100~Myr. We also fixed the oldest time bin, placed at ($0.85~t_\mathrm{univ} - t_\mathrm{univ}$) where $t_\mathrm{univ}$ is the age of the Universe at the observed redshift. The remaining five bins are spaced equally in logarithmic time between 100~Myr and $0.85~t_\mathrm{univ}$. 

We treated the effect of dust grains on starlight in the UV and optical wavelengths by using a variable dust attenuation law, and by constraining the strength of the UV bump according to the results of \citet{Kriek_2013ApJ...775L..16K}. This flexible attenuation law has two main components that model separately the dust attenuation in birth-clouds and from the diffuse dust in the ISM. The birth-cloud component attenuates the stellar emission from stars with an age up to 10~Myr, and is given by the following functional form:

\begin{equation} \label{eq:att_1}
\\\\\ \hat{\tau}_\mathrm{dust, 1} = \hat{\tau}_{1} \left(\lambda/5500~\mathrm{\AA}\right)^{-1}.
\end{equation}

\noindent The diffuse dust component has a flexible function that attenuates both stellar and nebular emission of a particular galaxy, using a powerlaw–modified starburst curve \citep{Calzetti_2000ApJ...533..682C}, extended with the \citet{Leitherer_2002ApJS..140..303L} curve. The following functional form is used, given in \citet{Noll_2009A&A...507.1793N}:

\begin{equation} \label{eq:att_2}
\\\\\ \hat{\tau}_\mathrm{dust, 2} = \frac{\hat{\tau}_{2}}{4.05} \left[k^\prime\left(\lambda\right) + D\left(\lambda\right)\right]\left(\frac{\lambda}{\lambda_V}\right)^{n},
\end{equation}

\noindent where $k^\prime\left(\lambda\right)$ is the original \citet{Calzetti_2000ApJ...533..682C} attenuation curve, and $D\left(\lambda\right)$ is a Lorentzian–like Drude profile characterizing the UV bump at 2175~\AA~in the attenuation curve. \citet{Barisic_2020ApJ...903..146B} presented a novel approach to constrain the attenuation curve from optical spectra, and measured the strength of the UV bumps for 260 individual star-forming galaxies in LEGA-C. In a future work, we plan to analyze the UV bumps from {\tt Prospector} and compare them with those of \citet{Barisic_2020ApJ...903..146B}.

The normalization of the birth-cloud and diffuse dust components, along with the powerlaw index ($n$), which are describing the shape of the attenuation curve for the diffuse dust component, are set as free parameters in our model \citep[see][for more details]{Leja_2017ApJ...837..170L}. Moreover, the inclusion of the MIPS~24~$\mu$m emission in the MIR has the added value of further constraining the total amount of dust attenuation in a particular galaxy, thanks to the energy balance principle within {\tt Prospector}. The energy balance principle in SED fitting dictates that all energy absorbed by dust in the restframe UV and optical is re-radiated in the MIR and FIR \citep{da_Cunha_2008MNRAS.388.1595D}. Hence, further assumptions need to be made for modeling the dust emission in the infrared regime.  

For the dust emission we used the \citet{Draine_2007ApJ...663..866D} templates, that were built upon the assumptions of a silicate-graphite-PAH (polycyclic aromatic hydrocarbon) model of interstellar dust \citep{Draine_1984ApJ...285...89D, Desert_1990A&A...237..215D, Siebenmorgen_1992A&A...259..614S, Dwek_1997ApJ...475..565D, Li_2001ApJ...554..778L, Li_2002ApJ...572..232L, Draine_2001ApJ...551..807D, Draine_2014ApJ...780..172D}. The shape of the IR SED is mainly controlled by three free parameters: (1) the minimum intensity value of the stellar radiation field that heats the dust grains, $U_\mathrm{min}$, (2) the fraction of the dust heated in photo-dissociation regions (PDR), $\gamma$, and (3) the mass fraction of PAHs, $q_\mathrm{PAH}$. In particular, $U_\mathrm{min}$ and $\gamma$ control the shape and location of the peak of the thermal dust emission in the FIR, while $q_\mathrm{PAH}$ controls the MIR emission originating from the strong PAH emission features. Since our photometry only includes the MIPS~24~$\mu$m band, we use informative fixed values for $U_\mathrm{min}$ and $\gamma$, while assuming a flat prior for $q_\mathrm{PAH}$. Constraining the IR priors of the dust emission is resulting in a minimal amount of bias, less than 0.15 dex both in the SFR and dust attenuation \citep[see Appendix C of][]{Leja_2017ApJ...837..170L}.

To describe the nebular continuum and line emission we used a {\tt CLOUDY} \citep{Ferland_2013RMxAA..49..137F} grid of models within {\tt FSPS} \citep[see][]{Byler_2017ApJ...840...44B}. There are two main parameters of the {\tt CLOUDY} grid in {\tt FSPS}: (1) the gas-phase metallicity, and (2) the ionization parameter, $U$, which is the ratio of ionizing photons to the total hydrogen density. We set both parameters free with a uniform prior distribution. Furthermore, each emission line in the observed spectrum is modeled with a Gaussian profile of a variable width and amplitude. {\tt Prospector} has an option to marginalize over the amplitude of each emission-line \citep{Johnson_2021ApJS..254...22J}, while it fits for the gas velocity dispersion, $\sigma_\mathrm{eline}$, assuming a flat prior distribution.  

Finally, {\tt Prospector} allows the simultaneous fitting of photometric and spectroscopic observations by including a flexible spectrophotometric calibration. Although the LEGA-C spectra are already calibrated using a subset of the UltraVISTA photometry (see Section~\ref{subsec:photometry}), we still used an outlier model \citep{Hogg_2010arXiv1008.4686H, Johnson_2021ApJS..254...22J, Tacchella_2022ApJ...926..134T} to describe any spectroscopic or photometric outliers, that are not masked. This outlier model assumes some possibility that any given photometric data point or spectral pixel is an outlier, $cp$ or $cs$ respectively. The assumed outliers have their uncertainties inflated by a factor of 50. Then, the likelihood is calculated by marginalizing over $cp$ and $cs$ for each data point. The last free parameter in our physical model is a constant factor ($\nu$) that multiplies the spectroscopic uncertainties. The purpose of $\nu$ is to account for possible under- or overestimates of the spectroscopic uncertainties. In general, if $\nu$ is very close to 1 then the spectroscopic noise is neither under- nor overestimated.

\subsection{Quality of the SED fits}

Here we conduct various different quality tests of the SED fits produced with {\tt Prospector}. In Section~\ref{subsubsec:chi2_red}, we quantify the goodness of the fits by presenting the reduced $\chi^2$ ($\chi^2_\nu$) distribution between the model spectrophotometric SEDs with {\tt Prospector} and the observations. In Section~\ref{subsubsec:jpost}, we show the joint posterior distributions of several physical quantities to highlight any degeneracies among them. We showcase the best-fit results of two handpicked galaxies, a star-forming and a quiescent. In Section~\ref{subsubsec:spectral_comp}, we perform a comparison between two age and metallicity-sensitive absorption features (\hda~and Fe4383) to highlight any subtle differences between the modeled and observed spectra. Moreover, in Section~\ref{subsubsec:specphot_gain} we demonstrate that when fitting photometry alone, a flat $Z_\star$ prior in linear space is better to be used than a flat $Z_\star$ prior in logarithmic space, and showcase the gain in fitting both photometry and spectroscopy. Finally, in Section~\ref{subsubsec:comp_phys_dupes} we compare the stellar properties for those galaxies in LEGA-C DR3 with duplicate observations.

\begin{figure}[t]
\centering
\includegraphics[width=\columnwidth]{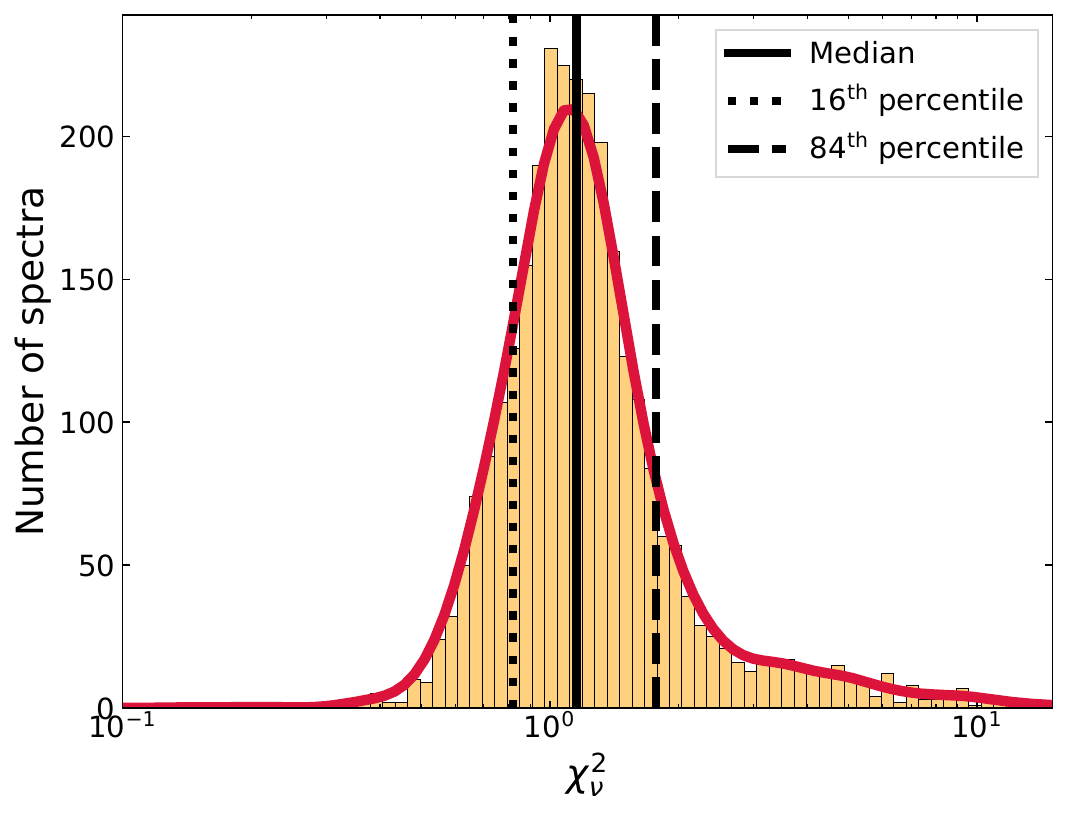}
\caption{Distribution of the reduced $\chi^2$ between the model spectrophotometric SEDs with {\tt Prospector} and the observations. The Kernel Density Estimate (KDE) distribution is shown in red, while the solid black line shows the median value. The dotted and dashed black lines indicate the $16^\mathrm{th}$ and $84^\mathrm{th}$ percentiles of the distribution, respectively.}
\label{fig:chisqr}
\end{figure}

\begin{figure*}[t]
    \centering
    \includegraphics[width=\textwidth]{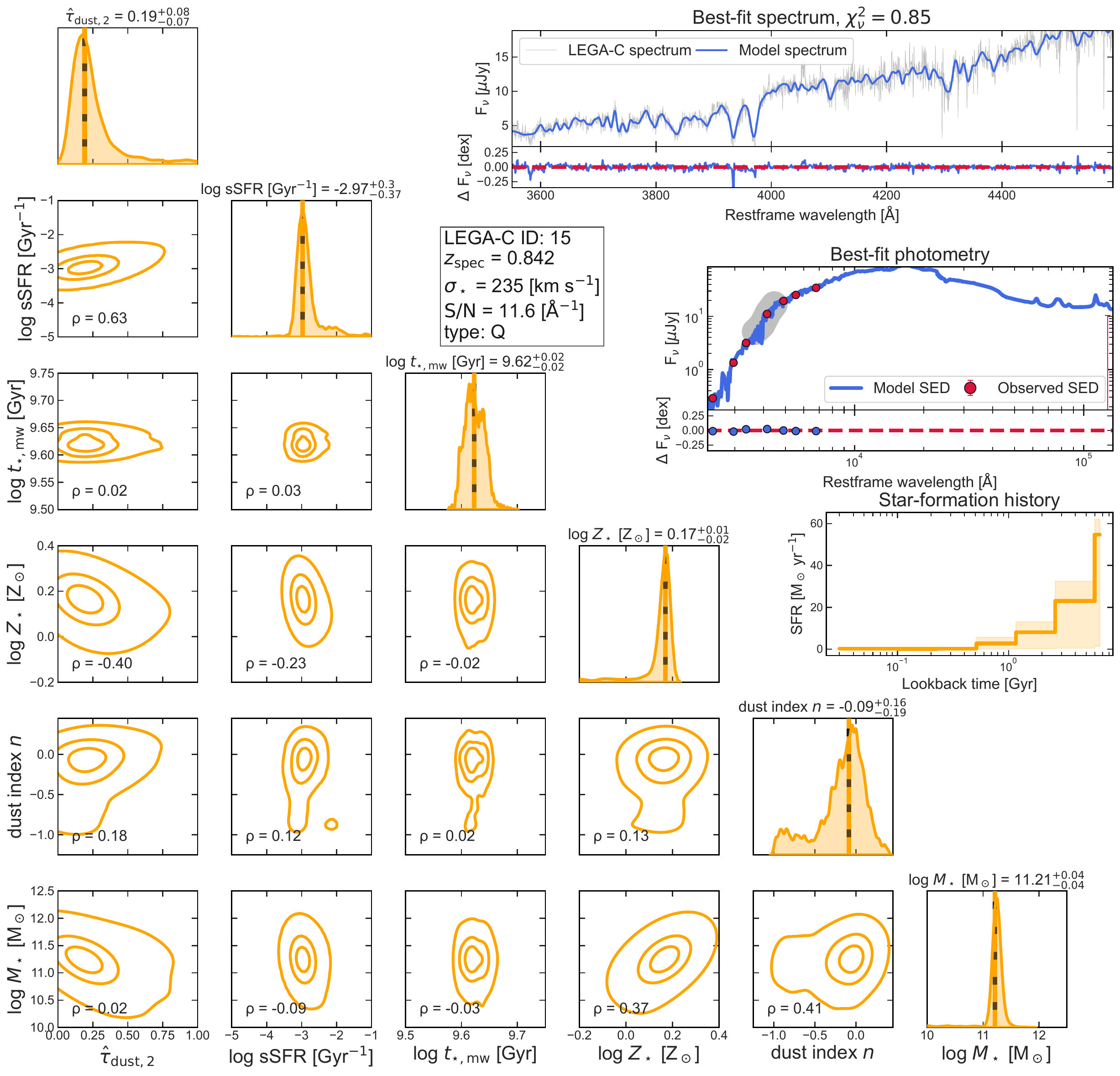}
    \caption{Joint posterior distributions of various physical quantities, best-fit spectrum, best-fit SED, and retrieved SFH, for an example quiescent galaxy at $z_\mathrm{spec} = 0.842$. In the cornerplot, we present the posterior distributions of $\hat{\tau}_\mathrm{ dust, 2}$, sSFR, $t_\mathrm{\star, mw}$, $Z_\star$, dust attenuation slope index ($n$), and $M_\star$. The contours enclose 20\%, 50\% and 80\% of the total data. The vertical black dashed line indicates the median value of each physical quantity. The Spearman's correlation coefficient, $\rho$, indicates the strength of the correlation between two posterior distributions. The inset in the right corner of the figure contains three panels. Top panel: comparison of the observed spectrum (gray line) with the fitted model (blue line). Middle panel: comparison of the observed photometry (red points) with the best-fit SED (blue line). The shaded gray region covers the wavelength range of the corresponding LEGA-C spectrum. The subpanels associated with the top and middle panels show the absolute residuals, defined as $\Delta F_{\nu} = \log \mathrm{obs}/\mathrm{mod}$. Bottom panel: the SFH posterior.}
    \label{fig:posteriors_q}
\end{figure*}

\begin{figure*}[t]
    \centering
    \includegraphics[width=18cm]{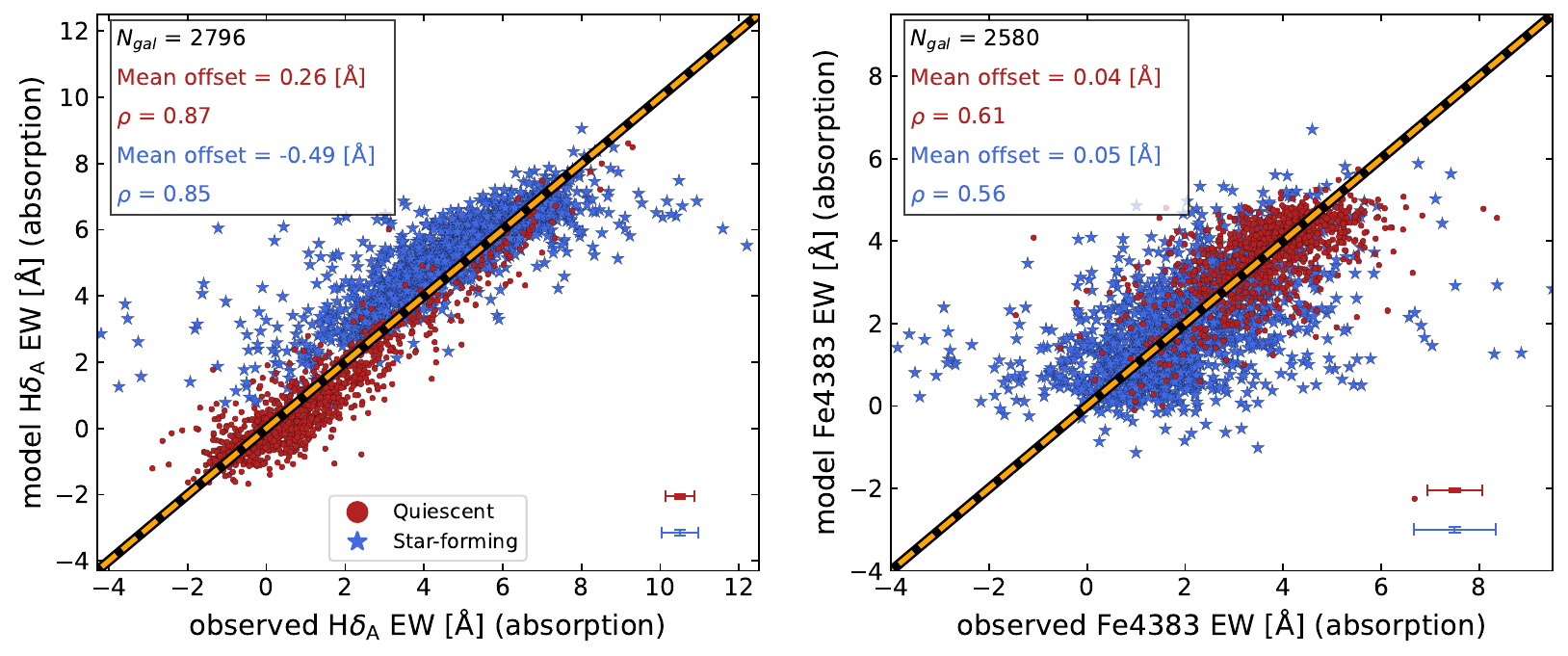}
    \caption{Modeled \hda~and Fe4383 absorption features from {\tt Prospector} fits compared to observations. Galaxies are color-coded by their $UVJ$ diagram classification as star-forming (blue stars) and quiescent (red points). The absorption-only models are compared to the observed values (corrected for emission). The dashed orange line shows the one-to-one relation. The statistics of the mean offsets and the Spearman's rank correlation coefficient ($\rho$) of each galaxy populations, are shown in the top-left corner of each panel. The typical uncertainties of each axis and galaxy population are shown in the bottom-right corner of each panel.}
    \label{fig:hd_fe4383_obs_vs_prd}
\end{figure*}

\subsubsection{Reduced $\chi^2$ distribution} \label{subsubsec:chi2_red}

We quantify the overall quality of the fits to the observed spectrophotometric SEDs of LEGA-C by calculating the reduced $\chi^2$ values through:

\begin{equation} \label{eq:chi2_red}
\\\\\ \chi^2_\nu = \frac{1}{N_\mathrm{dof}}\sum_{i=1} \frac{\left(O_i - P_i\right)^2}{\sigma_i^2}.
\end{equation}

Here $O_i$ represents the flux densities of the observed spectra, $P_i$ are the model spectra, $\sigma_i$ are the observed uncertainties, and $N_\mathrm{dof}$ is the degrees of freedom (dof), calculated as the wavelength elements (i.e., $N_\mathrm{\lambda,~spec}+N_\mathrm{\lambda,~phot}$) minus the free parameters in our model (i.e., 20 free parameters). The $\chi^2_\nu$ distribution is shown in Fig.~\ref{fig:chisqr}. We find that the peak of the $\chi^2_\nu$ distribution and the median value of the histogram are at the same value $\sim1.15$. The $16^\mathrm{th}$ percentile is at a $\chi^2_\nu$ value of 0.82 and the $84^\mathrm{th}$ is at 1.77. Out of the 2908 modeled spectra, 88\% have a $\chi^2_\nu\le2$. About 11\% of the modeled spectra have a $2<\chi^2_\nu<10$, while only a very small fraction of 1\% have $\chi^2_\nu\ge10$. Overall, we find an excellent agreement between the observed and the modeled spectra. However, the prominent tail of high-$\chi^2_\nu$ values may suggest that these fits are dominated by systematic uncertainties. 

\subsubsection{Joint posterior distributions} \label{subsubsec:jpost}

After fitting the spectrophotometric SEDs of the LEGA-C galaxies with {\tt Prospector}, we were able to derive estimates of several physical quantities, such as the specific star-formation rate (sSFR = $M_\star$/SFR), stellar mass ($M_\star$), stellar metallicity ($Z_\star$),\footnote{A constant stellar metallicity history is used here, assuming that all stars in a galaxy formed with the same metallicity.} mass-weighted stellar age ($t_\mathrm{\star, mw}$),\footnote{The mass-weighted stellar age is defined as the average age, weighted by mass}. and the dust opacity of the diffuse dust component in the $V$ band ($\hat{\tau}_\mathrm{dust, 2}$), to name a few. In fact, for every galaxy in our sample, {\tt Prospector} stored the posterior distributions of each free parameter in our physical model. From the posterior distributions we can assess if there are degeneracies among the estimated galaxy properties, and whether the posteriors pile up at the edges of the priors.

In Fig.~\ref{fig:posteriors_q} we show the resulting joint posterior distributions of the main physical quantities, of a handpicked quiescent galaxy with the LEGA-C ID 15. In Fig.~\ref{fig:posteriors_q}, we also present the observational data (photometry and spectroscopy), along with the best-fit model, best-fit values, and the resulted SFH. A similar information is depicted in Figure~\ref{fig:posteriors_sf} but for a handpicked star-forming galaxy with the LEGA-C ID 2808. Looking at the correlations between the joint posteriors we can uncover any degeneracies among the various estimated physical quantities. We determine whether two properties are correlated or not, by calculating the Spearman’s correlation coefficient ($\rho$). If $\rho$ is very close to -1 or 1, then the two properties are strongly anticorrelated or correlated, respectively. If $\rho$ is very close to 0, then there is no correlation between the two properties. The Spearman’s correlation coefficient $\rho$, of each joint posterior, is given in both Fig.~\ref{fig:posteriors_q} and Fig.~\ref{fig:posteriors_sf}.

For the quiescent galaxy in Fig.~\ref{fig:posteriors_q}, we find a very weak correlation or no correlation at all between the joint posteriors. An exception is the moderate correlation of $\hat{\tau}_\mathrm{dust, 2}$ with the sSFR ($\rho = 0.63$), and $Z_\star$ ($\rho = -0.40$). The moderate correlation between $\hat{\tau}_\mathrm{dust, 2}$ with sSFR hints at the underlying degeneracy of the dust attenuation with the SFH. Quiescent galaxies are more susceptible to these types of degeneracies because their red SEDs can result from a combination of factors, including an older stellar population, dust attenuation, and changes in metallicity, making it harder to distinguish between them.

For the star-forming galaxy in Fig.~\ref{fig:posteriors_sf}, we find that the posterior distributions of the main physical quantities are also very well converged. Most of the joint posteriors show either no correlation or a very weak correlation. The dust attenuation slope index $n$ hints at a positive correlation with $\hat{\tau}_\mathrm{dust, 2}$ and $M_\star$, but still remains weak ($\rho < 0.50$). A correlation between the attenuation slope and optical depth is expected \citep{Salim_2020ARA&A..58..529S, Nagaraj_2022ApJ...932...54N} and has been reported in several studies in the past that use attenuation curves derived from SED fitting \citep[e.g.,][]{Arnouts_2013A&A...558A..67A, Kriek_2013ApJ...775L..16K, Salmon_2016ApJ...827...20S, Battisti_2020ApJ...888..108B}. 

Overall, similar results were observed for most of the galaxies in our sample. Specifically, the median Spearman’s correlation coefficient $\rho$ of all the joint posteriors illustrated in Fig.\ref{fig:posteriors_q} and Fig.\ref{fig:posteriors_sf}, for the entire galaxy sample is $\langle \rho \rangle = 0.02^{+0.33}_{-0.07}$. We measure that the joint posteriors of $\hat{\tau}_\mathrm{dust, 2}$--sSFR, $\hat{\tau}_\mathrm{dust, 2}$--$n$, and $M_\star$--$n$ have the highest $\langle \rho \rangle$ values, that is $\langle \rho \rangle = 0.38^{+0.2}_{-0.18}$, $\langle \rho \rangle = 0.34^{+0.22}_{-0.29}$, and $\langle \rho \rangle = 0.39^{+0.17}_{-0.25}$, respectively. From the analysis of the joint posterior distributions we can conclude that we successfully managed to mitigate the dust–age–metallicity degeneracy and properly constrain the main galaxy properties.   

Looking now at the SEDs of both galaxies, it is evident that the model fits the photometric and spectroscopic data exceptionally well. We quantify the quality of the fits through the reduced $\chi^2$ (see Section~\ref{subsubsec:chi2_red}) and the absolute residual values ($\Delta F_\nu = \log \mathrm{obs}/\mathrm{mod}$). Galaxy 15 has a $\chi^2_\nu = 0.85$ and galaxy 2808 has a $\chi^2_\nu = 1.14$. These results are indicative of the goodness of the fits to the observational data. On average and in both cases, the absolute residual values $\Delta F_\nu$ remain within $\pm 0.09$~dex. Again, the exceptional fits to the observations by our model is reassuring of the proper constraining of the main galaxy properties in our sample.  

\subsubsection{Spectral measurements} \label{subsubsec:spectral_comp}

Following the studies of \citet{Wu_2018ApJ...855...85W} and \citet{Nersesian_2024A&A...681A..94N}, we make a comparison between the observed and model spectra in terms of their Lick indices \citep{Worthey_1994ApJS...94..687W}. A mismatch between the observed and synthetic spectral indices, can have a strong impact on the estimates of the physical properties. We primarily focus on the comparison of the strength of two age- and metallicity-sensitive features, the \hda~and Fe4383. We use the Lick indices and their associated uncertainties from the updated LEGA-C catalog \citep{van_der_Wel_2021ApJS..256...44V}, measured from the emission-corrected spectra (Gallazzi et al., in prep.). The updated Lick index measurements are consistent with the DR3 values, but the uncertainties are a factor 1.3 lower (for more information see Gallazzi et al., in prep.). 

As in \citet{Nersesian_2024A&A...681A..94N}, we measure the Lick indices of the modeled absorption spectra from {\tt Prospector}, with the {\tt python} package {\tt pyphot}.\footnote{\url{https://github.com/mfouesneau/pyphot}} The values of the absorption lines are derived from the best-fit spectrum. The corresponding 16$^\mathrm{th}$--84$^\mathrm{th}$ percentile uncertainties are estimated by drawing 500 spectra weighted by the {\tt dynesty} weights, and measuring the values of the absorption lines from these 500 spectra. In our sample, 2796 galaxies have a reliable \hda~measurement, and 2580 galaxies have a Fe4383 measurement.  

In Fig.~\ref{fig:hd_fe4383_obs_vs_prd}, we show the comparison between observed and modeled Lick indices of the two aforementioned absorption features. In both panels, we see a clear division between quiescent and star-forming galaxies. Typically, star-forming galaxies showcase a strong \hda~absorption line (\hda~$> 2$~\AA) in their spectra, while quiescent galaxies tend to be characterized by weak \hda~absorption (\hda~$< 2$~\AA) \citep[e.g.,][]{Kauffmann_2003MNRAS.341...33K, Kauffmann_2003MNRAS.341...54K}. Inversely, quiescent galaxies tend to be metal-rich (Fe4383~$> 2$~\AA), while star-forming galaxies tend to be metal-poor (low Fe4383). The bimodality seen in the observed spectral features is also reproduced in the distribution of modeled absorption line strengths \citep{Straatman_2018ApJS..239...27S, Wu_2018ApJ...855...85W, Nersesian_2024A&A...681A..94N}. 

In the left panel of Fig.~\ref{fig:hd_fe4383_obs_vs_prd} we present the comparison for the \hd~absorption line. We find a very strong correlation between the observed and modeled \hda~feature strength, for both galaxy populations ($\rho = 0.87$ for the quiescent and $\rho = 0.85$ for the star forming). Moreover, we only report small systematic offsets, -0.49~\AA~for the star-forming galaxies and 0.26~\AA~for the quiescent ones. \citet{Nersesian_2024A&A...681A..94N} showed that when predicting the spectral indices by fitting only the photometric SEDs, the \hda~values of star-forming galaxies follow a non-unity slope, with high-\hda~galaxies having underestimated \hda~values. Furthermore, \citet{Nersesian_2024A&A...681A..94N} reported a strong systematic offset of $\sim 0.85$~\AA~for the quiescent galaxies, reminiscent of the offset between simulated synthetic spectra and LEGA-C spectra analyzed by \citet{Wu_2021AJ....162..201W}. For the full sample, we measure a mean offset of 0.69~\AA~and a scatter $\sigma = 1.52$~\AA. The inclusion of spectroscopy in our fits resulted in a better constrain of the \hda~absorption line for both galaxy populations. The very good agreement between the observed and modeled \hda~is reassuring for the constrains on the stellar ages (both light-weighted and mass-weighed ages).

In the right panel of Fig.~\ref{fig:hd_fe4383_obs_vs_prd} we look for any systematic differences between the observed and the modeled Fe4383 feature. We find a moderate to strong correlation for both galaxy populations ($\rho \sim 0.60$) but with an increased scatter compared to the \hda~absorption line. \citet{Nersesian_2024A&A...681A..94N} measured a strong systematic offset for Fe4383 (1.21~\AA), from the best-fit spectra to the photometric SEDs. Here, thanks to the inclusion of spectroscopy in our SED fitting, we find that the median offsets from the unity slope are very small ($\le 0.05$~\AA), for both galaxy populations. We measure a mean offset of 0.28~\AA~and a scatter $\sigma = 1.28$~\AA~for the entire sample. This means that the metallicity estimates are better constrained when spectroscopic information is included in the SED fitting process. However, we should note that the increased scatter compared to \hda, and the moderate correlation for the Fe4383 feature strength is indicative of how difficult is to accurately measure the stellar metallicity, even when spectroscopy is included.   

To assess whether the scatter and correlation coefficient $\rho$ in Fig.~\ref{fig:hd_fe4383_obs_vs_prd} are due to random or systematic uncertainties, we calculated the standard error of the mean (SEM) using the following equation:

\begin{equation} \label{eq:standard_error}
\\\\\ \mathrm{SEM} = \sqrt{\frac{\sum_{i=1}^N \sigma_\mathrm{mod, i}^2}{N}},
\end{equation}

\noindent where $N$ represents the number of galaxies per Lick index, and $\sigma_\mathrm{mod, i}$ are the model uncertainties. We find that the SEM value for \hda~is 0.1~\AA, and for Fe4383 is 0.09~\AA. These SEM values are significantly smaller than the previously reported scatter $\sigma$ (i.e., 1.52~\AA~and 1.28~\AA, respectively). This suggests that the differences we observe between the model and observed indices are primarily driven by systematic effects rather than random uncertainties.

Lastly, we measure the Lick indices for 14 more absorption lines. Overall, we find similar strong correlation and very small systematic offsets for most of the measured indices from the model spectra, further supporting that more information is better to accurately measure the stellar properties of galaxies. We present the comparisons of the rest of the Lick indices in Appendix~\ref{ap:A}. 

\subsubsection{Gain in fitting both photometry and spectroscopy} \label{subsubsec:specphot_gain}

In this section, we would like to provide a detailed assessment on the benefits in fitting simultaneously photometric and spectroscopic observations of galaxies. Many studies use multiwavelength broadband photometry to infer the physical properties of galaxies from SED fitting. \citet{Pforr_2012MNRAS.422.3285P} showed that broadband photometry, with a coverage from the restframe UV to the restframe NIR, is sufficient to resolve the stellar properties without the need of spectroscopic observations. Furthermore, many photometric surveys, both at low and high redshift, have increased their spectral coverage in the optical and NIR wavelength regime by including a great number of broadband and narrowband filters. For instance, the COSMOS survey \citep{Scoville_2007ApJS..172....1S, Laigle_2016ApJS..224...24L, Weaver_2022ApJS..258...11W}, or the PAU/J-PAS \citep{Benitez_2009ApJ...691..241B, Abramo_2012MNRAS.423.3251A} contain more than 20 wavebands, that in principle should enable more reliable estimates of the stellar properties of galaxies. \citet{Nersesian_2024A&A...681A..94N} showed that, regardless of their spectral coverage, photometric SEDs cannot resolve the various spectral features that could potentially constrain the age and metallicity of stellar populations. The recent studies by \citet{Johnson_2021ApJS..254...22J} and \citet{Tacchella_2022ApJ...926..134T} provide significant evidence on the constraining of the galaxy parameters when combining spectra with photometry. 

To asses the gain in fitting simultaneously photometry and spectroscopy, we performed two more SED fitting runs with {\tt Prospector} in addition to our main (or fiducial) run that was presented in Section~\ref{subsec:physical_model}. These two new SED fitting runs include the photometry only, and spectroscopy only fits. The same physical model and prior functions were assumed in all those fits (see Table~\ref{tab:free_params_and_priors}), but with a few differences. Specifically, we have not used the dust emission model in the spectroscopy-only case, while in the photometry-only case, we changed the stellar metallicity prior function. Instead of a flat prior in logarithmic space, we used a flat prior in linear space. The reason is that when a flat prior in logarithmic space is given, lower metallicity values are given a higher weight. When fitting photometry alone, a broader parameter space needs to be explored, with broader steps for the stellar metallicities. Although, a mass-metallicity prior from empirical relations from Local Universe studies \citep[e.g.,][]{Bundy_2015ApJ...798....7B} could potentially improve the metallicity measurements, we argue that it could also impose a bias on the mass-metallicity relation at the redshift epoch we are studying.  

\begin{figure}[ht!]
    \centering
    \includegraphics[width=\columnwidth]{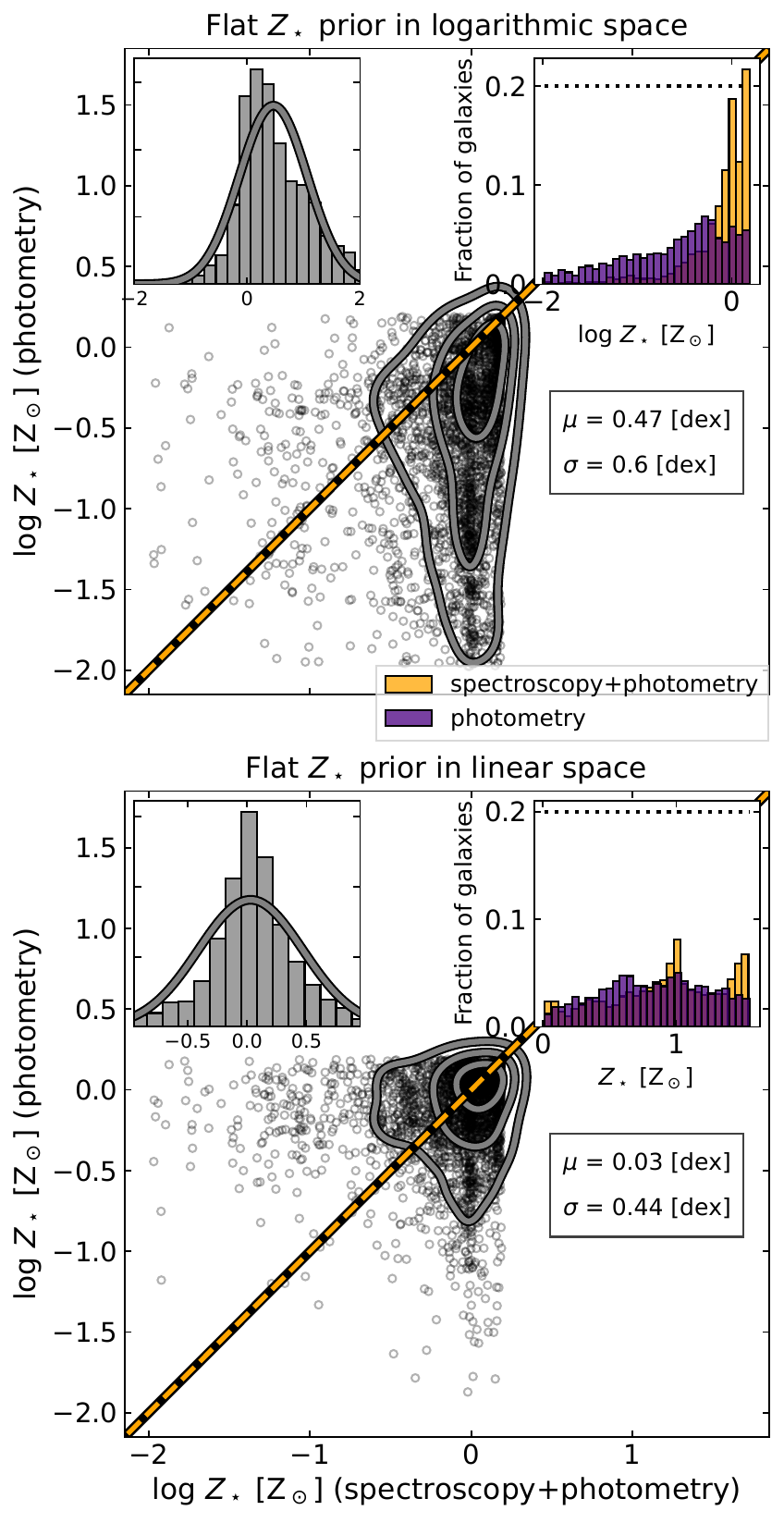}
    \caption{Comparison of stellar metallicities using different prior functions. The stellar metallicities obtained from spectrophotometric modeling ($x$-axis) are compared with those derived from photometry-only fits ($y$-axis), using a flat prior for $Z_\star$ in logarithmic space (top panel) and linear space (bottom panel). The dashed orange line represents the one-to-one relationship, while the contours enclose 20\%, 50\% and 80\% of the total data. The inset in the top-left corner shows the residual distribution, $\log \left(Z_\mathrm{\star, spec+phot} / Z_\mathrm{\star, phot}\right)$. Meanwhile, the inset in the top-right corner compares the fiducial $Z_\star$ from spectrophotometric modeling (orange distribution) with the photometry-only results (purple distribution). The dotted lines correspond to the prior probability distributions, scaled by an arbitrary factor for clarity. The mean offset and scatter between the two distributions are reported in each panel.}
    \label{fig:zstar_prior_comp}
\end{figure}

\begin{figure*}[t]
    \centering
    \includegraphics[width=\textwidth]{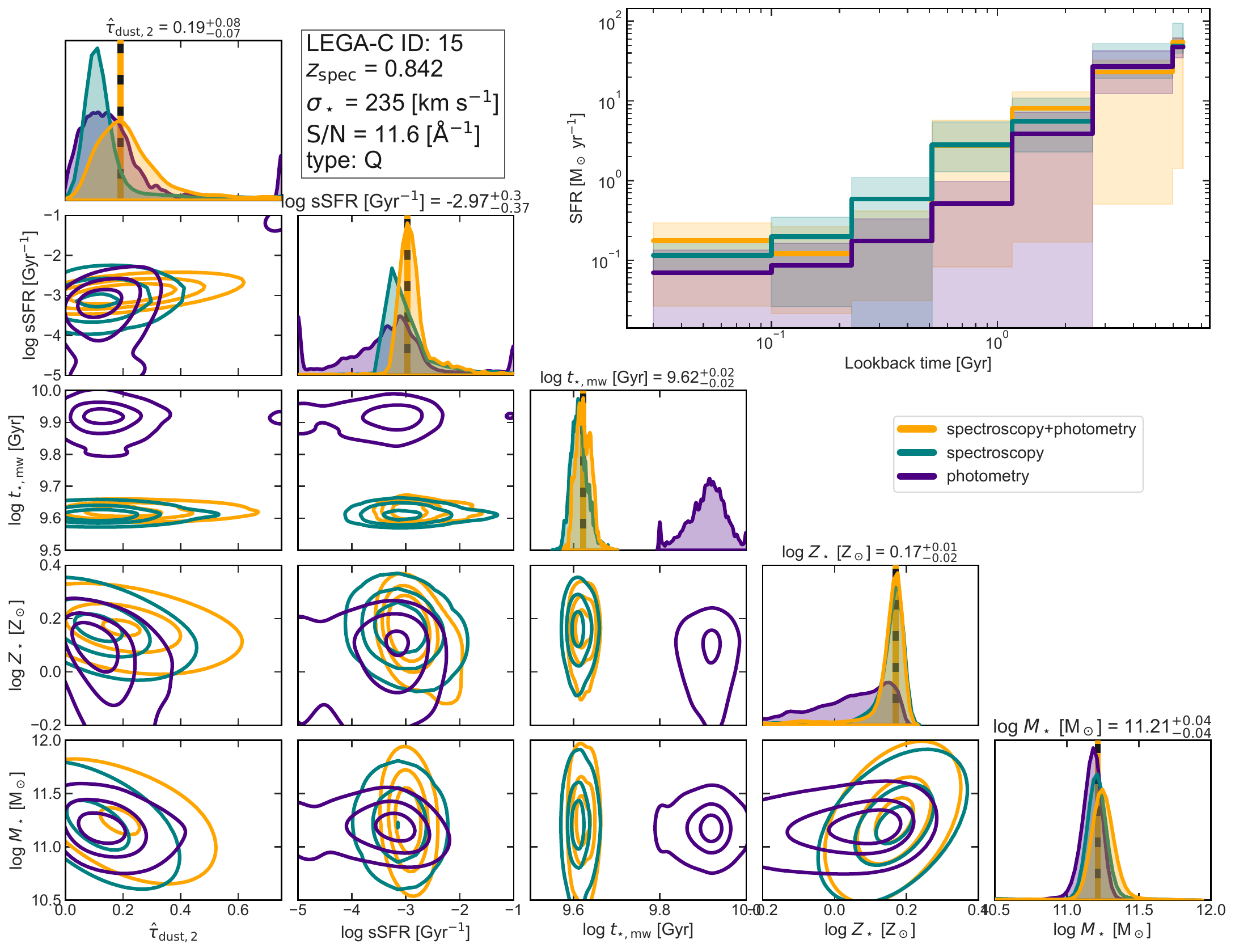}
    \caption{Joint posterior distributions of various physical quantities and SFHs retrieved from three different SED fitting runs, for an example quiescent galaxy at $z_\mathrm{spec} = 0.842$. In the cornerplot, we present the posterior distributions of $\hat{\tau}_\mathrm{dust, 2}$, sSFR, $t_\mathrm{\star, mw}$, $Z_\star$, and $M_\star$. We compare the posterior distributions of our fiducial run (spectroscopy$+$photometry, orange) with those from the spectroscopy-only (teal) and photometry-only (purple) run. The contours enclose 20\%, 50\% and 80\% of the total data, of each fitting run. The vertical black dashed line indicates the best-fit value of each physical quantity from our fiducial run. The inset in the right corner of the figure contains the SFH posteriors for each fitting run.}
    \label{fig:posteriors_q_}
\end{figure*}

In Fig.~\ref{fig:zstar_prior_comp}, we demonstrate that using a flat prior in linear space for $Z_\star$ in the photometry-only case results in measured metallicities that better align with those from our fiducial run. When a flat prior in linear space is used for $Z_\star$, the mean offset ($\mu$) from the fiducial $Z_\star$ is considerably reduced, from 0.47~dex to 0.03~dex. The scatter ($\sigma$) around the mean is also reduced by 0.16~dex. This improvement in the $Z_\star$ measurements does not necessarily imply a better constraint on the spectral absorption features. In fact, upon comparing the observed and model spectra from the photometry-only run, we still find significant discrepancies and offsets in terms of their \hda~and Fe4383 Lick indices, similar to the results shown in \citet{Nersesian_2024A&A...681A..94N}.

For completeness, we tested whether altering the $Z_\star$ prior function affects the results of the spectrophotometric modeling. We conducted an additional run by fitting the combined photometry and spectroscopy, using the same physical model as our fiducial run, but with the $Z_\star$ prior function sampled in linear space. We found a very good match compared to the stellar metallicities of the fiducial run, with a mean offset of 0.09~dex and a standard deviation of 0.25~dex. The measurement of the other physical properties appears to be largely unaffected by the change in the $Z_\star$ prior distribution. On average, there is only a small offset in all other properties, approximately 0.03 dex, with an average standard deviation of 0.3~dex. Thus, this adjustment in the $Z_\star$ prior function seems to have a negligible effect on the measurement of the other stellar properties.

In summary, we conclude that photometry-only fits with a linear $Z_\star$ prior do not yield meaningful constraints on $Z_\star$ for individual galaxies, though they effectively avoid a downward bias. While the precision remains limited, the accuracy is generally good. We also note that the results are highly sensitive to the choice of prior. However, when spectroscopy is included, the dependence on the prior is reduced--though not entirely eliminated--due to the additional information provided that helps constrain the stellar metallicity.

Now, we show how our results differ when fitting only photometry, only spectroscopy, and both photometry and spectroscopy together. Figures~\ref{fig:posteriors_q_}~and~\ref{fig:posteriors_sf_} compare the posterior distributions from the three SED fitting runs with {\tt Prospector}, of the quiescent (LEGA-C ID 15) and star-forming galaxy (LEGA-C ID 2808) respectively. From the posterior distributions shown in both figures we notice that the estimates of the physical parameters between our fiducial run and the spectroscopy-only run are consistent, especially for the stellar parameters ($Z_\star$, $t_\mathrm{\star, mw}$, sSFR). This is indicative of spectroscopy driving the properties of the stellar populations in our fitting. We also confirm that broadband SED fitting remains a reliable method for measuring stellar mass \citep{Gallazzi_2005MNRAS.362...41G, Conroy_2013ARA&A..51..393C}. Focusing on the dust attenuation $\hat{\tau}_\mathrm{dust, 2}$, we notice that the posteriors of all fitting runs return a similar value for the quiescent galaxy (low attenuation), whereas for the star-forming galaxy (high attenuation) the $\hat{\tau}_\mathrm{dust, 2}$ posterior distribution of our fiducial run lies in between the spectroscopy-only and photometry-only runs. This suggests that when estimating the properties of dusty galaxies using SED fitting, UV-MIR broadband photometry remains crucial for constraining dust properties.

On the contrary, photometry alone cannot constrain the mass-weighted stellar ages at all. In both examples, we find that the mass-weighted ages are mainly driven by spectroscopy, while the $t_\mathrm{\star, mw}$ posteriors of the photometry-only case return systematically older mass-weighted stellar ages. The older mass-weighted ages indicate a relatively longer formation phase for the LEGA-C galaxies. This is further confirmed by looking at the SFHs of these two galaxies, as reconstructed from the three different fitting runs. In both figures, we find that the SFH from the spectroscopy-only run is largely consistent with our fiducial run. For the quiescent galaxy (Fig.~\ref{fig:posteriors_q_}), the shape of the photometry-only SFH seems to be in good agreement with the shape of the other two SFHs, yet systematically lower by $\sim 0.22$~dex on average. In the case of the star-forming galaxy (Fig.~\ref{fig:posteriors_sf_}), we find a vast disagreement between the shape of the SFH of the photometry-only run and the SFH shapes of the other two runs, especially at a lookback time greater than 0.5~Gyr, when SFR is approximately lower by $\sim 0.21$~dex. This is not surprising, since the information provided by photometry is not sufficient to resolve the star-formation activity of a galaxy at earlier times. Overall, the consistently lower SFR in the photometry-only case align with the systematically older mass-weighted ages, suggesting a more extended formation period.

\begin{figure*}[t]
    \centering
    \includegraphics[width=\textwidth]{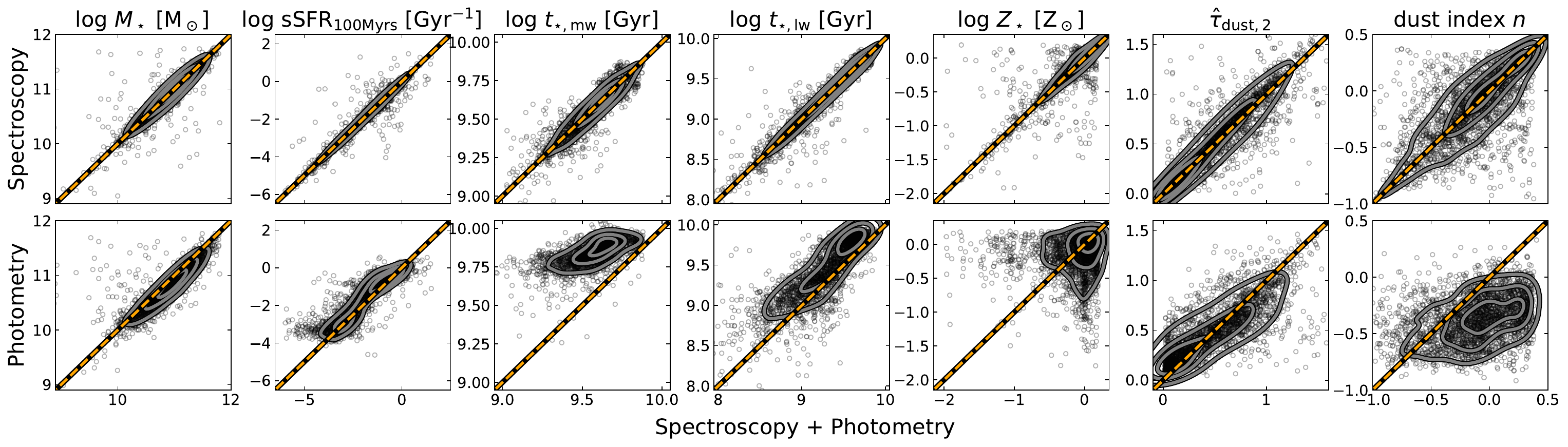}
    \caption{Comparison of the main global properties of our primary galaxy sample as retrieved from three different SED fitting runs. The properties presented here are the stellar mass ($M_\star$), the average specific SFR over the last 100~Myrs (\ssfr), the mass-weighted stellar age ($t_\mathrm{\star, mw}$), the light-weighted stellar age ($t_\mathrm{\star, lw}$) weighted by the bolometric luminosity, the stellar metallicity ($Z_\star$), the dust opacity of the diffuse dust component ($\hat{\tau}_\mathrm{dust, 2}$) in the $V$ band, and the dust attenuation slope index ($n$). We compare the distributions of our fiducial run with those from the spectroscopy-only (top row), and photometry-only (bottom row) run. Contours enclose 20\%, 50\% and 80\% of the total data. The dashed orange line shows the one-to-one relation.}
    \label{fig:sample_props_}
\end{figure*}

\begin{figure*}[h!]
    \centering
    \includegraphics[width=\textwidth]{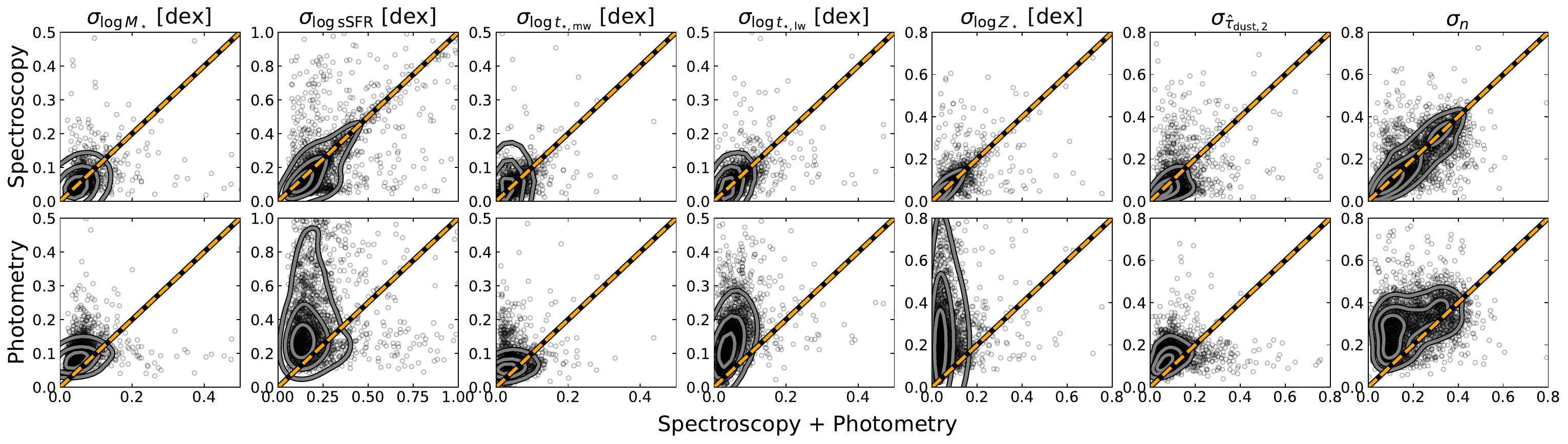}
    \caption{Comparison of the upper uncertainties of the main global properties of our primary galaxy sample from three different SED fitting runs. The properties presented here are the same as in Fig.~\ref{fig:sample_props_}. The top row shows the comparison of the uncertainties of our fiducial run with those from the spectroscopy-only run, and the bottom row shows the comparison with the photometry-only run. Contours enclose 20\%, 50\% and 80\% of the total data. The dashed orange line shows the one-to-one relation.}
    \label{fig:sample_props_errors}
\end{figure*}

A comparison between the derived physical properties of the whole sample, from the three fitting runs, can shed more light on the systematic differences. Figure~\ref{fig:sample_props_} shows the galaxy property distributions as retrieved from the analysis of the three fitting runs, when combining photometry and spectroscopy or when using only spectroscopy, and only photometry. The top row compares the parameters from our fiducial run to those from the spectroscopy-only run, while the bottom row compares them to the photometry-only run.   

The measured properties from the spectroscopy-only run are almost identical to our fiducial run, without evident offsets. The spectral index is notably the only parameter that presents a significant deviation from the mean ($\sigma = 0.25$~dex). Focusing on stellar metallicity, we observe an excellent agreement between the two fitting runs, though a few galaxies exhibit noteworthy deviations. Further inspection of these sources reveals that most of them are missing the \ion{Mg}{B} triplet at 5176~\AA, and the Fe lines at 5270~\AA, 5335~\AA, and 5406~\AA. This suggests that to accurately constrain stellar metallicity, it is necessary to include the Mg and all Fe lines (see Gallazzi et al., in prep. for a full discussion). 

\begin{figure*}[t]
    \centering
    \includegraphics[width=\textwidth]{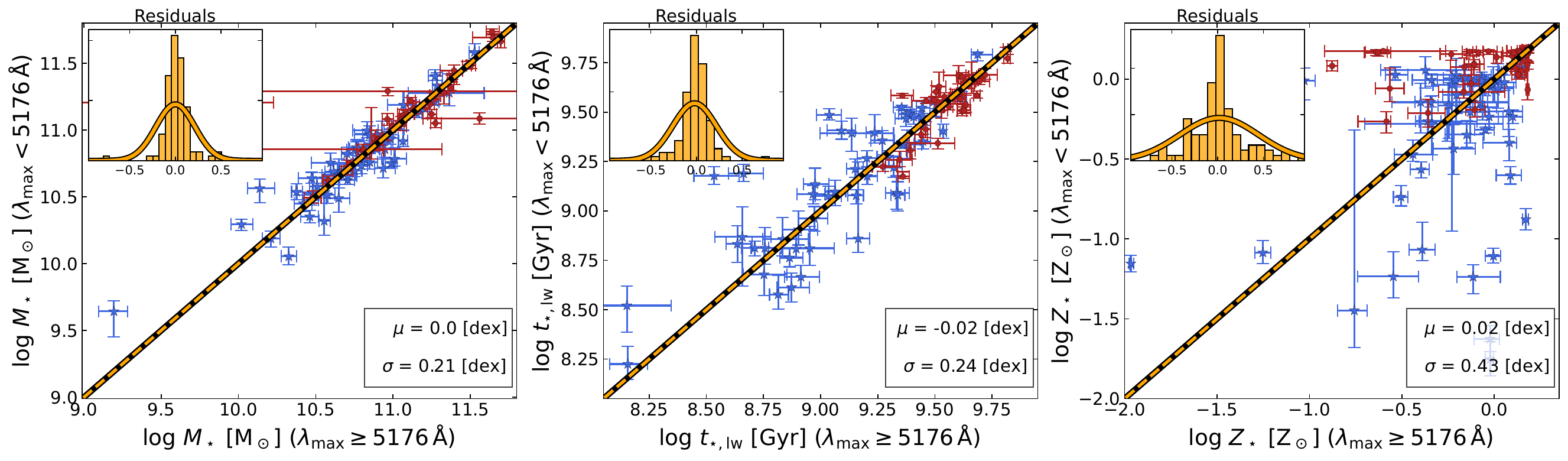}
    \caption{Comparison of the main stellar population properties of galaxies with duplicate observations in LEGA-C DR3. Galaxies are color-coded by their $UVJ$ diagram classification as star forming (blue stars) and quiescent (red points). The dashed orange line shows the one-to-one relation. The distribution of residuals is shown in the top-left corner of each panel, while the statistics of the mean offset ($\mu$) and variance ($\sigma$) are shown in the bottom-right corner.}
    \label{fig:comp_of_duplicates}
\end{figure*}

When comparing our fiducial run to the photometry-only run, we find small systematic differences for some of the parameters (e.g., $M_\star$, sSFR), while in other cases photometry seems to constrain the parameters quite well (e.g., $\hat{\tau}_\mathrm{dust, 2}$). Larger discrepancies are found for the ages of the stellar populations. We find a strong systematic offset for both the light-weight ages (0.2~dex) and the mass-weighted ages (0.28~dex). Fitting only the photometry results in the estimation of older stellar ages for the galaxies, or in other words, someone would infer that galaxies have built up half of their stellar mass at earlier epochs than the spectroscopy suggests.

Lastly, Fig.~\ref{fig:sample_props_errors} compares the estimated uncertainties when combining spectroscopy with photometry versus using spectroscopy alone (top row), and photometry alone (bottom row). The uncertainties of each galaxy in Fig.~\ref{fig:sample_props_errors} are computed from the difference between the $84^\mathrm{th}$ and $50^\mathrm{th}$ percentiles of the posteriors (upper uncertainty). We observe notable uncertainties in the sSFR, and to a certain extent, in the dust attenuation parameters. Regarding $Z_\star$, we find again that galaxies that are missing the Mg and Fe lines in their spectrum have larger uncertainties. 

When relying solely on photometry, significant uncertainties are observed in all parameters. In particular, larger uncertainties are found for the sSFR, light-weighted age, stellar metallicity, and dust attenuation index. The uncertainties in stellar mass and dust opacity, although comparable to those obtained from combined spectroscopy and photometry, remain systematically larger. From these results we can safely conclude and stress that pairing spectroscopy with photometry is required to provide meaningful constraints on age, metallicity, and SFH.   

\subsubsection{Comparing the stellar properties recovered from duplicate observations} \label{subsubsec:comp_phys_dupes}

Due to differences in the slit alignment, many of the LEGA-C galaxies end up having duplicate observations, often with differences in emission-line strength, kinematics, and wavelength coverage. In our sample we have 279 galaxies with duplicate observations, 160 star forming and 119 quiescent. An interesting exercise is to compare the stellar properties of these objects with duplicate observations, and test whether it is important to have a full spectral coverage or not. 

Figure~\ref{fig:comp_of_duplicates} shows a comparison of the three main stellar properties, namely $M_\star$, $t_\mathrm{\star, lw}$, and $Z_\star$, for galaxies with duplicate observations. We segregate the observations based on the wavelength coverage, and in particular whether the \ion{Mg}{B} absorption line is absent or not (i.e., $\lambda_\mathrm{max} < 5176~\mathrm{\AA}$ or $\lambda_\mathrm{max} \ge 5176~\mathrm{\AA}$). Out of the 279 galaxies with duplicate observations in our primary sample, 90 meet the wavelength coverage criterion, 52 star forming and 38 quiescent. Star-forming galaxies are depicted with blue stars, and quiescent galaxies with red points. We also calculate the absolute residuals of each parameter, the distribution of which is shown as an inset panel.

Overall, it is clear that there is an excellent agreement between the estimated properties, with the mean value of the residual distribution taking values very close to 0. This is particularly true for the quiescent galaxies, that show very little scatter. On the other hand, star-forming galaxies show more scatter, especially in their age and metallicity estimates. We also notice that the uncertainties for the observations with absent Mg and Fe absorption lines are in general larger. From this exercise we can conclude that regardless of the spectral coverage, the stellar mass and stellar age of galaxies can be systematically retrieved within 0.21~dex and 0.24~dex, respectively. Stellar metallicity is as expected more challenging, and to be able to constrain it properly a full spectral coverage is necessary (see also Gallazzi et al., in prep.).

\section{Fitting results} \label{sec:results}

In this section, we present the data products of SED fitting with {\tt Prospector}, for our fiducial run. Here, we would like to remind the readers that the following analysis was performed for the primary galaxy sample (2908 galaxies) as defined in Section~\ref{subsec:final_sample}. We showcase the distributions of the most important global properties of the galaxies in our sample, and provide the median values of those properties in bins of stellar mass (Section~\ref{subsec:sample_props}). Then, we briefly describe the scaling relations between stellar velocity dispersion ($\sigma_\star$), stellar age, and stellar metallicity (Section~\ref{subsec:scaling_relations}). Finally, we show the $UVJ$ and SFR--$M_\star$ diagrams color-coded by several key physical quantities, to highlight the underlying scaling relations between age, metallicity and dust (Section~\ref{subsec:uvj} and Section~\ref{subsec:sfms}). In Appendix~\ref{ap:B}, the main stellar properties derived in this work are compared with those from \citet{Cappellari_2023MNRAS.526.3273C}.

\subsection{Derived physical properties} \label{subsec:sample_props}

\begin{figure*}[t]
    \centering
    \includegraphics[width=\textwidth]{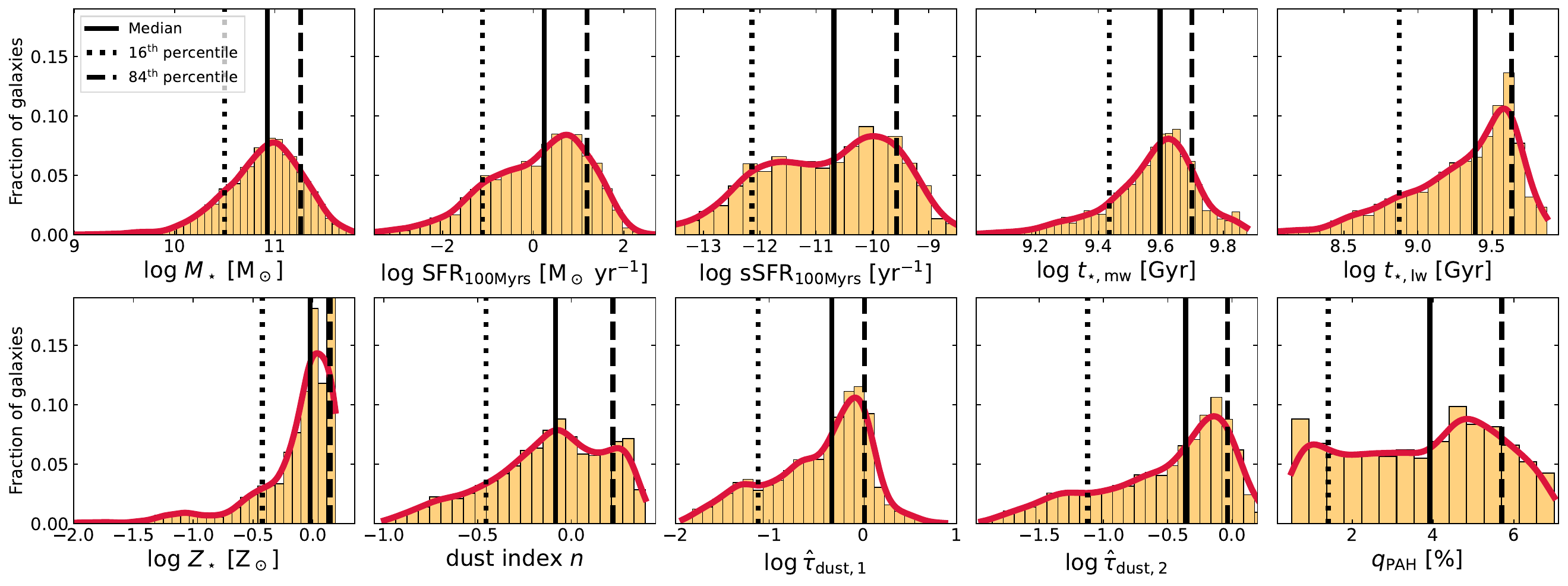}
    \caption{Distributions of the main global properties of our primary galaxy sample. The properties presented here are the stellar mass ($M_\star$), the average SFR over the last 100~Myrs (SFR$_\mathrm{100Myrs}$), the specific SFR$_\mathrm{100Myrs}$ (sSFR$_\mathrm{100Myrs}$), the mass-weighted stellar age ($t_\mathrm{\star, mw}$), the light-weighted stellar age ($t_\mathrm{\star, lw}$) weighted by the bolometric luminosity, the stellar metallicity ($Z_\star$), the dust attenuation slope index ($n$), the dust opacity of the birth-clouds ($\hat{\tau}_\mathrm{dust, 1}$) and diffuse dust component ($\hat{\tau}_\mathrm{dust, 2}$) in the $V$ band, and the PAH mass fraction ($q_\mathrm{PAH}$). The KDE distribution is shown in red, while the solid black line shows the median value. The black dotted and dashed lines indicate the 16$^\mathrm{th}$ and 84$^\mathrm{th}$ percentiles of each distribution, respectively.}
    \label{fig:sample_props}
\end{figure*}

\begin{table*}[t]
\caption{Median values of the main physical properties of our primary sample of LEGA-C galaxies, in different stellar mass bins.}
\begin{center}
\scalebox{0.85}{
\begin{threeparttable}
\begin{tabular}{lr||cccccccc|}
\hline 
\hline 
Mass bin & $N_\mathrm{obj}$ & $\log \langle M_\star \rangle$ &
$\log \langle$\sfr $\rangle$ &
$\log \langle$\ssfr $\rangle$ &
$\langle t_\mathrm{\star, mw} \rangle$ &
$\langle t_\mathrm{\star, lw} \rangle$ &
$\log \langle Z_\star \rangle$ &
$\langle \hat{\tau}_\mathrm{dust, 2} \rangle$ &
dust index $\langle n \rangle$ \\
$\log(M_\star)$ &  & [M$_{\odot}$] & [M$_{\odot}$/yr] & [yr$^{-1}$] & [Gyr] & [Gyr] & [Z$_{\odot}$] &  & \\
\hline 
\hline\\ 
\multicolumn{9}{c}{\textbf{Full sample}}\\ 
\hline 
$\left[9,10.5\right)$  & $469$  & $10.31 \pm 0.01$ & $0.61 \pm 0.05$  & $-9.65  \pm 0.05$ & $3.17 \pm 0.06$ & $0.87 \pm 0.06$ & $-0.13 \pm 0.02$ & $0.65 \pm 0.03$ & $-0.08 \pm 0.02$ \\ 
$\left[10.5,11\right)$  & $1233$  & $10.80 \pm 0.01$ & $0.32 \pm 0.03$  & $-10.46  \pm 0.04$ & $3.81 \pm 0.03$ & $2.10 \pm 0.04$ & $-0.04 \pm 0.01$ & $0.50 \pm 0.01$ & $-0.07 \pm 0.01$ \\ 
$\left[11,11.5\right)$  & $1103$  & $11.19 \pm 0.01$ & $-0.06 \pm 0.03$  & $-11.25  \pm 0.03$ & $4.23 \pm 0.03$ & $3.31 \pm 0.05$ & $0.02 \pm 0.01$ & $0.27 \pm 0.01$ & $-0.11 \pm 0.01$ \\ 
$\left[11.5,12\right)$  & $103$  & $11.59 \pm 0.01$ & $0.12 \pm 0.09$  & $-11.52  \pm 0.09$ & $4.35 \pm 0.10$ & $3.69 \pm 0.13$ & $0.13 \pm 0.02$ & $0.27 \pm 0.04$ & $-0.16 \pm 0.04$ \\ 
$\left[9,12\right)$  & $2908$  & $10.93 \pm 0.01$ & $0.25 \pm 0.02$  & $-10.68  \pm 0.02$ & $3.96 \pm 0.02$ & $2.44 \pm 0.03$ & $-0.02 \pm 0.01$ & $0.44 \pm 0.01$ & $-0.09 \pm 0.01$ \\ 
\hline 
\hline\\ 
\multicolumn{9}{c}{\textbf{Star-forming}}\\ 
\hline 
$\left[9,10.5\right)$  & $406$  & $10.31 \pm 0.01$ & $0.75 \pm 0.03$  & $-9.56  \pm 0.03$ & $3.05 \pm 0.06$ & $0.71 \pm 0.04$ & $-0.17 \pm 0.02$ & $0.68 \pm 0.03$ & $-0.05 \pm 0.02$ \\ 
$\left[10.5,11\right)$  & $778$  & $10.78 \pm 0.01$ & $0.81 \pm 0.02$  & $-9.99  \pm 0.02$ & $3.44 \pm 0.03$ & $1.39 \pm 0.04$ & $-0.08 \pm 0.01$ & $0.72 \pm 0.01$ & $0.02 \pm 0.01$ \\ 
$\left[11,11.5\right)$  & $481$  & $11.16 \pm 0.01$ & $0.89 \pm 0.03$  & $-10.31  \pm 0.03$ & $3.76 \pm 0.04$ & $2.08 \pm 0.05$ & $-0.04 \pm 0.02$ & $0.66 \pm 0.02$ & $0.11 \pm 0.01$ \\ 
$\left[11.5,12\right)$  & $35$  & $11.59 \pm 0.01$ & $1.06 \pm 0.13$  & $-10.54  \pm 0.13$ & $4.02 \pm 0.16$ & $2.74 \pm 0.20$ & $0.02 \pm 0.03$ & $0.54 \pm 0.08$ & $0.16 \pm 0.06$ \\ 
$\left[9,12\right)$  & $1700$  & $10.81 \pm 0.01$ & $0.80 \pm 0.02$  & $-9.98  \pm 0.02$ & $3.46 \pm 0.03$ & $1.39 \pm 0.03$ & $-0.08 \pm 0.01$ & $0.69 \pm 0.01$ & $0.02 \pm 0.01$ \\ 
\hline 
\hline\\ 
\multicolumn{9}{c}{\textbf{Quiescent}}\\ 
\hline 
$\left[9,10.5\right)$  & $63$  & $10.37 \pm 0.04$ & $-1.59 \pm 0.12$  & $-11.77  \pm 0.12$ & $4.68 \pm 0.15$ & $3.74 \pm 0.19$ & $-0.01 \pm 0.04$ & $0.33 \pm 0.15$ & $-0.27 \pm 0.04$ \\ 
$\left[10.5,11\right)$  & $455$  & $10.83 \pm 0.01$ & $-1.11 \pm 0.04$  & $-11.97  \pm 0.04$ & $4.52 \pm 0.05$ & $3.96 \pm 0.06$ & $0.02 \pm 0.01$ & $0.12 \pm 0.02$ & $-0.17 \pm 0.01$ \\ 
$\left[11,11.5\right)$  & $622$  & $11.21 \pm 0.01$ & $-0.66 \pm 0.03$  & $-11.89  \pm 0.03$ & $4.57 \pm 0.04$ & $4.01 \pm 0.05$ & $0.12 \pm 0.01$ & $0.11 \pm 0.01$ & $-0.24 \pm 0.01$ \\ 
$\left[11.5,12\right)$  & $68$  & $11.59 \pm 0.01$ & $-0.14 \pm 0.09$  & $-11.78  \pm 0.09$ & $4.63 \pm 0.11$ & $4.05 \pm 0.12$ & $0.16 \pm 0.03$ & $0.16 \pm 0.05$ & $-0.22 \pm 0.04$ \\ 
$\left[9,12\right)$  & $1208$  & $11.05 \pm 0.01$ & $-0.86 \pm 0.03$  & $-11.91  \pm 0.02$ & $4.55 \pm 0.03$ & $3.99 \pm 0.03$ & $0.09 \pm 0.01$ & $0.12 \pm 0.01$ & $-0.22 \pm 0.01$ \\ 
\hline \hline
\end{tabular}
\tablefoot{The properties presented here are the stellar mass ($M_\star$), the average SFR over the last 100~Myrs (SFR$_\mathrm{100Myrs}$), the mass-weighted stellar age ($t_\mathrm{\star, mw}$), the light-weighted stellar age ($t_\mathrm{\star, lw}$) weighted by the bolometric luminosity, the stellar metallicity ($Z_\star$), the diffuse dust component in the $V$ band ($\hat{\tau}_\mathrm{dust, 2}$), and the dust attenuation slope index ($n$). For every quantity, we provide the statistical uncertainties computed as $\sigma/\sqrt N$, where $N$ is the number of galaxies within the stellar mass bin.}
\end{threeparttable}}
\label{tab:phys_param}
\end{center}
\end{table*}

Figure~\ref{fig:sample_props} shows the distributions of the main global properties of our primary galaxy sample. Each property was estimated based on the median ($50^\mathrm{th}$ percentile) of its posterior distribution. The depicted properties are the stellar mass ($M_\star$), the average SFR over the last 100~Myrs (\sfr), the specific SFR$_\mathrm{100Myrs}$ ($\mathrm{sSFR}_\mathrm{100Myrs} = \mathrm{SFR}_\mathrm{100Myrs}/M_\star$), the mass-weighted stellar age ($t_\mathrm{\star, mw}$), the light-weighted stellar age ($t_\mathrm{\star, lw}$) weighted by the bolometric luminosity, the stellar metallicity ($Z_\star$), the diffuse dust component in the $V$ band ($\hat{\tau}_\mathrm{dust, 2}$), the dust attenuation slope index ($n$), and the PAH mass fraction ($q_\mathrm{PAH}$).\footnote{We should note here that the current estimates of $q_\mathrm{PAH}$ are solely based on the 24~$\mu$m emission (or restframe $\sim 12~\mu$m). Acquiring mid-infrared observations with JWST in the COSMOS field, can help constraining the PAH strength for galaxies at high-redshift.} covering the redshift range 0.6--1 (i.e., 2~Gyr in time). We measure a median stellar mass of $\log \langle M_\star / \mathrm{M}_\odot \rangle = 10.93$, with a wide range of SFRs, from passive galaxies ($\log \langle$ \sfr /$\mathrm{M}_{\odot}~\mathrm{yr}^{-1} \rangle < -0.86$) to highly star-forming ($\log \langle$ \sfr /$\mathrm{M}_{\odot}~\mathrm{yr}^{-1} \rangle > 0.80$). We measure a median value of $\log \langle$ \sfr /$\mathrm{M}_{\odot}~\mathrm{yr}^{-1} \rangle = 0.25$, indicating that most galaxies in our sample are actively forming new stars. Furthermore, we measure: $\langle t_\mathrm{\star, mw} \rangle = 3.96$~Gyr, and $\log \langle Z_\star / \mathrm{Z}_\odot \rangle = -0.02$. The median values of the main physical properties along with their statistical uncertainties computed as $\sigma/\sqrt N$, where $N$ is the number of galaxies within the stellar mass bin, are given in Table~\ref{tab:phys_param} for the primary sample (2908), star-forming (1700), and quiescent (1208) galaxies. In Table~\ref{tab:phys_param}, we also provide the median values by dividing our primary sample in four different mass bins.

A closer look at the median values of both galaxy populations in Table~\ref{tab:phys_param}, reveals that galaxies classified as quiescent, based on their $UVJ$ colors, are characterized by very low star formation ($\log \langle $\sfr /$\mathrm{M}_{\odot}~\mathrm{yr}^{-1} \rangle = -0.86$) and high metallicity ($\log \langle Z_\star / \mathrm{Z}_\odot \rangle = 0.09$), compared to star-forming galaxies. On average, quiescent galaxies in our sample are 1.09~Gyr older than the star-forming ones. In general, star-forming galaxies are characterized by significantly higher dust optical depths and shallower (grayer) attenuation slopes. As mentioned in a previous section (see Section~\ref{subsubsec:jpost}), a dependence between the attenuation slope and optical depth is expected \citep{Salim_2020ARA&A..58..529S}. Dust grains can either absorb or scatter the UV and optical starlight. At large optical depths (i.e., $\hat{\tau}_\mathrm{dust, 2} > 0.45$ or $A_V > 0.5$~mag) dust absorption is more dominant than scattering, resulting in a shallower (grayer) attenuation curve. As the optical depth decreases, scattering events start to dominate over absorption, resulting in a steeper attenuation curve.

Going back in Fig.~\ref{fig:sample_props} and focusing on the distribution of the \ssfr, we notice a primary peak in the distribution at high \ssfr~($\log \langle \mathrm{sSFR}_\mathrm{100Myrs} / \mathrm{yr}^{-1} \rangle = -10$), and a secondary peak at low \ssfr~($\log \langle \mathrm{sSFR}_\mathrm{100Myrs} / \mathrm{yr}^{-1} \rangle = -11.9$), whereas the median value of the full distribution is at $\log \langle \mathrm{sSFR}_\mathrm{100Myrs} / \mathrm{yr}^{-1} \rangle = -10.68$. Via the boundary between the two modes of the distribution, it is possible to define quiescence. Of course, there is no precise method to determine the exact boundary between quiescence and star formation in terms of sSFR. Several studies place the boundary at $\log \mathrm{sSFR} / \mathrm{yr}^{-1} \approx -10$ \citep[e.g.,][]{Whitaker_2017ApJ...838...19W, Wu_2018ApJ...868...37W, Leja_2019ApJ...880L...9L}, while others place the boundary at  $\log \mathrm{sSFR} / \mathrm{yr}^{-1} \approx -10.55$ \citep[e.g.,][]{Gallazzi_2014ApJ...788...72G} or at $\log \mathrm{sSFR} / \mathrm{yr}^{-1} \approx -11$ \citep{Brinchmann_2004MNRAS.351.1151B, Fontanot_2009MNRAS.397.1776F, Cecchi_2019ApJ...880L..14C, Donnari_2019MNRAS.485.4817D, Paspaliaris_2023A&A...669A..11P}. A $\log \mathrm{sSFR} / \mathrm{yr}^{-1} \approx -11$ seems to separate very well the two modes of the distribution of the \ssfr. Therefore, our galaxy sample can be alternatively divided into active and passive galaxies, depending on $\log \mathrm{sSFR}_\mathrm{100Myrs} / \mathrm{yr}^{-1}$ is above or below $-11$. According to this classification we find 1240 quiescent and 1668 star-forming galaxies. Very similar galaxy demographics were obtained through the $UVJ$ diagram and the definition of \citet{Muzzin_2013ApJ...777...18M} (see Section~\ref{subsec:final_sample}).

\subsection{Scaling relations} \label{subsec:scaling_relations}

\begin{figure}[t]
    \centering
    \includegraphics[width=\columnwidth]{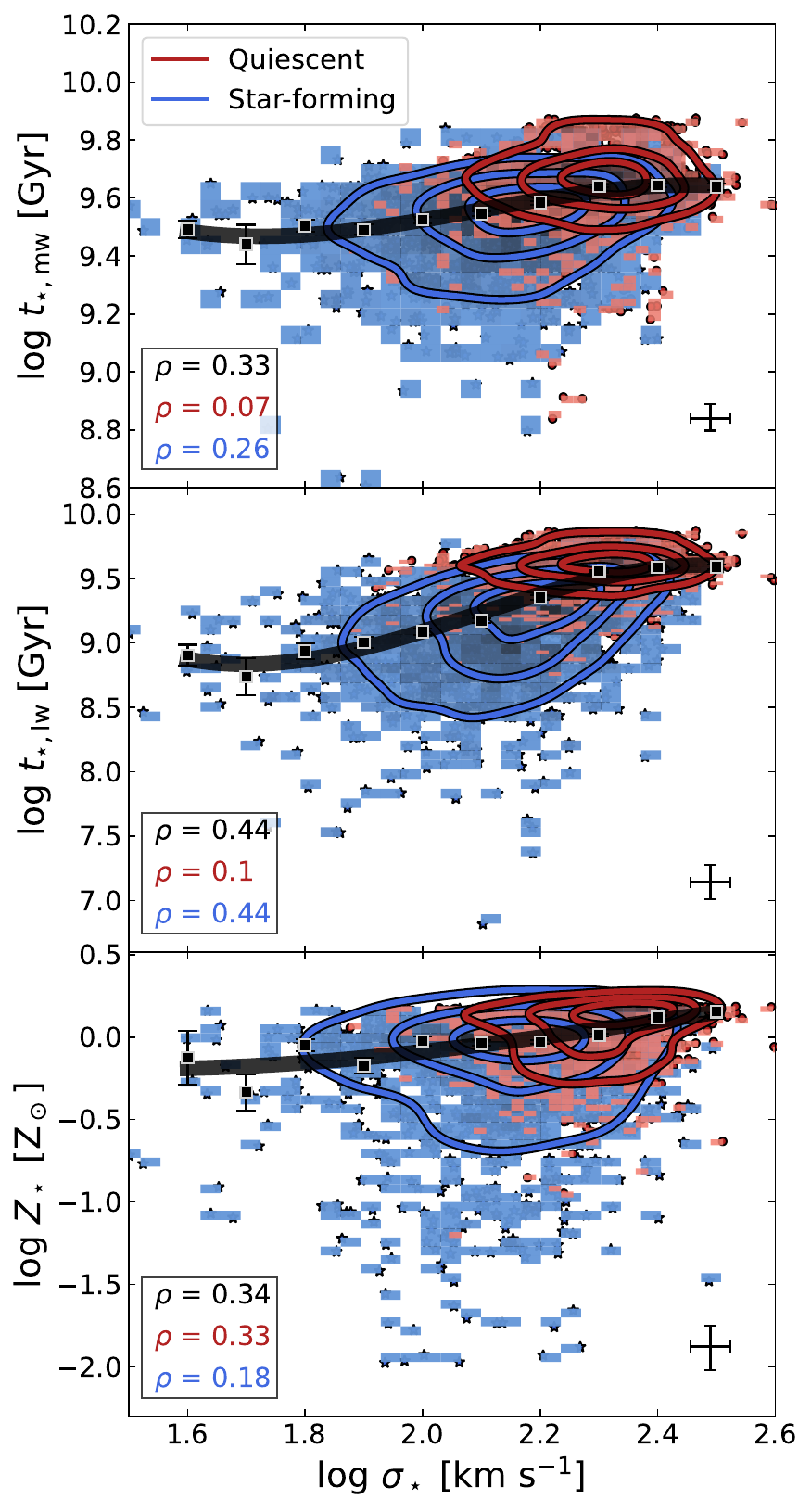}
    \caption{Mass-weighted age ($t_\mathrm{\star, mw}$), light-weighted ($t_\mathrm{\star, lw}$), and $Z_\star$ as a function of $\sigma_\star$. The star-forming and quiescent populations are indicated in blue and red respectively, separated based on their $UVJ$ colors. The Spearman’s rank correlation coefficient ($\rho$) of each galaxy population, is shown at the bottom-left corner of each panel, while the typical error bars are indicated in black at the bottom-right corner. The black lines are the $3^\mathrm{rd}$ order polynomial fits to the median values (black squares) in each 0.1~dex $\sigma_\star$ bin. The associated error bars identify the statistical uncertainties computed as $\sigma/\sqrt N$, where $N$ is the number of galaxies within each bin.}
    \label{fig:scaling_relations}
\end{figure}

Figure~\ref{fig:scaling_relations} shows the distributions of the mass-weighted age ($t_\mathrm{\star, mw}$), light-weighted age ($t_\mathrm{\star, lw}$), and $Z_\star$ as a function of the stellar velocity dispersion ($\sigma_\star$) from LEGA-C DR3. Galaxies are divided into star forming (blue distributions) and quiescent (red distributions) based on their location in the $UVJ$ plane. The black lines are the $3^\mathrm{rd}$ order polynomial fits to the median values in each 0.1~dex $\log \sigma_\star$ bin (black squares). The best fit scaling relations are given in Table~\ref{tab:spline}. 

The overall relationship between stellar ages and metallicity with $\sigma_\star$ shows a weak to moderate dependence. Examining the age--$\sigma_\star$ relation for the individual galaxy populations, we find a rather flat trend for quiescent galaxies, with a correlation coefficient $\rho \approx 0.1$. This weak correlation suggests that most quiescent galaxies are uniformly old, resulting in similar ages regardless of their stellar velocity dispersion. Star-forming galaxies exhibit a somewhat stronger age--$\sigma_\star$ trend, particularly between the light-weighted stellar age and $\sigma_\star$ ($\rho = 0.44$). Looking the sample as a whole, we find a positively increasing trend between stellar age and $\sigma_\star$, with a flattening at $\sigma_\star \ge 200$~km~s$^{-1}$, in agreement with previous studies of the LEGA-C sample \citep{Chauke_2018ApJ...861...13C, Cappellari_2023MNRAS.526.3273C}. 

The differences in the strength of the age--$\sigma_\star$ relation, based on the definition of mean stellar age, are intriguing. Light-weighted ages show greater variation among actively star-forming galaxies compared to mass-weighted ages. In contrast, the distribution of light-weighted ages for quiescent galaxies is more concentrated. This is expected, as light-weighted ages are directly measured from the photometric SEDs, and thus they are susceptible to outshining effects \citep[see review][and references therein]{Conroy_2013ARA&A..51..393C}.

\begin{table}[h!]
\caption{Scaling relations of Fig.~\ref{fig:scaling_relations}.}
\begin{center}
\scalebox{1.1}{
\begin{tabular}{lccccc}
\hline 
\hline
 $y$ & $\alpha_0$ & $\alpha_1$ & $\alpha_2$ & $\alpha_3$ \\
\hline
$\log(t_\mathrm{\star, mw}$/Gyr) & $16.99$ & $-11.38$ & $5.61$ & $-0.89$\\
$\log(t_\mathrm{\star, lw}$/Gyr) & $35.71$ & $-41.02$ & $20.34$ & $-3.25$ \\
$\log(Z_\star$/Z$_\odot$) & $1.81$ & $-3.08$ & $1.47$ & $-0.20$ \\
\hline \hline
\end{tabular}}
\tablefoot{The scaling relations are based on 3$^\mathrm{rd}$ order polynomial fits to the median values of $t_\mathrm{\star, mw}$, $t_\mathrm{\star, lw}$, and $Z_\star$ in each 0.1~dex $\log \sigma_\star$ bin. The scaling relations are given through $y = \alpha_0 + \alpha_1 \times \log(\sigma_\star) + \alpha_2 \times \log(\sigma_\star)^2 + \alpha_3 \times \log(\sigma_\star)^3$.}
\label{tab:spline}
\end{center}
\end{table}

Mass-weighted ages, however, provide a more fundamental measure of the integrated SFHs and are less affected by outshining. They are prone to biases from the assumed SFH prior distributions used in SED fitting, especially for actively star-forming galaxies \citep{Lee_2010ApJ...725.1644L, Wuyts_2011ApJ...738..106W}. The use of a nonparametric SFH, along with the high-quality spectroscopy available in LEGA-C DR3, should mitigate these biases in the mass-weighted ages, resulting in a more reliable and precise age--$\sigma_\star$ relation with lower uncertainty compared to light-weighted ages.

\begin{figure*}[ht!]
    \centering
    \includegraphics[width=\textwidth]{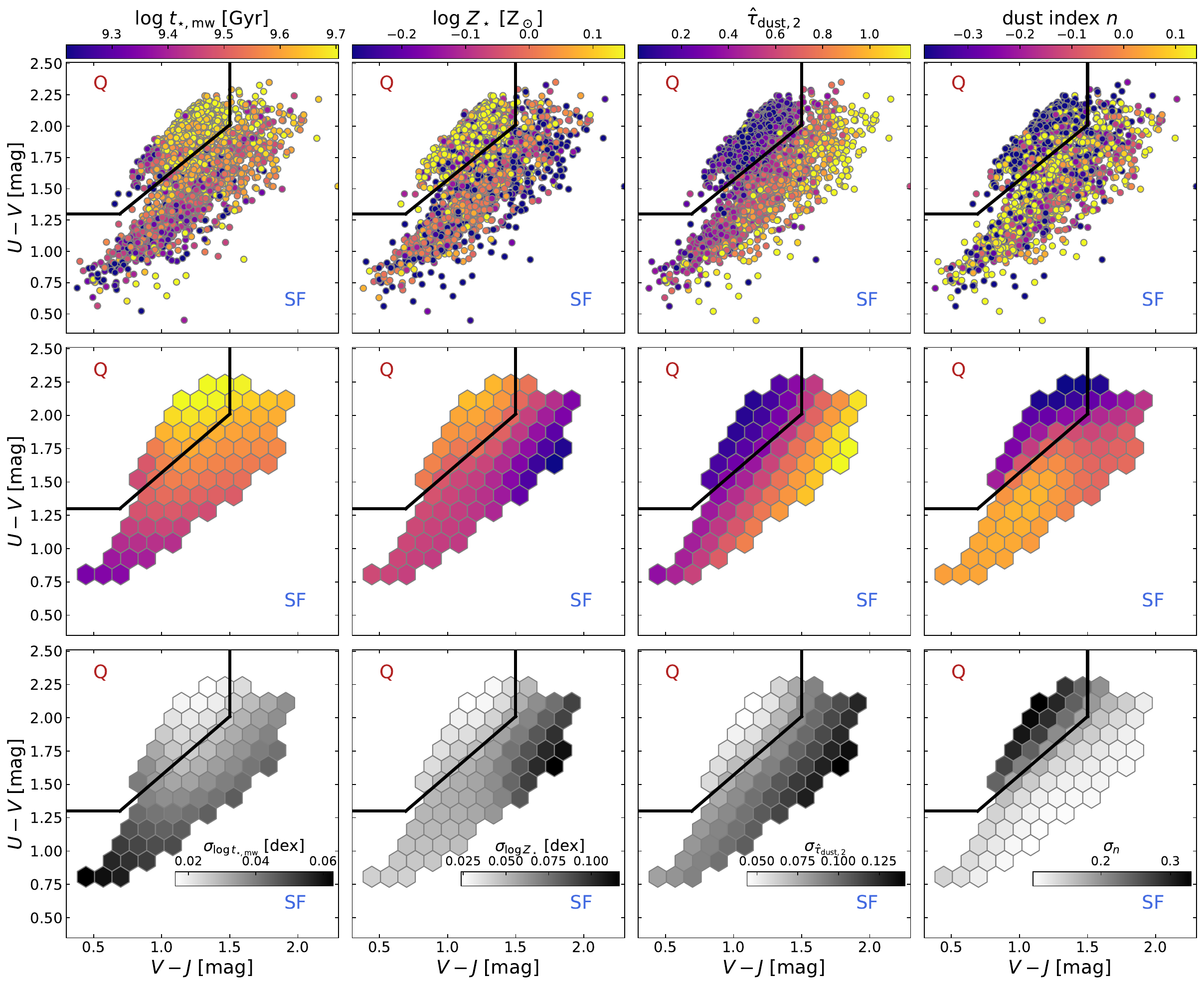}
    \caption{Mean stellar population and dust properties in the $UVJ$ diagram of our primary sample. From left to right: mass-weighted stellar age ($t_\mathrm{\star, mw}$), stellar metallicity ($Z_\star$), dust optical depth in the $V$ band ($\hat{\tau}_\mathrm{dust, 2}$), and dust attenuation slope index ($n$). We use the definition of \citet{Muzzin_2013ApJ...777...18M} to separate the galaxies into quiescent and star forming. The top row shows the individual data points, the middle row shows the average trends of the physical properties in the $UVJ$ diagrams with the {\tt LOESS} method, while the bottom row shows the average upper uncertainty in those properties. We required that each hexbin contains at least five galaxies.}
    \label{fig:uvj_diagram}
\end{figure*}

The metallicity--$\sigma_\star$ relation also presents a weak positive trend. Comparing the trends between star-forming and quiescent galaxies, we find that quiescent galaxies show a stronger, but still weak correlation ($\rho = 0.33$), compared to the star-forming ones ($\rho = 0.18$). At fixed $\sigma_\star$, quiescent galaxies are metal rich, while star-forming galaxies are metal poor. The scaling relations presented here indicate that massive galaxies with high $\sigma_\star$ tend to be older and have higher metallicity. In Gallazzi et al. (in prep.), we also analyze the difference between quiescent and star-forming galaxies either at fixed $M_\star$ or $\sigma_\star$, finding that quiescent galaxies are more metal-rich than star-forming ones, especially in the low-mass regime. These results are in excellent agreement with studies in the local Universe \citep[e.g.,][]{Scott_2017MNRAS.472.2833S, Li_2018MNRAS.476.1765L}, reinforcing the understanding that such galaxies have undergone significant star formation and chemical enrichment during an early epoch in their history. 

\subsection{The $UVJ$ color-color diagram} \label{subsec:uvj}

The various, complex physical processes that take place in galaxies give rise to the differences observed at UV and optical wavelengths. Interpreting the correlations between optical colors of galaxies is a challenging task. The combined effects of dust attenuation, star formation, and stellar evolution are very difficult to disentangle, leading to many degeneracies between these parameters. A simple yet powerful method to break these degeneracies is to combine an optical and an optical–NIR color. The restframe $UVJ$ color-color diagram has proven to be quite successful in that regard \citep{Labbe_2005ApJ...624L..81L, Wuyts_2007ApJ...655...51W, Williams_2009ApJ...691.1879W}. A $UVJ$ diagram not only can break the aforementioned degeneracies, but also can classify galaxies into quiescent and star forming. The $UVJ$ diagram is commonly used as a quiescence diagnostic in both observed galaxy samples \citep[e.g.,][]{Whitaker_2013ApJ...770L..39W, Straatman_2014ApJ...783L..14S, Straatman_2016ApJ...830...51S, Barro_2014ApJ...791...52B, Papovich_2015ApJ...803...26P, Papovich_2018ApJ...854...30P, Fang_2018ApJ...858..100F, Tan_2022ApJ...933...30T, Miller_2022ApJ...941L..37M, Antwi_Danso_2023ApJ...943..166A, Valentino_2023ApJ...947...20V} and in and cosmological hydrodynamical simulations \citep{Dave_2017MNRAS.471.1671D, Trayford_2017MNRAS.470..771T, Donnari_2019MNRAS.485.4817D, Akins_2022ApJ...929...94A, Nagaraj_2022ApJ...939...29N, Kurinchi_Vendhan_2024MNRAS.534.3974K, Baes_2024A&A...683A.181B}. 

Figure~\ref{fig:uvj_diagram} shows the $UVJ$ diagram of our primary galaxy sample color-coded by various dust and stellar properties. In particular, we color-coded the $UVJ$ plane with the following physical quantities, the dust attenuation slope index ($n$), dust optical depth in the $V$ band ($\hat{\tau}_\mathrm{dust, 2}$), mass-weighted stellar age ($t_\mathrm{\star, mw}$), and stellar metallicity ($Z_\star$). Instead of just showing scatter plots, which can sometimes conceal the underlying trends, we also derived the average trends of the physical properties in the $UVJ$ diagrams, as well as their average upper uncertainty (i.e., the difference between the $84^\mathrm{th}$ and $50^\mathrm{th}$ percentiles of the posteriors), with the Locally Weighted Regression ({\tt LOESS}) method \citep{Cleveland_doi:10.1080/01621459.1988.10478639} as implemented in the {\tt LOESS} routine\footnote{https://pypi.org/project/loess/} by \citet{Cappellari_2013MNRAS.432.1862C}. In the following figures, we use a linear local approximation, a smoothing factor of $f = 0.5$, and require that each hexbin contains at least five galaxies. The maps created with the {\tt LOESS} method in the $UVJ$ plane demonstrate that the effects of dust and stellar evolution can be separated. 

For the star-forming galaxies, we find a smooth increasing gradient of $t_\mathrm{\star, mw}$ toward the upper right corner of the $UVJ$ plane. Star-forming galaxies with bluer $U-V$ and $V-J$ colors are on average 0.13~dex younger than star-forming with redder $U-V$ and $V-J$ colors. On the other hand, we find a relatively weak trend with $Z_\star$, with the reddest $V-J$ color galaxies having the lowest $Z_\star$ values. Of course, we notice that the stellar metallicity estimates for these particular galaxies have the highest uncertainties (up to 0.1~dex). We also observe that star-forming galaxies with the lowest $Z_\star$ tend to have the highest $\hat{\tau}_\mathrm{dust, 2}$, suggesting a potential degeneracy between these two quantities. However, a strong degeneracy was not evident in the joint posterior distribution of these parameters (see discussion in Section~\ref{subsubsec:jpost}). 

In the upper right corner of the $UVJ$ plane, star-forming galaxies are characterized by a high dust attenuation and a relatively flat attenuation slope. This is the expected behavior since galaxies with high dust attenuation follow the attenuation curve and are red in both $U-V$ and $V-J$ colors. The flattening of the dust-attenuation curve is the result of high attenuation and a more complex geometry between the stellar and dust distributions, where more of the starlight is decoupled from dust \citep[e.g.,][]{Charlot_2000ApJ...539..718C, Narayanan_2018ApJ...869...70N, Salim_2020ARA&A..58..529S, Nersesian_2020A&A...637A..25N}. Surprisingly, at the intermediate-attenuation regime ($0.5 < \hat{\tau}_\mathrm{dust, 2} < 0.65$), toward the bottom left corner of the $UVJ$ plane, galaxies seem to sit to even flatter attenuation curves. The uncertainties on the dust attenuation seem to increase with increasing dust attenuation, while the average uncertainty of the dust index $n$ remains relatively low for the star-forming galaxies ($\sigma_n = 0.1$). 

Quiescent galaxies are relatively bluer in $V-J$ colors due to a very low dust attenuation that also produces very steep attenuation curves. The steepening of the dust-attenuation curve is likely a product of less dust scattering into the line of sight. This is consistent with the idea that the oldest and more passive galaxies are less dusty, and thus characterized by the lowest dust attenuation. Again, we find a smooth increasing gradient of $t_\mathrm{\star, mw}$ toward the upper right corner of the $UVJ$ plane, and also a moderate increasing trend with $Z_\star$. Our results are consistent with earlier findings on the direction of slow aging after quenching \citep{Whitaker_2013ApJ...770L..39W, Leja_2019ApJ...880L...9L, Belli_2019ApJ...874...17B, Beverage_2021ApJ...917L...1B, Tacchella_2022ApJ...926..134T, Beverage_2023ApJ...948..140B}. In general, the oldest quiescent galaxies are also the ones with the lowest dust attenuation, steepest dust slopes, and highest $Z_\star$. The uncertainties in the stellar and dust properties of quiescent galaxies are, on average, lower than those of star-forming galaxies. The only exception is the dust index $n$, which exhibits high uncertainty (up to $\sigma_n = 0.35$).

Lastly, a $UVJ$ quiescent selection yields a 10\%-30\% contamination from star-forming galaxies. We find that 7\% of $UVJ$-quiescent galaxies show signs of star formation (\ssfr$\ge 10^{-11}$ yr$^{-1}$). One possible explanation is that these galaxies are in transition from star forming to quiescence. This contamination fraction is consistent with previous studies using spectroscopic information \citep{Belli_2017ApJ...841L...6B, Schreiber_2018A&A...618A..85S} or SED fitting \citep{Moresco_2013A&A...558A..61M, Fang_2018ApJ...858..100F, Tacchella_2022ApJ...926..134T}. Another possible explanation for this contamination could be the dust attenuation law: nonquiescent galaxies (high-sSFR objects) are in the $UVJ$ quiescent region because they have a steep attenuation law with non negligible amounts of dust and star formation.

\begin{figure*}[t]
    \centering
    \includegraphics[width=\textwidth]{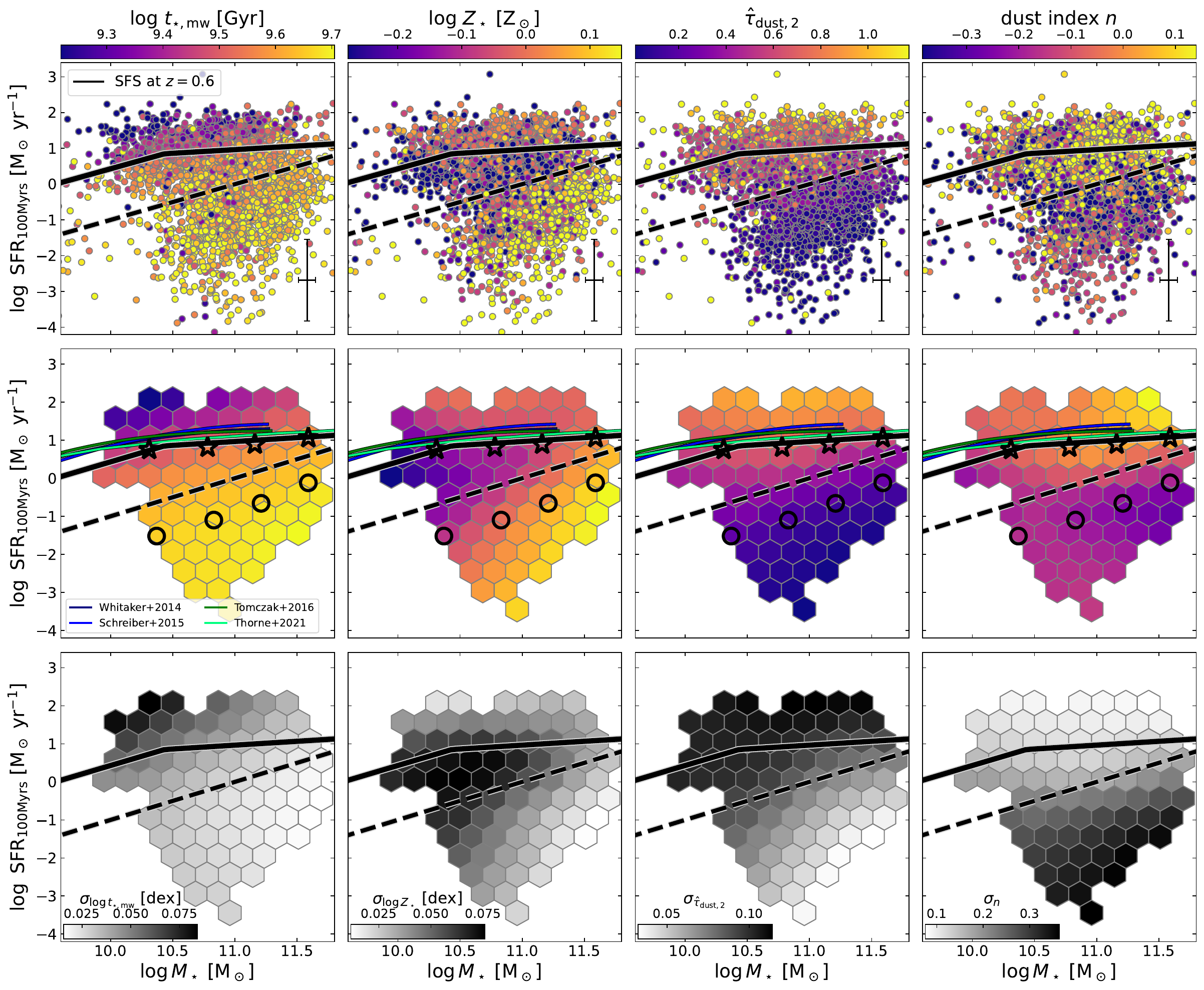}
    \caption{Mean stellar population and dust properties in the \sfr--$M_\star$ plane of our primary sample. Panels from left to right display: mass-weighted stellar age ($t_\mathrm{\star, mw}$), stellar metallicity ($Z_\star$), dust optical depth in the $V$ band ($\hat{\tau}_\mathrm{dust, 2}$), and dust attenuation slope index ($n$). Top row: Individual data points, with typical error bars shown in the bottom-right corner of each panel. Middle row: Average trends of the physical properties across the \sfr--$M_\star$ plane, derived using the {\tt LOESS} method. Bottom row: Average upper uncertainties for each property. Each hexbin includes a minimum of five galaxies. The SFS at $z = 0.60$ is indicated by the solid black line, following the definition of \citet{Leja_2022ApJ...936..165L}, while the dashed black line represents the boundary between star-forming and quiescent galaxies (sSFR = $10^{-11}$~yr$^{-1}$). In the middle row, black markers indicate the median \sfr values in bins of $M_\star$: open stars for star-forming galaxies and open circles for quiescent galaxies (see Table~\ref{tab:phys_param}). Additionally, results from previous studies at a similar redshift range are overplotted for comparison.}
    \label{fig:multi_ms}
\end{figure*}

\subsection{The \sfr--$M_\star$ relation} \label{subsec:sfms}

In this section we measure the relationship between \sfr~and $M_\star$, otherwise known as the star-forming sequence \citep[SFS;][]{Daddi_2007ApJ...670..156D, Noeske_2007ApJ...660L..43N, Elbaz_2007A&A...468...33E, Leja_2022ApJ...936..165L}. Galaxies form most of their mass on \citep{Leitner_2012ApJ...745..149L}, or have passed through \citep{Abramson_2015ApJ...801L..12A}, the SFS. Therefore, studying the SFS at different redshifts provides crucial information on the evolution of galaxies \citep[e.g.,][]{Noeske_2007ApJ...660L..43N, Whitaker_2012ApJ...754L..29W, Whitaker_2014ApJ...795..104W, Speagle_2014ApJS..214...15S, Schreiber_2015A&A...575A..74S, Renzini_2015ApJ...801L..29R, Tomczak_2014ApJ...783...85T, Iyer_2018ApJ...866..120I, Leslie_2020ApJ...899...58L, Thorne_2021MNRAS.505..540T,  Leja_2022ApJ...936..165L}. 

Figure~\ref{fig:multi_ms} presents the \sfr--$M_\star$ relation of our primary galaxy sample. As in Fig.~\ref{fig:uvj_diagram}, we color-coded the SFS by various dust and stellar properties, in order to highlight the underlying trends with stellar age, metallicity, and dust. Once more, we applied the {\tt LOESS} method to probe the average trends of the physical properties and their uncertainties on the \sfr--$M_\star$ plane. We also indicate the boundary between star formation and quiescence at a $\mathrm{sSFR} = 10^{-11}~\mathrm{Gyr}^{-1}$ (black dashed line), and overplot the SFS at $z = 0.60$ (solid black line) as prescribed in \citet{Leja_2022ApJ...936..165L}, as well as the results from other studies in the literature at a similar redshift range \citep{ Whitaker_2014ApJ...795..104W, Schreiber_2015A&A...575A..74S, Tomczak_2016ApJ...817..118T, Thorne_2021MNRAS.505..540T}. When necessary, these literature results were rescaled to a \citet{Chabrier_2003PASP..115..763C} IMF to match our study. 

Overall, we find a flat relationship between SFR and $M_\star$ for star-forming galaxies, with a weak correlation ($\rho = 0.09$). Quiescent galaxies form their own `red' sequence, showing a stronger correlation with stellar mass ($\rho = 0.39$). As expected, the median \sfr values in bins of $M_\star$ for star-forming galaxies (represented by open stars) align closely with the SFS defined by \citet{Leja_2022ApJ...936..165L}, as both studies derive SFRs and stellar masses simultaneously using {\tt Prospector}. However, previous studies report systematically higher SFS measurements, with offsets of approximately $\sim 0.15-0.4$~dex. Among these, the results from \citet{Thorne_2021MNRAS.505..540T} show better agreement with ours, particularly at higher stellar masses ($\log(M_\star/\mathrm{M}_\odot) \ge 10.5$), where their SFS values are typically $\sim0.15$~dex above those of {\tt Prospector}. \citet{Thorne_2021MNRAS.505..540T} also estimate SFRs and $M_\star$ through SED fitting. In contrast, studies such as \citet{Whitaker_2014ApJ...795..104W}, \citet{Schreiber_2015A&A...575A..74S}, and \citet{Tomczak_2016ApJ...817..118T} rely on broadband photometry for their SFR and $M_\star$ estimates. Deriving stellar properties, as done with SED fitting, tends to yield lower SFR estimates compared to traditional broadband formulae \citep[e.g.,][]{Nersesian_2019A&A...624A..80N, Leja_2022ApJ...936..165L}.

Focussing on the star-forming galaxies, we find that the average trends with age and metallicity are more pronounced in the \sfr--$M_\star$ plane, compared to what we found in the $UVJ$ diagrams. At fixed SFR, low-mass (high-mass) galaxies have lower (higher) stellar metallicity, while their stellar populations are on average younger (older). Again, we find that the average uncertainties over $Z_\star$ and $t_\mathrm{\star, mw}$ are relatively larger for galaxies with very low $Z_\star$ and $t_\mathrm{\star, mw}$, yet with values below 0.1~dex. Another interesting result is that the apparent $Z_\star$--$\hat{\tau}_\mathrm{dust, 2}$ degeneracy suggested by the $UVJ$ colors does not manifest in the \sfr--$M\star$ plane. In fact, the most active and massive star-forming galaxies are those with the highest dust attenuation, and with stellar metallicities very close to solar. The uncertainties on the dust attenuation for these galaxies remain high though ($\sigma_{\hat{\tau}_\mathrm{dust, 2}} \approx 0.12$).

Furthermore, we find a positive trend between $M_\star$ and the dust attenuation slope \citep[e.g.,][]{Reddy_2018ApJ...853...56R, Salim_2018ApJ...859...11S, Barisic_2020ApJ...903..146B}. The less massive star-forming galaxies with high SFR (i.e., high sSFR) tend to have a relatively steeper dust attenuation slope ($\langle n \rangle \approx 0$) than the more massive ones ($M_\star > 10^{11}~\mathrm{M}_\odot$; $\langle n \rangle > 0.1$). In fact, the attenuation slope seems to become shallower with decreasing sSFR among star-forming galaxies \citep[e.g.,][]{Salim_2018ApJ...859...11S}. Ultimately, observational \citep{Salmon_2016ApJ...827...20S, Salim_2018ApJ...859...11S, Nersesian_2020A&A...637A..25N}, theoretical \citep{Witt_2000ApJ...528..799W, Charlot_2000ApJ...539..718C, Calzetti_2001PASP..113.1449C, Chevallard_2013MNRAS.432.2061C}, and cosmological simulations \citep{Narayanan_2018ApJ...869...70N, Shen_2020MNRAS.495.4747S} all point to a complex star-to-dust geometry as the main driver of the flattening of the attenuation curve.  

Quiescent galaxies seem to display stronger trends with both stellar and dust properties. Similarly to our findings with the $UVJ$ diagrams, at a fixed \sfr, we find smooth increasing gradients for both stellar age and metallicity with increasing $M_\star$. As it has been previously shown, quiescent galaxies formed their mass early on $z\approx2-3$ \citep[e.g.,][]{Tacchella_2022ApJ...926..134T, Kaushal_2024ApJ...961..118K}, staying to the SFS for a very short period of time before ending up in the red sequence. At a fixed \sfr, quiescent galaxies also have steeper attenuation slopes than star-forming galaxies. At low dust optical depths, the steep dust-attenuation curve is likely the result of dust scattering, with more red light (isotropic scattering) and less blue light (forward scattering) escaping the galaxy \citep{Chevallard_2013MNRAS.432.2061C}. We find that $n$ becomes steeper with increasing stellar mass, whereas the dust optical depth is decreasing with \sfr~and increasing $M_\star$. Indeed, we observe that the most massive quiescent galaxies are characterized by the lowest optical depths and steepest attenuation slopes. 

In summary, we recover the well-known trends that on average quiescent galaxies sit on steeper attenuation slopes, have low dust attenuation, are older, and metal rich. As for the star-forming galaxies we find that they are on average younger, more dusty, and metal poor.

\section{Summary and conclusions} \label{sec:conclusions}

We presented a first analysis on the stellar properties of 2908 galaxies (1208 quiescent and 1700 star forming) at redshift $\sim 1$. We estimated the galaxy properties by employing the fully Bayesian SED fitting framework {\tt Prospector}. With {\tt Prospector}, we fitted the high spectral resolution spectroscopy and the broadband photometry together, drawn from the LEGA-C \citep{van_der_Wel_2021ApJS..256...44V} and UltraVISTA catalogs \citep{Muzzin_2013ApJS..206....8M, Muzzin_2013ApJ...777...18M} respectively. Our key findings are summarized below.

\begin{itemize}

\item We demonstrate the exceptional fits to the spectrophotometric observations by our {\tt Prospector} physical model, reassuring the proper constraining of the stellar and dust attenuation properties of our primary galaxy sample at a redshift range from $0.6 \le z \le 1$. We also show that pairing spectroscopy with photometry is essential to alleviate the dust–age–metallicity degeneracy. 

\item By comparing the observed and model spectra in terms of their Lick indices, we show that the inclusion of spectroscopy in the SED fitting results in the significant reduction of systematic offsets, previously reported by \citet{Nersesian_2024A&A...681A..94N} when fitting photometry alone. Mass-weighted and light-weighted ages are better constrained by using the spectroscopic information. Stellar metallicities are constrained by our spectroscopy, but precise measurements remain challenging (and impossible with photometry alone), particularly in the absence of Mg and Fe lines redward of 5000~\AA~in the observed spectrum. Nevertheless, spectroscopy significantly improves the measurements of the stellar population properties and is required to provide meaningful constraints on age, metallicity and other stellar population properties. 

\item We show that photometry-only fits with a linear $Z_\star$ prior do not yield meaningful constraints on $Z_\star$ for individual galaxies, though they effectively avoid a downward bias. While the precision remains limited, the accuracy is generally good. We also note that the results are highly sensitive to the choice of prior. However, when spectroscopy is included, the dependence on the prior is reduced--though not entirely eliminated--due to the additional information provided that helps constrain the stellar metallicity.

\item A comparison of the physical properties derived from fitting only photometry, only spectroscopy, and both photometry and spectroscopy together, revealed a strong systematic offset for both the light-weighted ages (0.2~dex) and the mass-weighted ages (0.28~dex) for the photometry-only case. This means that fitting photometry alone results in the estimation of older stellar ages for the galaxies. From this result we conclude that detailed constraints on the SFHs are only possible when fitting both photometry and spectroscopy. 

\item We confirm a positive trend between the light-weighted ages and the stellar velocity dispersion ($\sigma_\star$). The trend with $\sigma_\star$ is weaker for the mass-weighted ages and $Z_\star$. At fixed $\sigma_\star$, we find hints of a dependence of $Z_\star$ on stellar age. The scaling relations presented here indicate that massive galaxies, with high $\sigma_\star$ tend to be older and have higher metallicity, in agreement with studies of the Local Universe \citep[e.g.,][]{Scott_2017MNRAS.472.2833S, Li_2018MNRAS.476.1765L}, reinforcing the understanding that such galaxies have undergone significant star formation and chemical enrichment during an early epoch in their history.

\item Quiescent galaxies are characterized by very low star formation ($\log \langle $\sfr /$\mathrm{M}_{\odot}~\mathrm{yr}^{-1} \rangle = -0.85$) and high metallicity ($\log \langle Z_\star / \mathrm{Z}_\odot \rangle = 0.09$). On average, quiescent galaxies in our primary sample are 1.1~Gyr older than the star-forming ones. We also find that quiescent galaxies have steeper attenuation slopes than star-forming galaxies, with the most massive quiescent galaxies being characterized by the steepest attenuation slopes. At low dust optical depths, the steep dust-attenuation curve is likely the result of dust scattering than dust absorption \citep[e.g.,][]{Chevallard_2013MNRAS.432.2061C}. We find that $n$ becomes steeper with increasing stellar mass, whereas the dust optical depth is decreasing with \sfr~and increasing $M_\star$.

\item We find that low-mass (high-mass) star-forming galaxies have lower (higher) stellar metallicity, while their stellar populations are on average younger (older). Star-forming galaxies are also characterized by significantly higher dust optical depths and shallower (grayer) attenuation slopes. We find a positive trend between $M_\star$ and the dust attenuation slope. The less massive star-forming galaxies tend to have a relatively steeper dust attenuation slope ($\langle n \rangle \approx 0$) than the more massive ones ($M_\star > 10^{11}~\mathrm{M}_\odot$; $\langle n \rangle > 0.1$). In fact, the attenuation slope seems to become shallower with decreasing sSFR, suggesting a more complex star-to-dust geometries. 

\end{itemize}

\begin{acknowledgements}
We thank the anonymous referee for the valuable remarks and suggestions that helped us improve the paper. AN gratefully acknowledges the support of the Belgian Federal Science Policy Office (BELSPO) for the provision of financial support in the framework of the PRODEX Programme of the European Space Agency (ESA) under contract number 4000143347. AG acknowledges support from INAF-Minigrant-2022 "LEGA-C" 1.05.12.04.01. PFW acknowledge funding through the National Science and Technology Council grant 113-2112-M-002-027-MY2. This research made use of Astropy,\footnote{\url{http://www.astropy.org}} a community-developed core Python package for Astronomy \citep{Astropy_2013A&A...558A..33A, Astropy_2018AJ....156..123A}. 
\end{acknowledgements}



\bibliographystyle{aa}
\bibliography{References} 

\begin{thebibliography}{209}
\expandafter\ifx\csname natexlab\endcsname\relax\def\natexlab#1{#1}\fi

\bibitem[{{Abdurro'uf} {et~al.}(2021){Abdurro'uf}, {Lin}, {Wu}, \& {Akiyama}}]{Abdurrouf_2021ApJS..254...15A}
{Abdurro'uf}, {Lin}, Y.-T., {Wu}, P.-F., \& {Akiyama}, M. 2021, \apjs, 254, 15

\bibitem[{{Abramo} {et~al.}(2012){Abramo}, {Strauss}, {Lima}, {Hern{\'a}ndez-Monteagudo}, {Lazkoz}, {Moles}, {de Oliveira}, {Sendra}, {Sodr{\'e}}, \& {Storchi-Bergmann}}]{Abramo_2012MNRAS.423.3251A}
{Abramo}, L.~R., {Strauss}, M.~A., {Lima}, M., {et~al.} 2012, \mnras, 423, 3251

\bibitem[{{Abramson} {et~al.}(2015){Abramson}, {Gladders}, {Dressler}, {Oemler}, {Poggianti}, \& {Vulcani}}]{Abramson_2015ApJ...801L..12A}
{Abramson}, L.~E., {Gladders}, M.~D., {Dressler}, A., {et~al.} 2015, \apjl, 801, L12

\bibitem[{{Akins} {et~al.}(2022){Akins}, {Narayanan}, {Whitaker}, {Dav{\'e}}, {Lower}, {Bezanson}, {Feldmann}, \& {Kriek}}]{Akins_2022ApJ...929...94A}
{Akins}, H.~B., {Narayanan}, D., {Whitaker}, K.~E., {et~al.} 2022, \apj, 929, 94

\bibitem[{{Antwi-Danso} {et~al.}(2023){Antwi-Danso}, {Papovich}, {Leja}, {Marchesini}, {Marsan}, {Martis}, {Labb{\'e}}, {Muzzin}, {Glazebrook}, {Straatman}, \& {Tran}}]{Antwi_Danso_2023ApJ...943..166A}
{Antwi-Danso}, J., {Papovich}, C., {Leja}, J., {et~al.} 2023, \apj, 943, 166

\bibitem[{{Arnouts} {et~al.}(2013){Arnouts}, {Le Floc'h}, {Chevallard}, {Johnson}, {Ilbert}, {Treyer}, {Aussel}, {Capak}, {Sanders}, {Scoville}, {McCracken}, {Milliard}, {Pozzetti}, \& {Salvato}}]{Arnouts_2013A&A...558A..67A}
{Arnouts}, S., {Le Floc'h}, E., {Chevallard}, J., {et~al.} 2013, \aap, 558, A67

\bibitem[{{Astropy Collaboration} {et~al.}(2018){Astropy Collaboration}, {Price-Whelan}, {Sip{\H o}cz}, {G{\"u}nther}, {Lim}, {Crawford}, {Conseil}, {Shupe}, {Craig}, {Dencheva}, {Ginsburg}, {VanderPlas}, {Bradley}, {P{\'e}rez-Su{\'a}rez}, {de Val-Borro}, {Aldcroft}, {Cruz}, {Robitaille}, {Tollerud}, {Ardelean}, {Babej}, {Bach}, {Bachetti}, {Bakanov}, {Bamford}, {Barentsen}, {Barmby}, {Baumbach}, {Berry}, {Biscani}, {Boquien}, {Bostroem}, {Bouma}, {Brammer}, {Bray}, {Breytenbach}, {Buddelmeijer}, {Burke}, {Calderone}, {Cano Rodr{\'{\i}}guez}, {Cara}, {Cardoso}, {Cheedella}, {Copin}, {Corrales}, {Crichton}, {D'Avella}, {Deil}, {Depagne}, {Dietrich}, {Donath}, {Droettboom}, {Earl}, {Erben}, {Fabbro}, {Ferreira}, {Finethy}, {Fox}, {Garrison}, {Gibbons}, {Goldstein}, {Gommers}, {Greco}, {Greenfield}, {Groener}, {Grollier}, {Hagen}, {Hirst}, {Homeier}, {Horton}, {Hosseinzadeh}, {Hu}, {Hunkeler}, {Ivezi{\'c}}, {Jain}, {Jenness}, {Kanarek}, {Kendrew}, {Kern}, {Kerzendorf}, {Khvalko}, {King}, {Kirkby}, {Kulkarni},
  {Kumar}, {Lee}, {Lenz}, {Littlefair}, {Ma}, {Macleod}, {Mastropietro}, {McCully}, {Montagnac}, {Morris}, {Mueller}, {Mumford}, {Muna}, {Murphy}, {Nelson}, {Nguyen}, {Ninan}, {N{\"o}the}, {Ogaz}, {Oh}, {Parejko}, {Parley}, {Pascual}, {Patil}, {Patil}, {Plunkett}, {Prochaska}, {Rastogi}, {Reddy Janga}, {Sabater}, {Sakurikar}, {Seifert}, {Sherbert}, {Sherwood-Taylor}, {Shih}, {Sick}, {Silbiger}, {Singanamalla}, {Singer}, {Sladen}, {Sooley}, {Sornarajah}, {Streicher}, {Teuben}, {Thomas}, {Tremblay}, {Turner}, {Terr{\'o}n}, {van Kerkwijk}, {de la Vega}, {Watkins}, {Weaver}, {Whitmore}, {Woillez}, {Zabalza}, \& {Astropy Contributors}}]{Astropy_2018AJ....156..123A}
{Astropy Collaboration}, {Price-Whelan}, A.~M., {Sip{\H o}cz}, B.~M., {et~al.} 2018, \aj, 156, 123

\bibitem[{{Astropy Collaboration} {et~al.}(2013){Astropy Collaboration}, {Robitaille}, {Tollerud}, {Greenfield}, {Droettboom}, {Bray}, {Aldcroft}, {Davis}, {Ginsburg}, {Price-Whelan}, {Kerzendorf}, {Conley}, {Crighton}, {Barbary}, {Muna}, {Ferguson}, {Grollier}, {Parikh}, {Nair}, {Unther}, {Deil}, {Woillez}, {Conseil}, {Kramer}, {Turner}, {Singer}, {Fox}, {Weaver}, {Zabalza}, {Edwards}, {Azalee Bostroem}, {Burke}, {Casey}, {Crawford}, {Dencheva}, {Ely}, {Jenness}, {Labrie}, {Lim}, {Pierfederici}, {Pontzen}, {Ptak}, {Refsdal}, {Servillat}, \& {Streicher}}]{Astropy_2013A&A...558A..33A}
{Astropy Collaboration}, {Robitaille}, T.~P., {Tollerud}, E.~J., {et~al.} 2013, \aap, 558, A33

\bibitem[{{Baes} {et~al.}(2024){Baes}, {Gebek}, {Tr{\v{c}}ka}, {Camps}, {van der Wel}, {Abdurro'uf}, {Andreadis}, {Tulu}, {Emana}, {Fritz}, {Kelly}, {Kova{\v{c}}i{\'c}}, {La Marca}, {Martorano}, {Mosenkov}, {Nersesian}, {Rodriguez-Gomez}, {Tortora}, {Vander Meulen}, \& {Wang}}]{Baes_2024A&A...683A.181B}
{Baes}, M., {Gebek}, A., {Tr{\v{c}}ka}, A., {et~al.} 2024, \aap, 683, A181

\bibitem[{{Baldry} {et~al.}(2004){Baldry}, {Glazebrook}, {Brinkmann}, {Ivezi{\'c}}, {Lupton}, {Nichol}, \& {Szalay}}]{Baldry_2004ApJ...600..681B}
{Baldry}, I.~K., {Glazebrook}, K., {Brinkmann}, J., {et~al.} 2004, \apj, 600, 681

\bibitem[{{Bari{\v{s}}i{\'c}} {et~al.}(2020){Bari{\v{s}}i{\'c}}, {Pacifici}, {van der Wel}, {Straatman}, {Bell}, {Bezanson}, {Brammer}, {D'Eugenio}, {Franx}, {van Houdt}, {Maseda}, {Muzzin}, {Sobral}, \& {Wu}}]{Barisic_2020ApJ...903..146B}
{Bari{\v{s}}i{\'c}}, I., {Pacifici}, C., {van der Wel}, A., {et~al.} 2020, \apj, 903, 146

\bibitem[{{Barrientos Acevedo} {et~al.}(2023){Barrientos Acevedo}, {van der Wel}, {Baes}, {Grand}, {Kapoor}, {Camps}, {de Graaff}, {Straatman}, \& {Bezanson}}]{Barrientos_Acevedo_2023MNRAS.524..907B}
{Barrientos Acevedo}, D., {van der Wel}, A., {Baes}, M., {et~al.} 2023, \mnras, 524, 907

\bibitem[{{Barro} {et~al.}(2014){Barro}, {Faber}, {P{\'e}rez-Gonz{\'a}lez}, {Pacifici}, {Trump}, {Koo}, {Wuyts}, {Guo}, {Bell}, {Dekel}, {Porter}, {Primack}, {Ferguson}, {Ashby}, {Caputi}, {Ceverino}, {Croton}, {Fazio}, {Giavalisco}, {Hsu}, {Kocevski}, {Koekemoer}, {Kurczynski}, {Kollipara}, {Lee}, {McIntosh}, {McGrath}, {Moody}, {Somerville}, {Papovich}, {Salvato}, {Santini}, {Tal}, {van der Wel}, {Williams}, {Willner}, \& {Zolotov}}]{Barro_2014ApJ...791...52B}
{Barro}, G., {Faber}, S.~M., {P{\'e}rez-Gonz{\'a}lez}, P.~G., {et~al.} 2014, \apj, 791, 52

\bibitem[{{Battisti} {et~al.}(2020){Battisti}, {Cunha}, {Shivaei}, \& {Calzetti}}]{Battisti_2020ApJ...888..108B}
{Battisti}, A.~J., {Cunha}, E.~d., {Shivaei}, I., \& {Calzetti}, D. 2020, \apj, 888, 108

\bibitem[{{Bell} \& {de Jong}(2001)}]{Bell_2001ApJ...550..212B}
{Bell}, E.~F. \& {de Jong}, R.~S. 2001, \apj, 550, 212

\bibitem[{{Bell} {et~al.}(2004){Bell}, {Wolf}, {Meisenheimer}, {Rix}, {Borch}, {Dye}, {Kleinheinrich}, {Wisotzki}, \& {McIntosh}}]{Bell_2004ApJ...608..752B}
{Bell}, E.~F., {Wolf}, C., {Meisenheimer}, K., {et~al.} 2004, \apj, 608, 752

\bibitem[{{Belli} {et~al.}(2017){Belli}, {Genzel}, {F{\"o}rster Schreiber}, {Wisnioski}, {Wilman}, {Wuyts}, {Mendel}, {Beifiori}, {Bender}, {Brammer}, {Burkert}, {Chan}, {Davies}, {Davies}, {Fabricius}, {Fossati}, {Galametz}, {Lang}, {Lutz}, {Momcheva}, {Nelson}, {Saglia}, {Tacconi}, {Tadaki}, {{\"U}bler}, \& {van Dokkum}}]{Belli_2017ApJ...841L...6B}
{Belli}, S., {Genzel}, R., {F{\"o}rster Schreiber}, N.~M., {et~al.} 2017, \apjl, 841, L6

\bibitem[{{Belli} {et~al.}(2015){Belli}, {Newman}, \& {Ellis}}]{Belli_2015ApJ...799..206B}
{Belli}, S., {Newman}, A.~B., \& {Ellis}, R.~S. 2015, \apj, 799, 206

\bibitem[{{Belli} {et~al.}(2019){Belli}, {Newman}, \& {Ellis}}]{Belli_2019ApJ...874...17B}
{Belli}, S., {Newman}, A.~B., \& {Ellis}, R.~S. 2019, \apj, 874, 17

\bibitem[{{Ben{\'\i}tez} {et~al.}(2009){Ben{\'\i}tez}, {Gazta{\~n}aga}, {Miquel}, {Castander}, {Moles}, {Crocce}, {Fern{\'a}ndez-Soto}, {Fosalba}, {Ballesteros}, {Campa}, {Cardiel-Sas}, {Castilla}, {Crist{\'o}bal-Hornillos}, {Delfino}, {Fern{\'a}ndez}, {Fern{\'a}ndez-Sopuerta}, {Garc{\'\i}a-Bellido}, {Lobo}, {Mart{\'\i}nez}, {Ortiz}, {Pacheco}, {Paredes}, {Pons-Border{\'\i}a}, {S{\'a}nchez}, {S{\'a}nchez}, {Varela}, \& {de Vicente}}]{Benitez_2009ApJ...691..241B}
{Ben{\'\i}tez}, N., {Gazta{\~n}aga}, E., {Miquel}, R., {et~al.} 2009, \apj, 691, 241

\bibitem[{{Beverage} {et~al.}(2021){Beverage}, {Kriek}, {Conroy}, {Bezanson}, {Franx}, \& {van der Wel}}]{Beverage_2021ApJ...917L...1B}
{Beverage}, A.~G., {Kriek}, M., {Conroy}, C., {et~al.} 2021, \apjl, 917, L1

\bibitem[{{Beverage} {et~al.}(2023){Beverage}, {Kriek}, {Conroy}, {Sandford}, {Bezanson}, {Franx}, {van der Wel}, \& {Weisz}}]{Beverage_2023ApJ...948..140B}
{Beverage}, A.~G., {Kriek}, M., {Conroy}, C., {et~al.} 2023, \apj, 948, 140

\bibitem[{{Bezanson} {et~al.}(2018){Bezanson}, {van der Wel}, {Pacifici}, {Noeske}, {Bari{\v{s}}i{\'c}}, {Bell}, {Brammer}, {Calhau}, {Chauke}, {van Dokkum}, {Franx}, {Gallazzi}, {van Houdt}, {Labb{\'e}}, {Maseda}, {Mu{\~n}os-Mateos}, {Muzzin}, {van de Sande}, {Sobral}, {Straatman}, \& {Wu}}]{Bezanson_2018ApJ...858...60B}
{Bezanson}, R., {van der Wel}, A., {Pacifici}, C., {et~al.} 2018, \apj, 858, 60

\bibitem[{{Boquien} {et~al.}(2019){Boquien}, {Burgarella}, {Roehlly}, {Buat}, {Ciesla}, {Corre}, {Inoue}, \& {Salas}}]{Boquien_2019A&A...622A.103B}
{Boquien}, M., {Burgarella}, D., {Roehlly}, Y., {et~al.} 2019, \aap, 622, A103

\bibitem[{{Brammer} {et~al.}(2012){Brammer}, {van Dokkum}, {Franx}, {Fumagalli}, {Patel}, {Rix}, {Skelton}, {Kriek}, {Nelson}, {Schmidt}, {Bezanson}, {da Cunha}, {Erb}, {Fan}, {F{\"o}rster Schreiber}, {Illingworth}, {Labb{\'e}}, {Leja}, {Lundgren}, {Magee}, {Marchesini}, {McCarthy}, {Momcheva}, {Muzzin}, {Quadri}, {Steidel}, {Tal}, {Wake}, {Whitaker}, \& {Williams}}]{Brammer_2012ApJS..200...13B}
{Brammer}, G.~B., {van Dokkum}, P.~G., {Franx}, M., {et~al.} 2012, \apjs, 200, 13

\bibitem[{{Brinchmann} {et~al.}(2004){Brinchmann}, {Charlot}, {White}, {Tremonti}, {Kauffmann}, {Heckman}, \& {Brinkmann}}]{Brinchmann_2004MNRAS.351.1151B}
{Brinchmann}, J., {Charlot}, S., {White}, S.~D.~M., {et~al.} 2004, \mnras, 351, 1151

\bibitem[{{Bruzual} \& {Charlot}(2003)}]{Bruzual_2003MNRAS.344.1000B}
{Bruzual}, G. \& {Charlot}, S. 2003, \mnras, 344, 1000

\bibitem[{{Bryant} {et~al.}(2015){Bryant}, {Owers}, {Robotham}, {Croom}, {Driver}, {Drinkwater}, {Lorente}, {Cortese}, {Scott}, {Colless}, {Schaefer}, {Taylor}, {Konstantopoulos}, {Allen}, {Baldry}, {Barnes}, {Bauer}, {Bland-Hawthorn}, {Bloom}, {Brooks}, {Brough}, {Cecil}, {Couch}, {Croton}, {Davies}, {Ellis}, {Fogarty}, {Foster}, {Glazebrook}, {Goodwin}, {Green}, {Gunawardhana}, {Hampton}, {Ho}, {Hopkins}, {Kewley}, {Lawrence}, {Leon-Saval}, {Leslie}, {McElroy}, {Lewis}, {Liske}, {L{\'o}pez-S{\'a}nchez}, {Mahajan}, {Medling}, {Metcalfe}, {Meyer}, {Mould}, {Obreschkow}, {O'Toole}, {Pracy}, {Richards}, {Shanks}, {Sharp}, {Sweet}, {Thomas}, {Tonini}, \& {Walcher}}]{Bryant_2015MNRAS.447.2857B}
{Bryant}, J.~J., {Owers}, M.~S., {Robotham}, A.~S.~G., {et~al.} 2015, \mnras, 447, 2857

\bibitem[{{Bundy} {et~al.}(2015){Bundy}, {Bershady}, {Law}, {Yan}, {Drory}, {MacDonald}, {Wake}, {Cherinka}, {S{\'a}nchez-Gallego}, {Weijmans}, {Thomas}, {Tremonti}, {Masters}, {Coccato}, {Diamond-Stanic}, {Arag{\'o}n-Salamanca}, {Avila-Reese}, {Badenes}, {Falc{\'o}n-Barroso}, {Belfiore}, {Bizyaev}, {Blanc}, {Bland-Hawthorn}, {Blanton}, {Brownstein}, {Byler}, {Cappellari}, {Conroy}, {Dutton}, {Emsellem}, {Etherington}, {Frinchaboy}, {Fu}, {Gunn}, {Harding}, {Johnston}, {Kauffmann}, {Kinemuchi}, {Klaene}, {Knapen}, {Leauthaud}, {Li}, {Lin}, {Maiolino}, {Malanushenko}, {Malanushenko}, {Mao}, {Maraston}, {McDermid}, {Merrifield}, {Nichol}, {Oravetz}, {Pan}, {Parejko}, {Sanchez}, {Schlegel}, {Simmons}, {Steele}, {Steinmetz}, {Thanjavur}, {Thompson}, {Tinker}, {van den Bosch}, {Westfall}, {Wilkinson}, {Wright}, {Xiao}, \& {Zhang}}]{Bundy_2015ApJ...798....7B}
{Bundy}, K., {Bershady}, M.~A., {Law}, D.~R., {et~al.} 2015, \apj, 798, 7

\bibitem[{{Byler} {et~al.}(2017){Byler}, {Dalcanton}, {Conroy}, \& {Johnson}}]{Byler_2017ApJ...840...44B}
{Byler}, N., {Dalcanton}, J.~J., {Conroy}, C., \& {Johnson}, B.~D. 2017, \apj, 840, 44

\bibitem[{{Byrne} {et~al.}(2022){Byrne}, {Stanway}, {Eldridge}, {McSwiney}, \& {Townsend}}]{Byrne_2022MNRAS.512.5329B}
{Byrne}, C.~M., {Stanway}, E.~R., {Eldridge}, J.~J., {McSwiney}, L., \& {Townsend}, O.~T. 2022, \mnras, 512, 5329

\bibitem[{{Calzetti}(2001)}]{Calzetti_2001PASP..113.1449C}
{Calzetti}, D. 2001, \pasp, 113, 1449

\bibitem[{{Calzetti} {et~al.}(2000){Calzetti}, {Armus}, {Bohlin}, {Kinney}, {Koornneef}, \& {Storchi-Bergmann}}]{Calzetti_2000ApJ...533..682C}
{Calzetti}, D., {Armus}, L., {Bohlin}, R.~C., {et~al.} 2000, \apj, 533, 682

\bibitem[{{Cappellari}(2017)}]{Cappellari_2017MNRAS.466..798C}
{Cappellari}, M. 2017, \mnras, 466, 798

\bibitem[{{Cappellari}(2023)}]{Cappellari_2023MNRAS.526.3273C}
{Cappellari}, M. 2023, \mnras, 526, 3273

\bibitem[{{Cappellari} \& {Emsellem}(2004)}]{Cappellari_2004PASP..116..138C}
{Cappellari}, M. \& {Emsellem}, E. 2004, \pasp, 116, 138

\bibitem[{{Cappellari} {et~al.}(2011){Cappellari}, {Emsellem}, {Krajnovi{\'c}}, {McDermid}, {Scott}, {Verdoes Kleijn}, {Young}, {Alatalo}, {Bacon}, {Blitz}, {Bois}, {Bournaud}, {Bureau}, {Davies}, {Davis}, {de Zeeuw}, {Duc}, {Khochfar}, {Kuntschner}, {Lablanche}, {Morganti}, {Naab}, {Oosterloo}, {Sarzi}, {Serra}, \& {Weijmans}}]{Cappellari_2011MNRAS.413..813C}
{Cappellari}, M., {Emsellem}, E., {Krajnovi{\'c}}, D., {et~al.} 2011, \mnras, 413, 813

\bibitem[{{Cappellari} {et~al.}(2013){Cappellari}, {McDermid}, {Alatalo}, {Blitz}, {Bois}, {Bournaud}, {Bureau}, {Crocker}, {Davies}, {Davis}, {de Zeeuw}, {Duc}, {Emsellem}, {Khochfar}, {Krajnovi{\'c}}, {Kuntschner}, {Morganti}, {Naab}, {Oosterloo}, {Sarzi}, {Scott}, {Serra}, {Weijmans}, \& {Young}}]{Cappellari_2013MNRAS.432.1862C}
{Cappellari}, M., {McDermid}, R.~M., {Alatalo}, K., {et~al.} 2013, \mnras, 432, 1862

\bibitem[{{Carnall} {et~al.}(2018){Carnall}, {McLure}, {Dunlop}, \& {Dav{\'e}}}]{Carnall_2018MNRAS.480.4379C}
{Carnall}, A.~C., {McLure}, R.~J., {Dunlop}, J.~S., \& {Dav{\'e}}, R. 2018, \mnras, 480, 4379

\bibitem[{{Cecchi} {et~al.}(2019){Cecchi}, {Bolzonella}, {Cimatti}, \& {Girelli}}]{Cecchi_2019ApJ...880L..14C}
{Cecchi}, R., {Bolzonella}, M., {Cimatti}, A., \& {Girelli}, G. 2019, \apjl, 880, L14

\bibitem[{{Chabrier}(2003)}]{Chabrier_2003PASP..115..763C}
{Chabrier}, G. 2003, \pasp, 115, 763

\bibitem[{{Charlot} \& {Fall}(2000)}]{Charlot_2000ApJ...539..718C}
{Charlot}, S. \& {Fall}, S.~M. 2000, \apj, 539, 718

\bibitem[{{Chauke} {et~al.}(2018){Chauke}, {van der Wel}, {Pacifici}, {Bezanson}, {Wu}, {Gallazzi}, {Noeske}, {Straatman}, {Mu{\~n}os-Mateos}, {Franx}, {Bari{\v{s}}i{\'c}}, {Bell}, {Brammer}, {Calhau}, {van Houdt}, {Labb{\'e}}, {Maseda}, {Muzzin}, {Rix}, \& {Sobral}}]{Chauke_2018ApJ...861...13C}
{Chauke}, P., {van der Wel}, A., {Pacifici}, C., {et~al.} 2018, \apj, 861, 13

\bibitem[{{Chaves-Montero} \& {Hearin}(2020)}]{Chaves_Montero_2020MNRAS.495.2088C}
{Chaves-Montero}, J. \& {Hearin}, A. 2020, \mnras, 495, 2088

\bibitem[{{Chevallard} \& {Charlot}(2016)}]{Chevallard_2016MNRAS.462.1415C}
{Chevallard}, J. \& {Charlot}, S. 2016, \mnras, 462, 1415

\bibitem[{{Chevallard} {et~al.}(2013){Chevallard}, {Charlot}, {Wandelt}, \& {Wild}}]{Chevallard_2013MNRAS.432.2061C}
{Chevallard}, J., {Charlot}, S., {Wandelt}, B., \& {Wild}, V. 2013, \mnras, 432, 2061

\bibitem[{{Choi} {et~al.}(2014){Choi}, {Conroy}, {Moustakas}, {Graves}, {Holden}, {Brodwin}, {Brown}, \& {van Dokkum}}]{Choi_2014ApJ...792...95C}
{Choi}, J., {Conroy}, C., {Moustakas}, J., {et~al.} 2014, \apj, 792, 95

\bibitem[{{Choi} {et~al.}(2016){Choi}, {Dotter}, {Conroy}, {Cantiello}, {Paxton}, \& {Johnson}}]{Choi_2016ApJ...823..102C}
{Choi}, J., {Dotter}, A., {Conroy}, C., {et~al.} 2016, \apj, 823, 102

\bibitem[{{Cid Fernandes} {et~al.}(2005){Cid Fernandes}, {Mateus}, {Sodr{\'e}}, {Stasi{\'n}ska}, \& {Gomes}}]{Cid_Fernandes_2005MNRAS.358..363C}
{Cid Fernandes}, R., {Mateus}, A., {Sodr{\'e}}, L., {Stasi{\'n}ska}, G., \& {Gomes}, J.~M. 2005, \mnras, 358, 363

\bibitem[{Cleveland \& Devlin(1988)}]{Cleveland_doi:10.1080/01621459.1988.10478639}
Cleveland, W.~S. \& Devlin, S.~J. 1988, Journal of the American Statistical Association, 83, 596

\bibitem[{{Cole} {et~al.}(2020){Cole}, {Bezanson}, {van der Wel}, {Bell}, {D'Eugenio}, {Franx}, {Gallazzi}, {van Houdt}, {Muzzin}, {Pacifici}, {van de Sande}, {Sobral}, {Straatman}, \& {Wu}}]{Cole_2020ApJ...890L..25C}
{Cole}, J., {Bezanson}, R., {van der Wel}, A., {et~al.} 2020, \apjl, 890, L25

\bibitem[{{Conroy}(2013)}]{Conroy_2013ARA&A..51..393C}
{Conroy}, C. 2013, \araa, 51, 393

\bibitem[{{Conroy} \& {Gunn}(2010)}]{Conroy_2010ApJ...712..833C}
{Conroy}, C. \& {Gunn}, J.~E. 2010, \apj, 712, 833

\bibitem[{{Conroy} {et~al.}(2009){Conroy}, {Gunn}, \& {White}}]{Conroy_2009ApJ...699..486C}
{Conroy}, C., {Gunn}, J.~E., \& {White}, M. 2009, \apj, 699, 486

\bibitem[{{Csizi} {et~al.}(2024){Csizi}, {Tortorelli}, {Siudek}, {Gr{\"u}n}, {Renard}, {Tallada-Cresp{\'\i}}, {S{\'a}nchez}, {Miquel}, {Padilla}, {Garc{\'\i}a-Bellido}, {Gazta{\~n}aga}, {Casas}, {Serrano}, {De Vicente}, {Fernandez}, {Eriksen}, {Manzoni}, {Baugh}, {Carretero}, \& {Castander}}]{Csizi_2024A&A...689A..37C}
{Csizi}, B., {Tortorelli}, L., {Siudek}, M., {et~al.} 2024, \aap, 689, A37

\bibitem[{{da Cunha} {et~al.}(2008){da Cunha}, {Charlot}, \& {Elbaz}}]{da_Cunha_2008MNRAS.388.1595D}
{da Cunha}, E., {Charlot}, S., \& {Elbaz}, D. 2008, \mnras, 388, 1595

\bibitem[{{Daddi} {et~al.}(2007){Daddi}, {Dickinson}, {Morrison}, {Chary}, {Cimatti}, {Elbaz}, {Frayer}, {Renzini}, {Pope}, {Alexander}, {Bauer}, {Giavalisco}, {Huynh}, {Kurk}, \& {Mignoli}}]{Daddi_2007ApJ...670..156D}
{Daddi}, E., {Dickinson}, M., {Morrison}, G., {et~al.} 2007, \apj, 670, 156

\bibitem[{{Dav{\'e}} {et~al.}(2017){Dav{\'e}}, {Rafieferantsoa}, \& {Thompson}}]{Dave_2017MNRAS.471.1671D}
{Dav{\'e}}, R., {Rafieferantsoa}, M.~H., \& {Thompson}, R.~J. 2017, \mnras, 471, 1671

\bibitem[{{de Graaff} {et~al.}(2021){de Graaff}, {Bezanson}, {Franx}, {van der Wel}, {Holden}, {van de Sande}, {Bell}, {D'Eugenio}, {Maseda}, {Muzzin}, {Sobral}, {Straatman}, \& {Wu}}]{de_Graaff_2021ApJ...913..103D}
{de Graaff}, A., {Bezanson}, R., {Franx}, M., {et~al.} 2021, \apj, 913, 103

\bibitem[{{Desert} {et~al.}(1990){Desert}, {Boulanger}, \& {Puget}}]{Desert_1990A&A...237..215D}
{Desert}, F.-X., {Boulanger}, F., \& {Puget}, J.~L. 1990, \aap, 237, 215

\bibitem[{{D'Eugenio} {et~al.}(2023{\natexlab{a}}){D'Eugenio}, {van der Wel}, {Derkenne}, {van Houdt}, {Bezanson}, {Taylor}, {van de Sande}, {Baker}, {Bell}, {Bland-Hawthorn}, {Bluck}, {Brough}, {Bryant}, {Colless}, {Cortese}, {Croom}, {van Dokkum}, {Fisher}, {Foster}, {Fraser-McKelvie}, {Gallazzi}, {de Graaff}, {Groves}, {del P. Lagos}, {Looser}, {Maiolino}, {Maseda}, {Mendel}, {Nersesian}, {Pacifici}, {Piotrowska}, {Poci}, {Remus}, {Sharma}, {Sweet}, {Thater}, {Tran}, {{\"U}bler}, {Valenzuela}, {Wisnioski}, \& {Zibetti}}]{D_Eugenio_2023MNRAS.525.2765D}
{D'Eugenio}, F., {van der Wel}, A., {Derkenne}, C., {et~al.} 2023{\natexlab{a}}, \mnras, 525, 2765

\bibitem[{{D'Eugenio} {et~al.}(2023{\natexlab{b}}){D'Eugenio}, {van der Wel}, {Piotrowska}, {Bezanson}, {Taylor}, {van de Sande}, {Baker}, {Bell}, {Bellstedt}, {Bland-Hawthorn}, {Bluck}, {Brough}, {Bryant}, {Colless}, {Cortese}, {Croom}, {Derkenne}, {van Dokkum}, {Fisher}, {Foster}, {Gallazzi}, {de Graaff}, {Groves}, {van Houdt}, {del P. Lagos}, {Looser}, {Maiolino}, {Maseda}, {Mendel}, {Nersesian}, {Pacifici}, {Poci}, {Remus}, {Sweet}, {Thater}, {Tran}, {{\"U}bler}, {Valenzuela}, {Wisnioski}, \& {Zibetti}}]{D_Eugenio_2023MNRAS.525.2789D}
{D'Eugenio}, F., {van der Wel}, A., {Piotrowska}, J.~M., {et~al.} 2023{\natexlab{b}}, \mnras, 525, 2789

\bibitem[{{Devriendt} {et~al.}(1999){Devriendt}, {Guiderdoni}, \& {Sadat}}]{Devriendt_1999A&A...350..381D}
{Devriendt}, J.~E.~G., {Guiderdoni}, B., \& {Sadat}, R. 1999, \aap, 350, 381

\bibitem[{{Donnari} {et~al.}(2019){Donnari}, {Pillepich}, {Nelson}, {Vogelsberger}, {Genel}, {Weinberger}, {Marinacci}, {Springel}, \& {Hernquist}}]{Donnari_2019MNRAS.485.4817D}
{Donnari}, M., {Pillepich}, A., {Nelson}, D., {et~al.} 2019, \mnras, 485, 4817

\bibitem[{{Draine} {et~al.}(2014){Draine}, {Aniano}, {Krause}, {Groves}, {Sandstrom}, {Braun}, {Leroy}, {Klaas}, {Linz}, {Rix}, {Schinnerer}, {Schmiedeke}, \& {Walter}}]{Draine_2014ApJ...780..172D}
{Draine}, B.~T., {Aniano}, G., {Krause}, O., {et~al.} 2014, \apj, 780, 172

\bibitem[{{Draine} {et~al.}(2007){Draine}, {Dale}, {Bendo}, {Gordon}, {Smith}, {Armus}, {Engelbracht}, {Helou}, {Kennicutt}, {Li}, {Roussel}, {Walter}, {Calzetti}, {Moustakas}, {Murphy}, {Rieke}, {Bot}, {Hollenbach}, {Sheth}, \& {Teplitz}}]{Draine_2007ApJ...663..866D}
{Draine}, B.~T., {Dale}, D.~A., {Bendo}, G., {et~al.} 2007, \apj, 663, 866

\bibitem[{{Draine} \& {Lee}(1984)}]{Draine_1984ApJ...285...89D}
{Draine}, B.~T. \& {Lee}, H.~M. 1984, \apj, 285, 89

\bibitem[{{Draine} \& {Li}(2001)}]{Draine_2001ApJ...551..807D}
{Draine}, B.~T. \& {Li}, A. 2001, \apj, 551, 807

\bibitem[{{Driver} {et~al.}(2011){Driver}, {Hill}, {Kelvin}, {Robotham}, {Liske}, {Norberg}, {Baldry}, {Bamford}, {Hopkins}, {Loveday}, {Peacock}, {Andrae}, {Bland-Hawthorn}, {Brough}, {Brown}, {Cameron}, {Ching}, {Colless}, {Conselice}, {Croom}, {Cross}, {de Propris}, {Dye}, {Drinkwater}, {Ellis}, {Graham}, {Grootes}, {Gunawardhana}, {Jones}, {van Kampen}, {Maraston}, {Nichol}, {Parkinson}, {Phillipps}, {Pimbblet}, {Popescu}, {Prescott}, {Roseboom}, {Sadler}, {Sansom}, {Sharp}, {Smith}, {Taylor}, {Thomas}, {Tuffs}, {Wijesinghe}, {Dunne}, {Frenk}, {Jarvis}, {Madore}, {Meyer}, {Seibert}, {Staveley-Smith}, {Sutherland}, \& {Warren}}]{Driver_2011MNRAS.413..971D}
{Driver}, S.~P., {Hill}, D.~T., {Kelvin}, L.~S., {et~al.} 2011, \mnras, 413, 971

\bibitem[{{Dwek} {et~al.}(1997){Dwek}, {Arendt}, {Fixsen}, {Sodroski}, {Odegard}, {Weiland}, {Reach}, {Hauser}, {Kelsall}, {Moseley}, {Silverberg}, {Shafer}, {Ballester}, {Bazell}, \& {Isaacman}}]{Dwek_1997ApJ...475..565D}
{Dwek}, E., {Arendt}, R.~G., {Fixsen}, D.~J., {et~al.} 1997, \apj, 475, 565

\bibitem[{{Elbaz} {et~al.}(2007){Elbaz}, {Daddi}, {Le Borgne}, {Dickinson}, {Alexander}, {Chary}, {Starck}, {Brand t}, {Kitzbichler}, {MacDonald}, {Nonino}, {Popesso}, {Stern}, \& {Vanzella}}]{Elbaz_2007A&A...468...33E}
{Elbaz}, D., {Daddi}, E., {Le Borgne}, D., {et~al.} 2007, \aap, 468, 33

\bibitem[{{Eldridge} {et~al.}(2017){Eldridge}, {Stanway}, {Xiao}, {McClelland}, {Taylor}, {Ng}, {Greis}, \& {Bray}}]{Eldridge_2017PASA...34...58E}
{Eldridge}, J.~J., {Stanway}, E.~R., {Xiao}, L., {et~al.} 2017, \pasa, 34, e058

\bibitem[{{Faber} {et~al.}(2007){Faber}, {Willmer}, {Wolf}, {Koo}, {Weiner}, {Newman}, {Im}, {Coil}, {Conroy}, {Cooper}, {Davis}, {Finkbeiner}, {Gerke}, {Gebhardt}, {Groth}, {Guhathakurta}, {Harker}, {Kaiser}, {Kassin}, {Kleinheinrich}, {Konidaris}, {Kron}, {Lin}, {Luppino}, {Madgwick}, {Meisenheimer}, {Noeske}, {Phillips}, {Sarajedini}, {Schiavon}, {Simard}, {Szalay}, {Vogt}, \& {Yan}}]{Faber_2007ApJ...665..265F}
{Faber}, S.~M., {Willmer}, C.~N.~A., {Wolf}, C., {et~al.} 2007, \apj, 665, 265

\bibitem[{{Fang} {et~al.}(2018){Fang}, {Faber}, {Koo}, {Rodr{\'\i}guez-Puebla}, {Guo}, {Barro}, {Behroozi}, {Brammer}, {Chen}, {Dekel}, {Ferguson}, {Gawiser}, {Giavalisco}, {Kartaltepe}, {Kocevski}, {Koekemoer}, {McGrath}, {McIntosh}, {Newman}, {Pacifici}, {Pandya}, {P{\'e}rez-Gonz{\'a}lez}, {Primack}, {Salmon}, {Trump}, {Weiner}, {Willner}, {Acquaviva}, {Dahlen}, {Finkelstein}, {Finlator}, {Fontana}, {Galametz}, {Grogin}, {Gruetzbauch}, {Johnson}, {Mobasher}, {Papovich}, {Pforr}, {Salvato}, {Santini}, {van der Wel}, {Wiklind}, \& {Wuyts}}]{Fang_2018ApJ...858..100F}
{Fang}, J.~J., {Faber}, S.~M., {Koo}, D.~C., {et~al.} 2018, \apj, 858, 100

\bibitem[{{Ferland} {et~al.}(2013){Ferland}, {Porter}, {van Hoof}, {Williams}, {Abel}, {Lykins}, {Shaw}, {Henney}, \& {Stancil}}]{Ferland_2013RMxAA..49..137F}
{Ferland}, G.~J., {Porter}, R.~L., {van Hoof}, P.~A.~M., {et~al.} 2013, \rmxaa, 49, 137

\bibitem[{{Fontanot} {et~al.}(2009){Fontanot}, {De Lucia}, {Monaco}, {Somerville}, \& {Santini}}]{Fontanot_2009MNRAS.397.1776F}
{Fontanot}, F., {De Lucia}, G., {Monaco}, P., {Somerville}, R.~S., \& {Santini}, P. 2009, \mnras, 397, 1776

\bibitem[{{Gallazzi} \& {Bell}(2009)}]{Gallazzi_2009ApJS..185..253G}
{Gallazzi}, A. \& {Bell}, E.~F. 2009, \apjs, 185, 253

\bibitem[{{Gallazzi} {et~al.}(2014){Gallazzi}, {Bell}, {Zibetti}, {Brinchmann}, \& {Kelson}}]{Gallazzi_2014ApJ...788...72G}
{Gallazzi}, A., {Bell}, E.~F., {Zibetti}, S., {Brinchmann}, J., \& {Kelson}, D.~D. 2014, \apj, 788, 72

\bibitem[{{Gallazzi} {et~al.}(2006){Gallazzi}, {Charlot}, {Brinchmann}, \& {White}}]{Gallazzi_2006MNRAS.370.1106G}
{Gallazzi}, A., {Charlot}, S., {Brinchmann}, J., \& {White}, S. D.~M. 2006, \mnras, 370, 1106

\bibitem[{{Gallazzi} {et~al.}(2005){Gallazzi}, {Charlot}, {Brinchmann}, {White}, \& {Tremonti}}]{Gallazzi_2005MNRAS.362...41G}
{Gallazzi}, A., {Charlot}, S., {Brinchmann}, J., {White}, S. D.~M., \& {Tremonti}, C.~A. 2005, \mnras, 362, 41

\bibitem[{{Gomes} \& {Papaderos}(2017)}]{Gomes_2017A&A...603A..63G}
{Gomes}, J.~M. \& {Papaderos}, P. 2017, \aap, 603, A63

\bibitem[{{Hogg}(1999)}]{Hogg_1999astro.ph..5116H}
{Hogg}, D.~W. 1999, arXiv e-prints, astro

\bibitem[{{Hogg} {et~al.}(2010){Hogg}, {Bovy}, \& {Lang}}]{Hogg_2010arXiv1008.4686H}
{Hogg}, D.~W., {Bovy}, J., \& {Lang}, D. 2010, arXiv e-prints, arXiv:1008.4686

\bibitem[{{Ilbert} {et~al.}(2010){Ilbert}, {Salvato}, {Le Floc'h}, {Aussel}, {Capak}, {McCracken}, {Mobasher}, {Kartaltepe}, {Scoville}, {Sanders}, {Arnouts}, {Bundy}, {Cassata}, {Kneib}, {Koekemoer}, {Le F{\`e}vre}, {Lilly}, {Surace}, {Taniguchi}, {Tasca}, {Thompson}, {Tresse}, {Zamojski}, {Zamorani}, \& {Zucca}}]{Ilbert_2010ApJ...709..644I}
{Ilbert}, O., {Salvato}, M., {Le Floc'h}, E., {et~al.} 2010, \apj, 709, 644

\bibitem[{{Iyer} \& {Gawiser}(2017)}]{Iyer_2017ApJ...838..127I}
{Iyer}, K. \& {Gawiser}, E. 2017, \apj, 838, 127

\bibitem[{{Iyer} {et~al.}(2018){Iyer}, {Gawiser}, {Dav{\'e}}, {Davis}, {Finkelstein}, {Kodra}, {Koekemoer}, {Kurczynski}, {Newman}, {Pacifici}, \& {Somerville}}]{Iyer_2018ApJ...866..120I}
{Iyer}, K., {Gawiser}, E., {Dav{\'e}}, R., {et~al.} 2018, \apj, 866, 120

\bibitem[{{Iyer} {et~al.}(2019){Iyer}, {Gawiser}, {Faber}, {Ferguson}, {Kartaltepe}, {Koekemoer}, {Pacifici}, \& {Somerville}}]{Iyer_2019ApJ...879..116I}
{Iyer}, K.~G., {Gawiser}, E., {Faber}, S.~M., {et~al.} 2019, \apj, 879, 116

\bibitem[{{Johnson} {et~al.}(2021){Johnson}, {Leja}, {Conroy}, \& {Speagle}}]{Johnson_2021ApJS..254...22J}
{Johnson}, B.~D., {Leja}, J., {Conroy}, C., \& {Speagle}, J.~S. 2021, \apjs, 254, 22

\bibitem[{{J{\o}rgensen}(1999)}]{Jorgensen_1999MNRAS.306..607J}
{J{\o}rgensen}, I. 1999, \mnras, 306, 607

\bibitem[{{Kauffmann} {et~al.}(2003{\natexlab{a}}){Kauffmann}, {Heckman}, {White}, {Charlot}, {Tremonti}, {Brinchmann}, {Bruzual}, {Peng}, {Seibert}, {Bernardi}, {Blanton}, {Brinkmann}, {Castander}, {Cs{\'a}bai}, {Fukugita}, {Ivezic}, {Munn}, {Nichol}, {Padmanabhan}, {Thakar}, {Weinberg}, \& {York}}]{Kauffmann_2003MNRAS.341...33K}
{Kauffmann}, G., {Heckman}, T.~M., {White}, S. D.~M., {et~al.} 2003{\natexlab{a}}, \mnras, 341, 33

\bibitem[{{Kauffmann} {et~al.}(2003{\natexlab{b}}){Kauffmann}, {Heckman}, {White}, {Charlot}, {Tremonti}, {Peng}, {Seibert}, {Brinkmann}, {Nichol}, {SubbaRao}, \& {York}}]{Kauffmann_2003MNRAS.341...54K}
{Kauffmann}, G., {Heckman}, T.~M., {White}, S. D.~M., {et~al.} 2003{\natexlab{b}}, \mnras, 341, 54

\bibitem[{{Kaushal} {et~al.}(2024){Kaushal}, {Nersesian}, {Bezanson}, {van der Wel}, {Leja}, {Carnall}, {Gallazzi}, {Zibetti}, {Khullar}, {Franx}, {Muzzin}, {de Graaff}, {Pacifici}, {Whitaker}, {Bell}, \& {Martorano}}]{Kaushal_2024ApJ...961..118K}
{Kaushal}, Y., {Nersesian}, A., {Bezanson}, R., {et~al.} 2024, \apj, 961, 118

\bibitem[{{Koposov} {et~al.}(2022){Koposov}, {Speagle}, {Barbary}, {Ashton}, {Bennett}, {Buchner}, {Scheffler}, {Cook}, {Talbot}, {Guillochon}, {Cubillos}, {Asensio Ramos}, {Johnson}, {Lang}, {Ilya}, {Dartiailh}, {Nitz}, {McCluskey}, {Archibald}, {Deil}, {Foreman-Mackey}, {Goldstein}, {Tollerud}, {Leja}, {Kirk}, {Pitkin}, {Sheehan}, {Cargile}, {Ruskin23}, \& {Angus}}]{Koposov_2022zndo...7388523K}
{Koposov}, S., {Speagle}, J., {Barbary}, K., {et~al.} 2022, {joshspeagle/dynesty: v2.0.3}

\bibitem[{{Kriek} \& {Conroy}(2013)}]{Kriek_2013ApJ...775L..16K}
{Kriek}, M. \& {Conroy}, C. 2013, \apjl, 775, L16

\bibitem[{{Kriek} {et~al.}(2015){Kriek}, {Shapley}, {Reddy}, {Siana}, {Coil}, {Mobasher}, {Freeman}, {de Groot}, {Price}, {Sanders}, {Shivaei}, {Brammer}, {Momcheva}, {Skelton}, {van Dokkum}, {Whitaker}, {Aird}, {Azadi}, {Kassis}, {Bullock}, {Conroy}, {Dav{\'e}}, {Kere{\v{s}}}, \& {Krumholz}}]{Kriek_2015ApJS..218...15K}
{Kriek}, M., {Shapley}, A.~E., {Reddy}, N.~A., {et~al.} 2015, \apjs, 218, 15

\bibitem[{{Kuntschner}(2000)}]{Kuntschner_2000MNRAS.315..184K}
{Kuntschner}, H. 2000, \mnras, 315, 184

\bibitem[{{Kurinchi-Vendhan} {et~al.}(2024){Kurinchi-Vendhan}, {Farcy}, {Hirschmann}, \& {Valentino}}]{Kurinchi_Vendhan_2024MNRAS.534.3974K}
{Kurinchi-Vendhan}, S., {Farcy}, M., {Hirschmann}, M., \& {Valentino}, F. 2024, \mnras, 534, 3974

\bibitem[{{Labb{\'e}} {et~al.}(2005){Labb{\'e}}, {Huang}, {Franx}, {Rudnick}, {Barmby}, {Daddi}, {van Dokkum}, {Fazio}, {F{\"o}rster Schreiber}, {Moorwood}, {Rix}, {R{\"o}ttgering}, {Trujillo}, \& {van der Werf}}]{Labbe_2005ApJ...624L..81L}
{Labb{\'e}}, I., {Huang}, J., {Franx}, M., {et~al.} 2005, \apjl, 624, L81

\bibitem[{{Lacerda} {et~al.}(2022){Lacerda}, {S{\'a}nchez}, {Mej{\'\i}a-Narv{\'a}ez}, {Camps-Fari{\~n}a}, {Espinosa-Ponce}, {Barrera-Ballesteros}, {Ibarra-Medel}, \& {Lugo-Aranda}}]{Lacerda_2022NewA...9701895L}
{Lacerda}, E. A.~D., {S{\'a}nchez}, S.~F., {Mej{\'\i}a-Narv{\'a}ez}, A., {et~al.} 2022, \na, 97, 101895

\bibitem[{{Laigle} {et~al.}(2016){Laigle}, {McCracken}, {Ilbert}, {Hsieh}, {Davidzon}, {Capak}, {Hasinger}, {Silverman}, {Pichon}, {Coupon}, {Aussel}, {Le Borgne}, {Caputi}, {Cassata}, {Chang}, {Civano}, {Dunlop}, {Fynbo}, {Kartaltepe}, {Koekemoer}, {Le F{\`e}vre}, {Le Floc'h}, {Leauthaud}, {Lilly}, {Lin}, {Marchesi}, {Milvang-Jensen}, {Salvato}, {Sanders}, {Scoville}, {Smolcic}, {Stockmann}, {Taniguchi}, {Tasca}, {Toft}, {Vaccari}, \& {Zabl}}]{Laigle_2016ApJS..224...24L}
{Laigle}, C., {McCracken}, H.~J., {Ilbert}, O., {et~al.} 2016, \apjs, 224, 24

\bibitem[{{Le F{\`e}vre} {et~al.}(2003){Le F{\`e}vre}, {Saisse}, {Mancini}, {Brau-Nogue}, {Caputi}, {Castinel}, {D'Odorico}, {Garilli}, {Kissler-Patig}, {Lucuix}, {Mancini}, {Pauget}, {Sciarretta}, {Scodeggio}, {Tresse}, \& {Vettolani}}]{Le_Fevre_2003SPIE.4841.1670L}
{Le F{\`e}vre}, O., {Saisse}, M., {Mancini}, D., {et~al.} 2003, in Society of Photo-Optical Instrumentation Engineers (SPIE) Conference Series, Vol. 4841, Instrument Design and Performance for Optical/Infrared Ground-based Telescopes, ed. M.~{Iye} \& A.~F.~M. {Moorwood}, 1670--1681

\bibitem[{{Lee} {et~al.}(2010){Lee}, {Ferguson}, {Somerville}, {Wiklind}, \& {Giavalisco}}]{Lee_2010ApJ...725.1644L}
{Lee}, S.-K., {Ferguson}, H.~C., {Somerville}, R.~S., {Wiklind}, T., \& {Giavalisco}, M. 2010, \apj, 725, 1644

\bibitem[{{Leitherer} {et~al.}(2002){Leitherer}, {Li}, {Calzetti}, \& {Heckman}}]{Leitherer_2002ApJS..140..303L}
{Leitherer}, C., {Li}, I.-H., {Calzetti}, D., \& {Heckman}, T.~M. 2002, \apjs, 140, 303

\bibitem[{{Leitner}(2012)}]{Leitner_2012ApJ...745..149L}
{Leitner}, S.~N. 2012, \apj, 745, 149

\bibitem[{{Leja} {et~al.}(2019{\natexlab{a}}){Leja}, {Carnall}, {Johnson}, {Conroy}, \& {Speagle}}]{Leja_2019ApJ...876....3L}
{Leja}, J., {Carnall}, A.~C., {Johnson}, B.~D., {Conroy}, C., \& {Speagle}, J.~S. 2019{\natexlab{a}}, \apj, 876, 3

\bibitem[{{Leja} {et~al.}(2018){Leja}, {Johnson}, {Conroy}, \& {van Dokkum}}]{Leja_2018ApJ...854...62L}
{Leja}, J., {Johnson}, B.~D., {Conroy}, C., \& {van Dokkum}, P. 2018, \apj, 854, 62

\bibitem[{{Leja} {et~al.}(2019{\natexlab{b}}){Leja}, {Johnson}, {Conroy}, {van Dokkum}, {Speagle}, {Brammer}, {Momcheva}, {Skelton}, {Whitaker}, \& {Franx}}]{Leja_2019ApJ...877..140L}
{Leja}, J., {Johnson}, B.~D., {Conroy}, C., {et~al.} 2019{\natexlab{b}}, \apj, 877, 140

\bibitem[{{Leja} {et~al.}(2017){Leja}, {Johnson}, {Conroy}, {van Dokkum}, \& {Byler}}]{Leja_2017ApJ...837..170L}
{Leja}, J., {Johnson}, B.~D., {Conroy}, C., {van Dokkum}, P.~G., \& {Byler}, N. 2017, \apj, 837, 170

\bibitem[{{Leja} {et~al.}(2022){Leja}, {Speagle}, {Ting}, {Johnson}, {Conroy}, {Whitaker}, {Nelson}, {van Dokkum}, \& {Franx}}]{Leja_2022ApJ...936..165L}
{Leja}, J., {Speagle}, J.~S., {Ting}, Y.-S., {et~al.} 2022, \apj, 936, 165

\bibitem[{{Leja} {et~al.}(2019{\natexlab{c}}){Leja}, {Tacchella}, \& {Conroy}}]{Leja_2019ApJ...880L...9L}
{Leja}, J., {Tacchella}, S., \& {Conroy}, C. 2019{\natexlab{c}}, \apjl, 880, L9

\bibitem[{{Leslie} {et~al.}(2020){Leslie}, {Schinnerer}, {Liu}, {Magnelli}, {Algera}, {Karim}, {Davidzon}, {Gozaliasl}, {Jim{\'e}nez-Andrade}, {Lang}, {Sargent}, {Novak}, {Groves}, {Smol{\v{c}}i{\'c}}, {Zamorani}, {Vaccari}, {Battisti}, {Vardoulaki}, {Peng}, \& {Kartaltepe}}]{Leslie_2020ApJ...899...58L}
{Leslie}, S.~K., {Schinnerer}, E., {Liu}, D., {et~al.} 2020, \apj, 899, 58

\bibitem[{{Lewis} {et~al.}(2024){Lewis}, {Andrews}, {Bezanson}, {Maseda}, {Bell}, {Dav{\'e}}, {D'Eugenio}, {Franx}, {Gallazzi}, {de Graaff}, {Kaushal}, {Nersesian}, {Newman}, {van der Wel}, \& {Wu}}]{Lewis_2024ApJ...964...59L}
{Lewis}, Z.~J., {Andrews}, B.~H., {Bezanson}, R., {et~al.} 2024, \apj, 964, 59

\bibitem[{{Li} \& {Draine}(2001)}]{Li_2001ApJ...554..778L}
{Li}, A. \& {Draine}, B.~T. 2001, \apj, 554, 778

\bibitem[{{Li} \& {Draine}(2002)}]{Li_2002ApJ...572..232L}
{Li}, A. \& {Draine}, B.~T. 2002, \apj, 572, 232

\bibitem[{{Li} {et~al.}(2018){Li}, {Mao}, {Cappellari}, {Ge}, {Long}, {Li}, {Mo}, {Li}, {Zheng}, {Bundy}, {Thomas}, {Brownstein}, {Roman Lopes}, {Law}, \& {Drory}}]{Li_2018MNRAS.476.1765L}
{Li}, H., {Mao}, S., {Cappellari}, M., {et~al.} 2018, \mnras, 476, 1765

\bibitem[{{Lower} {et~al.}(2020){Lower}, {Narayanan}, {Leja}, {Johnson}, {Conroy}, \& {Dav{\'e}}}]{Lower_2020ApJ...904...33L}
{Lower}, S., {Narayanan}, D., {Leja}, J., {et~al.} 2020, \apj, 904, 33

\bibitem[{{Maraston}(2005)}]{Maraston_2005MNRAS.362..799M}
{Maraston}, C. 2005, \mnras, 362, 799

\bibitem[{{Maraston} {et~al.}(2020){Maraston}, {Hill}, {Thomas}, {Yan}, {Chen}, {Lian}, {Parikh}, {Neumann}, {Meneses-Goytia}, {Bershady}, {Drory}, {Bizyaev}, {Concas}, {Brownstein}, {Lazarz}, {Stringfellow}, \& {Stassun}}]{Maraston_2020MNRAS.496.2962M}
{Maraston}, C., {Hill}, L., {Thomas}, D., {et~al.} 2020, \mnras, 496, 2962

\bibitem[{{Maraston} \& {Str{\"o}mb{\"a}ck}(2011)}]{Maraston_2011MNRAS.418.2785M}
{Maraston}, C. \& {Str{\"o}mb{\"a}ck}, G. 2011, \mnras, 418, 2785

\bibitem[{{McCracken} {et~al.}(2012){McCracken}, {Milvang-Jensen}, {Dunlop}, {Franx}, {Fynbo}, {Le F{\`e}vre}, {Holt}, {Caputi}, {Goranova}, {Buitrago}, {Emerson}, {Freudling}, {Hudelot}, {L{\'o}pez-Sanjuan}, {Magnard}, {Mellier}, {M{\o}ller}, {Nilsson}, {Sutherland}, {Tasca}, \& {Zabl}}]{McCracken_2012A&A...544A.156M}
{McCracken}, H.~J., {Milvang-Jensen}, B., {Dunlop}, J., {et~al.} 2012, \aap, 544, A156

\bibitem[{{McDermid} {et~al.}(2015){McDermid}, {Alatalo}, {Blitz}, {Bournaud}, {Bureau}, {Cappellari}, {Crocker}, {Davies}, {Davis}, {de Zeeuw}, {Duc}, {Emsellem}, {Khochfar}, {Krajnovi{\'c}}, {Kuntschner}, {Morganti}, {Naab}, {Oosterloo}, {Sarzi}, {Scott}, {Serra}, {Weijmans}, \& {Young}}]{McDermid_2015MNRAS.448.3484M}
{McDermid}, R.~M., {Alatalo}, K., {Blitz}, L., {et~al.} 2015, \mnras, 448, 3484

\bibitem[{{Miller} {et~al.}(2022){Miller}, {Whitaker}, {Nelson}, {van Dokkum}, {Bezanson}, {Brammer}, {Heintz}, {Leja}, {Suess}, \& {Weaver}}]{Miller_2022ApJ...941L..37M}
{Miller}, T.~B., {Whitaker}, K.~E., {Nelson}, E.~J., {et~al.} 2022, \apjl, 941, L37

\bibitem[{{Momcheva} {et~al.}(2016){Momcheva}, {Brammer}, {van Dokkum}, {Skelton}, {Whitaker}, {Nelson}, {Fumagalli}, {Maseda}, {Leja}, {Franx}, {Rix}, {Bezanson}, {Da Cunha}, {Dickey}, {F{\"o}rster Schreiber}, {Illingworth}, {Kriek}, {Labb{\'e}}, {Ulf Lange}, {Lundgren}, {Magee}, {Marchesini}, {Oesch}, {Pacifici}, {Patel}, {Price}, {Tal}, {Wake}, {van der Wel}, \& {Wuyts}}]{Momcheva_2016ApJS..225...27M}
{Momcheva}, I.~G., {Brammer}, G.~B., {van Dokkum}, P.~G., {et~al.} 2016, \apjs, 225, 27

\bibitem[{{Moresco} {et~al.}(2013){Moresco}, {Pozzetti}, {Cimatti}, {Zamorani}, {Bolzonella}, {Lamareille}, {Mignoli}, {Zucca}, {Lilly}, {Carollo}, {Contini}, {Kneib}, {Le F{\`e}vre}, {Mainieri}, {Renzini}, {Scodeggio}, {Bardelli}, {Bongiorno}, {Caputi}, {Cucciati}, {de la Torre}, {de Ravel}, {Franzetti}, {Garilli}, {Iovino}, {Kampczyk}, {Knobel}, {Kova{\v{c}}}, {Le Borgne}, {Le Brun}, {Maier}, {Pell{\'o}}, {Peng}, {Perez-Montero}, {Presotto}, {Silverman}, {Tanaka}, {Tasca}, {Tresse}, {Vergani}, {Barnes}, {Bordoloi}, {Cappi}, {Diener}, {Koekemoer}, {Le Floc'h}, {L{\'o}pez-Sanjuan}, {McCracken}, {Nair}, {Oesch}, {Scarlata}, {Scoville}, \& {Welikala}}]{Moresco_2013A&A...558A..61M}
{Moresco}, M., {Pozzetti}, L., {Cimatti}, A., {et~al.} 2013, \aap, 558, A61

\bibitem[{{Muzzin} {et~al.}(2013{\natexlab{a}}){Muzzin}, {Marchesini}, {Stefanon}, {Franx}, {McCracken}, {Milvang-Jensen}, {Dunlop}, {Fynbo}, {Brammer}, {Labb{\'e}}, \& {van Dokkum}}]{Muzzin_2013ApJ...777...18M}
{Muzzin}, A., {Marchesini}, D., {Stefanon}, M., {et~al.} 2013{\natexlab{a}}, \apj, 777, 18

\bibitem[{{Muzzin} {et~al.}(2013{\natexlab{b}}){Muzzin}, {Marchesini}, {Stefanon}, {Franx}, {Milvang-Jensen}, {Dunlop}, {Fynbo}, {Brammer}, {Labb{\'e}}, \& {van Dokkum}}]{Muzzin_2013ApJS..206....8M}
{Muzzin}, A., {Marchesini}, D., {Stefanon}, M., {et~al.} 2013{\natexlab{b}}, \apjs, 206, 8

\bibitem[{{Nagaraj} {et~al.}(2022{\natexlab{a}}){Nagaraj}, {Forbes}, {Leja}, {Foreman-Mackey}, \& {Hayward}}]{Nagaraj_2022ApJ...932...54N}
{Nagaraj}, G., {Forbes}, J.~C., {Leja}, J., {Foreman-Mackey}, D., \& {Hayward}, C.~C. 2022{\natexlab{a}}, \apj, 932, 54

\bibitem[{{Nagaraj} {et~al.}(2022{\natexlab{b}}){Nagaraj}, {Forbes}, {Leja}, {Foreman-Mackey}, \& {Hayward}}]{Nagaraj_2022ApJ...939...29N}
{Nagaraj}, G., {Forbes}, J.~C., {Leja}, J., {Foreman-Mackey}, D., \& {Hayward}, C.~C. 2022{\natexlab{b}}, \apj, 939, 29

\bibitem[{{Narayanan} {et~al.}(2018){Narayanan}, {Conroy}, {Dav{\'e}}, {Johnson}, \& {Popping}}]{Narayanan_2018ApJ...869...70N}
{Narayanan}, D., {Conroy}, C., {Dav{\'e}}, R., {Johnson}, B.~D., \& {Popping}, G. 2018, \apj, 869, 70

\bibitem[{{Nelson} {et~al.}(2019){Nelson}, {Pillepich}, {Springel}, {Pakmor}, {Weinberger}, {Genel}, {Torrey}, {Vogelsberger}, {Marinacci}, \& {Hernquist}}]{Nelson_2019MNRAS.490.3234N}
{Nelson}, D., {Pillepich}, A., {Springel}, V., {et~al.} 2019, \mnras, 490, 3234

\bibitem[{{Nersesian} {et~al.}(2024){Nersesian}, {van der Wel}, {Gallazzi}, {Leja}, {Bezanson}, {Bell}, {D'Eugenio}, {de Graaff}, {Kaushal}, {Martorano}, {Maseda}, \& {Zibetti}}]{Nersesian_2024A&A...681A..94N}
{Nersesian}, A., {van der Wel}, A., {Gallazzi}, A., {et~al.} 2024, \aap, 681, A94

\bibitem[{{Nersesian} {et~al.}(2020){Nersesian}, {Verstocken}, {Viaene}, {Baes}, {Xilouris}, {Bianchi}, {Casasola}, {Clark}, {Davies}, {De Looze}, {De Vis}, {Dobbels}, {Fritz}, {Galametz}, {Galliano}, {Jones}, {Madden}, {Mosenkov}, {Tr{\v{c}}ka}, \& {Ysard}}]{Nersesian_2020A&A...637A..25N}
{Nersesian}, A., {Verstocken}, S., {Viaene}, S., {et~al.} 2020, \aap, 637, A25

\bibitem[{{Nersesian} {et~al.}(2019){Nersesian}, {Xilouris}, {Bianchi}, {Galliano}, {Jones}, {Baes}, {Casasola}, {Cassar{\`a}}, {Clark}, {Davies}, {Decleir}, {Dobbels}, {De Looze}, {De Vis}, {Fritz}, {Galametz}, {Madden}, {Mosenkov}, {Tr{\v c}ka}, {Verstocken}, {Viaene}, \& {Lianou}}]{Nersesian_2019A&A...624A..80N}
{Nersesian}, A., {Xilouris}, E.~M., {Bianchi}, S., {et~al.} 2019, \aap, 624, A80

\bibitem[{{Noeske} {et~al.}(2007){Noeske}, {Weiner}, {Faber}, {Papovich}, {Koo}, {Somerville}, {Bundy}, {Conselice}, {Newman}, {Schiminovich}, {Le Floc'h}, {Coil}, {Rieke}, {Lotz}, {Primack}, {Barmby}, {Cooper}, {Davis}, {Ellis}, {Fazio}, {Guhathakurta}, {Huang}, {Kassin}, {Martin}, {Phillips}, {Rich}, {Small}, {Willmer}, \& {Wilson}}]{Noeske_2007ApJ...660L..43N}
{Noeske}, K.~G., {Weiner}, B.~J., {Faber}, S.~M., {et~al.} 2007, \apjl, 660, L43

\bibitem[{{Noll} {et~al.}(2009){Noll}, {Burgarella}, {Giovannoli}, {Buat}, {Marcillac}, \& {Mu{\~n}oz-Mateos}}]{Noll_2009A&A...507.1793N}
{Noll}, S., {Burgarella}, D., {Giovannoli}, E., {et~al.} 2009, \aap, 507, 1793

\bibitem[{{Ocvirk} {et~al.}(2006){Ocvirk}, {Pichon}, {Lan{\c{c}}on}, \& {Thi{\'e}baut}}]{Ocvirk_2006MNRAS.365...46O}
{Ocvirk}, P., {Pichon}, C., {Lan{\c{c}}on}, A., \& {Thi{\'e}baut}, E. 2006, \mnras, 365, 46

\bibitem[{{Pacifici} {et~al.}(2023){Pacifici}, {Iyer}, {Mobasher}, {da Cunha}, {Acquaviva}, {Burgarella}, {Calistro Rivera}, {Carnall}, {Chang}, {Chartab}, {Cooke}, {Fairhurst}, {Kartaltepe}, {Leja}, {Ma{\l}ek}, {Salmon}, {Torelli}, {Vidal-Garc{\'\i}a}, {Boquien}, {Brammer}, {Brown}, {Capak}, {Chevallard}, {Circosta}, {Croton}, {Davidzon}, {Dickinson}, {Duncan}, {Faber}, {Ferguson}, {Fontana}, {Guo}, {Haeussler}, {Hemmati}, {Jafariyazani}, {Kassin}, {Larson}, {Lee}, {Mantha}, {Marchi}, {Nayyeri}, {Newman}, {Pandya}, {Pforr}, {Reddy}, {Sanders}, {Shah}, {Shahidi}, {Stevans}, {Triani}, {Tyler}, {Vanderhoof}, {de la Vega}, {Wang}, \& {Weston}}]{Pacifici_2023ApJ...944..141P}
{Pacifici}, C., {Iyer}, K.~G., {Mobasher}, B., {et~al.} 2023, \apj, 944, 141

\bibitem[{{Papovich} {et~al.}(2018){Papovich}, {Kawinwanichakij}, {Quadri}, {Glazebrook}, {Labb{\'e}}, {Tran}, {Forrest}, {Kacprzak}, {Spitler}, {Straatman}, \& {Tomczak}}]{Papovich_2018ApJ...854...30P}
{Papovich}, C., {Kawinwanichakij}, L., {Quadri}, R.~F., {et~al.} 2018, \apj, 854, 30

\bibitem[{{Papovich} {et~al.}(2015){Papovich}, {Labb{\'e}}, {Quadri}, {Tilvi}, {Behroozi}, {Bell}, {Glazebrook}, {Spitler}, {Straatman}, {Tran}, {Cowley}, {Dav{\'e}}, {Dekel}, {Dickinson}, {Ferguson}, {Finkelstein}, {Gawiser}, {Inami}, {Faber}, {Kacprzak}, {Kawinwanichakij}, {Kocevski}, {Koekemoer}, {Koo}, {Kurczynski}, {Lotz}, {Lu}, {Lucas}, {McIntosh}, {Mehrtens}, {Mobasher}, {Monson}, {Morrison}, {Nanayakkara}, {Persson}, {Salmon}, {Simons}, {Tomczak}, {van Dokkum}, {Weiner}, \& {Willner}}]{Papovich_2015ApJ...803...26P}
{Papovich}, C., {Labb{\'e}}, I., {Quadri}, R., {et~al.} 2015, \apj, 803, 26

\bibitem[{{Paspaliaris} {et~al.}(2023){Paspaliaris}, {Xilouris}, {Nersesian}, {Bianchi}, {Georgantopoulos}, {Masoura}, {Magdis}, \& {Plionis}}]{Paspaliaris_2023A&A...669A..11P}
{Paspaliaris}, E.~D., {Xilouris}, E.~M., {Nersesian}, A., {et~al.} 2023, \aap, 669, A11

\bibitem[{{Paxton} {et~al.}(2011){Paxton}, {Bildsten}, {Dotter}, {Herwig}, {Lesaffre}, \& {Timmes}}]{Paxton_2011ApJS..192....3P}
{Paxton}, B., {Bildsten}, L., {Dotter}, A., {et~al.} 2011, \apjs, 192, 3

\bibitem[{{Paxton} {et~al.}(2013){Paxton}, {Cantiello}, {Arras}, {Bildsten}, {Brown}, {Dotter}, {Mankovich}, {Montgomery}, {Stello}, {Timmes}, \& {Townsend}}]{Paxton_2013ApJS..208....4P}
{Paxton}, B., {Cantiello}, M., {Arras}, P., {et~al.} 2013, \apjs, 208, 4

\bibitem[{{Paxton} {et~al.}(2015){Paxton}, {Marchant}, {Schwab}, {Bauer}, {Bildsten}, {Cantiello}, {Dessart}, {Farmer}, {Hu}, {Langer}, {Townsend}, {Townsley}, \& {Timmes}}]{Paxton_2015ApJS..220...15P}
{Paxton}, B., {Marchant}, P., {Schwab}, J., {et~al.} 2015, \apjs, 220, 15

\bibitem[{{Paxton} {et~al.}(2018){Paxton}, {Schwab}, {Bauer}, {Bildsten}, {Blinnikov}, {Duffell}, {Farmer}, {Goldberg}, {Marchant}, {Sorokina}, {Thoul}, {Townsend}, \& {Timmes}}]{Paxton_2018ApJS..234...34P}
{Paxton}, B., {Schwab}, J., {Bauer}, E.~B., {et~al.} 2018, \apjs, 234, 34

\bibitem[{{Pernet} {et~al.}(2024){Pernet}, {Boecker}, \& {Mart{\'\i}n-Navarro}}]{Pernet_2024A&A...687L..14P}
{Pernet}, E., {Boecker}, A., \& {Mart{\'\i}n-Navarro}, I. 2024, \aap, 687, L14

\bibitem[{{Pforr} {et~al.}(2012){Pforr}, {Maraston}, \& {Tonini}}]{Pforr_2012MNRAS.422.3285P}
{Pforr}, J., {Maraston}, C., \& {Tonini}, C. 2012, \mnras, 422, 3285

\bibitem[{{Pozzetti} \& {Mannucci}(2000)}]{Pozzetti_2000MNRAS.317L..17P}
{Pozzetti}, L. \& {Mannucci}, F. 2000, \mnras, 317, L17

\bibitem[{{Reddy} {et~al.}(2018){Reddy}, {Oesch}, {Bouwens}, {Montes}, {Illingworth}, {Steidel}, {van Dokkum}, {Atek}, {Carollo}, {Cibinel}, {Holden}, {Labb{\'e}}, {Magee}, {Morselli}, {Nelson}, \& {Wilkins}}]{Reddy_2018ApJ...853...56R}
{Reddy}, N.~A., {Oesch}, P.~A., {Bouwens}, R.~J., {et~al.} 2018, \apj, 853, 56

\bibitem[{{Renzini} \& {Peng}(2015)}]{Renzini_2015ApJ...801L..29R}
{Renzini}, A. \& {Peng}, Y.-j. 2015, \apjl, 801, L29

\bibitem[{{Robotham} {et~al.}(2020){Robotham}, {Bellstedt}, {Lagos}, {Thorne}, {Davies}, {Driver}, \& {Bravo}}]{Robotham_2020MNRAS.495..905R}
{Robotham}, A.~S.~G., {Bellstedt}, S., {Lagos}, C. d.~P., {et~al.} 2020, \mnras, 495, 905

\bibitem[{{Salim}(2014)}]{Salim_2014SerAJ.189....1S}
{Salim}, S. 2014, Serbian Astronomical Journal, 189, 1

\bibitem[{{Salim} {et~al.}(2018){Salim}, {Boquien}, \& {Lee}}]{Salim_2018ApJ...859...11S}
{Salim}, S., {Boquien}, M., \& {Lee}, J.~C. 2018, \apj, 859, 11

\bibitem[{{Salim} \& {Narayanan}(2020)}]{Salim_2020ARA&A..58..529S}
{Salim}, S. \& {Narayanan}, D. 2020, \araa, 58, 529

\bibitem[{{Salmon} {et~al.}(2016){Salmon}, {Papovich}, {Long}, {Willner}, {Finkelstein}, {Ferguson}, {Dickinson}, {Duncan}, {Faber}, {Hathi}, {Koekemoer}, {Kurczynski}, {Newman}, {Pacifici}, {P{\'e}rez-Gonz{\'a}lez}, \& {Pforr}}]{Salmon_2016ApJ...827...20S}
{Salmon}, B., {Papovich}, C., {Long}, J., {et~al.} 2016, \apj, 827, 20

\bibitem[{{Salpeter}(1955)}]{Salpeter_1955ApJ...121..161S}
{Salpeter}, E.~E. 1955, \apj, 121, 161

\bibitem[{{S{\'a}nchez} {et~al.}(2016){S{\'a}nchez}, {P{\'e}rez}, {S{\'a}nchez-Bl{\'a}zquez}, {Gonz{\'a}lez}, {Ros{\'a}les-Ortega}, {Cano-D{\'\i}az}, {L{\'o}pez-Cob{\'a}}, {Marino}, {Gil de Paz}, {Moll{\'a}}, {L{\'o}pez-S{\'a}nchez}, {Ascasibar}, \& {Barrera-Ballesteros}}]{Sanchez_2016RMxAA..52...21S}
{S{\'a}nchez}, S.~F., {P{\'e}rez}, E., {S{\'a}nchez-Bl{\'a}zquez}, P., {et~al.} 2016, \rmxaa, 52, 21

\bibitem[{{S{\'a}nchez} {et~al.}(2013){S{\'a}nchez}, {Rosales-Ortega}, {Jungwiert}, {Iglesias-P{\'a}ramo}, {V{\'\i}lchez}, {Marino}, {Walcher}, {Husemann}, {Mast}, {Monreal-Ibero}, {Cid Fernandes}, {P{\'e}rez}, {Gonz{\'a}lez Delgado}, {Garc{\'\i}a-Benito}, {Galbany}, {van de Ven}, {Jahnke}, {Flores}, {Bland-Hawthorn}, {L{\'o}pez-S{\'a}nchez}, {Stanishev}, {Miralles-Caballero}, {D{\'\i}az}, {S{\'a}nchez-Blazquez}, {Moll{\'a}}, {Gallazzi}, {Papaderos}, {Gomes}, {Gruel}, {P{\'e}rez}, {Ruiz-Lara}, {Florido}, {de Lorenzo-C{\'a}ceres}, {Mendez-Abreu}, {Kehrig}, {Roth}, {Ziegler}, {Alves}, {Wisotzki}, {Kupko}, {Quirrenbach}, {Bomans}, \& {Califa Collaboration}}]{Sanchez_2013A&A...554A..58S}
{S{\'a}nchez}, S.~F., {Rosales-Ortega}, F.~F., {Jungwiert}, B., {et~al.} 2013, \aap, 554, A58

\bibitem[{{S{\'a}nchez-Bl{\'a}zquez} {et~al.}(2006){S{\'a}nchez-Bl{\'a}zquez}, {Peletier}, {Jim{\'e}nez-Vicente}, {Cardiel}, {Cenarro}, {Falc{\'o}n-Barroso}, {Gorgas}, {Selam}, \& {Vazdekis}}]{Sanchez_Blazquez_2006MNRAS.371..703S}
{S{\'a}nchez-Bl{\'a}zquez}, P., {Peletier}, R.~F., {Jim{\'e}nez-Vicente}, J., {et~al.} 2006, \mnras, 371, 703

\bibitem[{{Sanders} {et~al.}(2007){Sanders}, {Salvato}, {Aussel}, {Ilbert}, {Scoville}, {Surace}, {Frayer}, {Sheth}, {Helou}, {Brooke}, {Bhattacharya}, {Yan}, {Kartaltepe}, {Barnes}, {Blain}, {Calzetti}, {Capak}, {Carilli}, {Carollo}, {Comastri}, {Daddi}, {Ellis}, {Elvis}, {Fall}, {Franceschini}, {Giavalisco}, {Hasinger}, {Impey}, {Koekemoer}, {Le F{\`e}vre}, {Lilly}, {Liu}, {McCracken}, {Mobasher}, {Renzini}, {Rich}, {Schinnerer}, {Shopbell}, {Taniguchi}, {Thompson}, {Urry}, \& {Williams}}]{Sanders_2007ApJS..172...86S}
{Sanders}, D.~B., {Salvato}, M., {Aussel}, H., {et~al.} 2007, \apjs, 172, 86

\bibitem[{{Schlegel} {et~al.}(1998){Schlegel}, {Finkbeiner}, \& {Davis}}]{Schlegel_1998ApJ...500..525S}
{Schlegel}, D.~J., {Finkbeiner}, D.~P., \& {Davis}, M. 1998, \apj, 500, 525

\bibitem[{{Schreiber} {et~al.}(2018){Schreiber}, {Glazebrook}, {Nanayakkara}, {Kacprzak}, {Labb{\'e}}, {Oesch}, {Yuan}, {Tran}, {Papovich}, {Spitler}, \& {Straatman}}]{Schreiber_2018A&A...618A..85S}
{Schreiber}, C., {Glazebrook}, K., {Nanayakkara}, T., {et~al.} 2018, \aap, 618, A85

\bibitem[{{Schreiber} {et~al.}(2015){Schreiber}, {Pannella}, {Elbaz}, {B{\'e}thermin}, {Inami}, {Dickinson}, {Magnelli}, {Wang}, {Aussel}, {Daddi}, {Juneau}, {Shu}, {Sargent}, {Buat}, {Faber}, {Ferguson}, {Giavalisco}, {Koekemoer}, {Magdis}, {Morrison}, {Papovich}, {Santini}, \& {Scott}}]{Schreiber_2015A&A...575A..74S}
{Schreiber}, C., {Pannella}, M., {Elbaz}, D., {et~al.} 2015, \aap, 575, A74

\bibitem[{{Scodeggio} {et~al.}(2018){Scodeggio}, {Guzzo}, {Garilli}, {Granett}, {Bolzonella}, {de la Torre}, {Abbas}, {Adami}, {Arnouts}, {Bottini}, {Cappi}, {Coupon}, {Cucciati}, {Davidzon}, {Franzetti}, {Fritz}, {Iovino}, {Krywult}, {Le Brun}, {Le F{\`e}vre}, {Maccagni}, {Ma{\l}ek}, {Marchetti}, {Marulli}, {Polletta}, {Pollo}, {Tasca}, {Tojeiro}, {Vergani}, {Zanichelli}, {Bel}, {Branchini}, {De Lucia}, {Ilbert}, {McCracken}, {Moutard}, {Peacock}, {Zamorani}, {Burden}, {Fumana}, {Jullo}, {Marinoni}, {Mellier}, {Moscardini}, \& {Percival}}]{Scodeggio_2018A&A...609A..84S}
{Scodeggio}, M., {Guzzo}, L., {Garilli}, B., {et~al.} 2018, \aap, 609, A84

\bibitem[{{Scott} {et~al.}(2017){Scott}, {Brough}, {Croom}, {Davies}, {van de Sande}, {Allen}, {Bland-Hawthorn}, {Bryant}, {Cortese}, {D'Eugenio}, {Federrath}, {Ferreras}, {Goodwin}, {Groves}, {Konstantopoulos}, {Lawrence}, {Medling}, {Moffett}, {Owers}, {Richards}, {Robotham}, {Tonini}, \& {Yi}}]{Scott_2017MNRAS.472.2833S}
{Scott}, N., {Brough}, S., {Croom}, S.~M., {et~al.} 2017, \mnras, 472, 2833

\bibitem[{{Scoville} {et~al.}(2007){Scoville}, {Aussel}, {Brusa}, {Capak}, {Carollo}, {Elvis}, {Giavalisco}, {Guzzo}, {Hasinger}, {Impey}, {Kneib}, {LeFevre}, {Lilly}, {Mobasher}, {Renzini}, {Rich}, {Sanders}, {Schinnerer}, {Schminovich}, {Shopbell}, {Taniguchi}, \& {Tyson}}]{Scoville_2007ApJS..172....1S}
{Scoville}, N., {Aussel}, H., {Brusa}, M., {et~al.} 2007, \apjs, 172, 1

\bibitem[{{Shen} {et~al.}(2020){Shen}, {Vogelsberger}, {Nelson}, {Pillepich}, {Tacchella}, {Marinacci}, {Torrey}, {Hernquist}, \& {Springel}}]{Shen_2020MNRAS.495.4747S}
{Shen}, X., {Vogelsberger}, M., {Nelson}, D., {et~al.} 2020, \mnras, 495, 4747

\bibitem[{{Siebenmorgen} \& {Kruegel}(1992)}]{Siebenmorgen_1992A&A...259..614S}
{Siebenmorgen}, R. \& {Kruegel}, E. 1992, \aap, 259, 614

\bibitem[{{Silva} {et~al.}(1998){Silva}, {Granato}, {Bressan}, \& {Danese}}]{Silva_1998ApJ...509..103S}
{Silva}, L., {Granato}, G.~L., {Bressan}, A., \& {Danese}, L. 1998, \apj, 509, 103

\bibitem[{{Skilling}(2004)}]{Skilling_2004AIPC..735..395S}
{Skilling}, J. 2004, in American Institute of Physics Conference Series, Vol. 735, Bayesian Inference and Maximum Entropy Methods in Science and Engineering: 24th International Workshop on Bayesian Inference and Maximum Entropy Methods in Science and Engineering, ed. R.~{Fischer}, R.~{Preuss}, \& U.~V. {Toussaint}, 395--405

\bibitem[{{Skilling}(2006)}]{Skilling_2006AIPC..872..321S}
{Skilling}, J. 2006, in American Institute of Physics Conference Series, Vol. 872, Bayesian Inference and Maximum Entropy Methods In Science and Engineering, ed. A.~{Mohammad-Djafari}, 321--330

\bibitem[{{Speagle} {et~al.}(2014){Speagle}, {Steinhardt}, {Capak}, \& {Silverman}}]{Speagle_2014ApJS..214...15S}
{Speagle}, J.~S., {Steinhardt}, C.~L., {Capak}, P.~L., \& {Silverman}, J.~D. 2014, \apjs, 214, 15

\bibitem[{{Spilker} {et~al.}(2018){Spilker}, {Bezanson}, {Bari{\v{s}}i{\'c}}, {Bell}, {Lagos}, {Maseda}, {Muzzin}, {Pacifici}, {Sobral}, {Straatman}, {van der Wel}, {van Dokkum}, {Weiner}, {Whitaker}, {Williams}, \& {Wu}}]{Spilker_2018ApJ...860..103S}
{Spilker}, J., {Bezanson}, R., {Bari{\v{s}}i{\'c}}, I., {et~al.} 2018, \apj, 860, 103

\bibitem[{{Stanway} \& {Eldridge}(2018)}]{Stanway_2018MNRAS.479...75S}
{Stanway}, E.~R. \& {Eldridge}, J.~J. 2018, \mnras, 479, 75

\bibitem[{{Straatman} {et~al.}(2014){Straatman}, {Labb{\'e}}, {Spitler}, {Allen}, {Altieri}, {Brammer}, {Dickinson}, {van Dokkum}, {Inami}, {Glazebrook}, {Kacprzak}, {Kawinwanichakij}, {Kelson}, {McCarthy}, {Mehrtens}, {Monson}, {Murphy}, {Papovich}, {Persson}, {Quadri}, {Rees}, {Tomczak}, {Tran}, \& {Tilvi}}]{Straatman_2014ApJ...783L..14S}
{Straatman}, C. M.~S., {Labb{\'e}}, I., {Spitler}, L.~R., {et~al.} 2014, \apjl, 783, L14

\bibitem[{{Straatman} {et~al.}(2016){Straatman}, {Spitler}, {Quadri}, {Labb{\'e}}, {Glazebrook}, {Persson}, {Papovich}, {Tran}, {Brammer}, {Cowley}, {Tomczak}, {Nanayakkara}, {Alcorn}, {Allen}, {Broussard}, {van Dokkum}, {Forrest}, {van Houdt}, {Kacprzak}, {Kawinwanichakij}, {Kelson}, {Lee}, {McCarthy}, {Mehrtens}, {Monson}, {Murphy}, {Rees}, {Tilvi}, \& {Whitaker}}]{Straatman_2016ApJ...830...51S}
{Straatman}, C. M.~S., {Spitler}, L.~R., {Quadri}, R.~F., {et~al.} 2016, \apj, 830, 51

\bibitem[{{Straatman} {et~al.}(2018){Straatman}, {van der Wel}, {Bezanson}, {Pacifici}, {Gallazzi}, {Wu}, {Noeske}, {Bari{\v{s}}i{\'c}}, {Bell}, {Brammer}, {Calhau}, {Chauke}, {Franx}, {van Houdt}, {Labb{\'e}}, {Maseda}, {Mu{\~n}oz-Mateos}, {Muzzin}, {van de Sande}, {Sobral}, \& {Spilker}}]{Straatman_2018ApJS..239...27S}
{Straatman}, C. M.~S., {van der Wel}, A., {Bezanson}, R., {et~al.} 2018, \apjs, 239, 27

\bibitem[{{Strateva} {et~al.}(2001){Strateva}, {Ivezi{\'c}}, {Knapp}, {Narayanan}, {Strauss}, {Gunn}, {Lupton}, {Schlegel}, {Bahcall}, {Brinkmann}, {Brunner}, {Budav{\'a}ri}, {Csabai}, {Castander}, {Doi}, {Fukugita}, {Gy{\H{o}}ry}, {Hamabe}, {Hennessy}, {Ichikawa}, {Kunszt}, {Lamb}, {McKay}, {Okamura}, {Racusin}, {Sekiguchi}, {Schneider}, {Shimasaku}, \& {York}}]{Strateva_2001AJ....122.1861S}
{Strateva}, I., {Ivezi{\'c}}, {\v{Z}}., {Knapp}, G.~R., {et~al.} 2001, \aj, 122, 1861

\bibitem[{{Tacchella} {et~al.}(2022){Tacchella}, {Conroy}, {Faber}, {Johnson}, {Leja}, {Barro}, {Cunningham}, {Deason}, {Guhathakurta}, {Guo}, {Hernquist}, {Koo}, {McKinnon}, {Rockosi}, {Speagle}, {van Dokkum}, \& {Yesuf}}]{Tacchella_2022ApJ...926..134T}
{Tacchella}, S., {Conroy}, C., {Faber}, S.~M., {et~al.} 2022, \apj, 926, 134

\bibitem[{{Tan} {et~al.}(2022){Tan}, {Muzzin}, {Marsan}, {Sok}, {Alcorn}, {Matharu}, {Shipley}, {Marchesini}, {Nedkova}, {Martis}, {van der Wel}, \& {Whitaker}}]{Tan_2022ApJ...933...30T}
{Tan}, V. Y.~Y., {Muzzin}, A., {Marsan}, Z.~C., {et~al.} 2022, \apj, 933, 30

\bibitem[{{Taniguchi} {et~al.}(2007){Taniguchi}, {Scoville}, {Murayama}, {Sanders}, {Mobasher}, {Aussel}, {Capak}, {Ajiki}, {Miyazaki}, {Komiyama}, {Shioya}, {Nagao}, {Sasaki}, {Koda}, {Carilli}, {Giavalisco}, {Guzzo}, {Hasinger}, {Impey}, {LeFevre}, {Lilly}, {Renzini}, {Rich}, {Schinnerer}, {Shopbell}, {Kaifu}, {Karoji}, {Arimoto}, {Okamura}, \& {Ohta}}]{Taniguchi_2007ApJS..172....9T}
{Taniguchi}, Y., {Scoville}, N., {Murayama}, T., {et~al.} 2007, \apjs, 172, 9

\bibitem[{{Thorne} {et~al.}(2021){Thorne}, {Robotham}, {Davies}, {Bellstedt}, {Driver}, {Bravo}, {Bremer}, {Holwerda}, {Hopkins}, {Lagos}, {Phillipps}, {Siudek}, {Taylor}, \& {Wright}}]{Thorne_2021MNRAS.505..540T}
{Thorne}, J.~E., {Robotham}, A. S.~G., {Davies}, L. J.~M., {et~al.} 2021, \mnras, 505, 540

\bibitem[{{Tojeiro} {et~al.}(2007){Tojeiro}, {Heavens}, {Jimenez}, \& {Panter}}]{Tojeiro_2007MNRAS.381.1252T}
{Tojeiro}, R., {Heavens}, A.~F., {Jimenez}, R., \& {Panter}, B. 2007, \mnras, 381, 1252

\bibitem[{{Tomczak} {et~al.}(2016){Tomczak}, {Quadri}, {Tran}, {Labb{\'e}}, {Straatman}, {Papovich}, {Glazebrook}, {Allen}, {Brammer}, {Cowley}, {Dickinson}, {Elbaz}, {Inami}, {Kacprzak}, {Morrison}, {Nanayakkara}, {Persson}, {Rees}, {Salmon}, {Schreiber}, {Spitler}, \& {Whitaker}}]{Tomczak_2016ApJ...817..118T}
{Tomczak}, A.~R., {Quadri}, R.~F., {Tran}, K.-V.~H., {et~al.} 2016, \apj, 817, 118

\bibitem[{{Tomczak} {et~al.}(2014){Tomczak}, {Quadri}, {Tran}, {Labb{\'e}}, {Straatman}, {Papovich}, {Glazebrook}, {Allen}, {Brammer}, {Kacprzak}, {Kawinwanichakij}, {Kelson}, {McCarthy}, {Mehrtens}, {Monson}, {Persson}, {Spitler}, {Tilvi}, \& {van Dokkum}}]{Tomczak_2014ApJ...783...85T}
{Tomczak}, A.~R., {Quadri}, R.~F., {Tran}, K.-V.~H., {et~al.} 2014, \apj, 783, 85

\bibitem[{{Trager} {et~al.}(2000{\natexlab{a}}){Trager}, {Faber}, {Worthey}, \& {Gonz{\'a}lez}}]{Trager_2000AJ....120..165T}
{Trager}, S.~C., {Faber}, S.~M., {Worthey}, G., \& {Gonz{\'a}lez}, J.~J. 2000{\natexlab{a}}, \aj, 120, 165

\bibitem[{{Trager} {et~al.}(2000{\natexlab{b}}){Trager}, {Faber}, {Worthey}, \& {Gonz{\'a}lez}}]{Trager_2000AJ....119.1645T}
{Trager}, S.~C., {Faber}, S.~M., {Worthey}, G., \& {Gonz{\'a}lez}, J.~J. 2000{\natexlab{b}}, \aj, 119, 1645

\bibitem[{{Trayford} {et~al.}(2017){Trayford}, {Camps}, {Theuns}, {Baes}, {Bower}, {Crain}, {Gunawardhana}, {Schaller}, {Schaye}, \& {Frenk}}]{Trayford_2017MNRAS.470..771T}
{Trayford}, J.~W., {Camps}, P., {Theuns}, T., {et~al.} 2017, \mnras, 470, 771

\bibitem[{{Valentino} {et~al.}(2023){Valentino}, {Brammer}, {Gould}, {Kokorev}, {Fujimoto}, {Jespersen}, {Vijayan}, {Weaver}, {Ito}, {Tanaka}, {Ilbert}, {Magdis}, {Whitaker}, {Faisst}, {Gallazzi}, {Gillman}, {Gim{\'e}nez-Arteaga}, {G{\'o}mez-Guijarro}, {Kubo}, {Heintz}, {Hirschmann}, {Oesch}, {Onodera}, {Rizzo}, {Lee}, {Strait}, \& {Toft}}]{Valentino_2023ApJ...947...20V}
{Valentino}, F., {Brammer}, G., {Gould}, K. M.~L., {et~al.} 2023, \apj, 947, 20

\bibitem[{{van der Wel} {et~al.}(2021){van der Wel}, {Bezanson}, {D'Eugenio}, {Straatman}, {Franx}, {van Houdt}, {Maseda}, {Gallazzi}, {Wu}, {Pacifici}, {Barisic}, {Brammer}, {Munoz-Mateos}, {Vervalcke}, {Zibetti}, {Sobral}, {de Graaff}, {Calhau}, {Kaushal}, {Muzzin}, {Bell}, \& {van Dokkum}}]{van_der_Wel_2021ApJS..256...44V}
{van der Wel}, A., {Bezanson}, R., {D'Eugenio}, F., {et~al.} 2021, \apjs, 256, 44

\bibitem[{{van der Wel} {et~al.}(2016){van der Wel}, {Noeske}, {Bezanson}, {Pacifici}, {Gallazzi}, {Franx}, {Mu{\~n}oz-Mateos}, {Bell}, {Brammer}, {Charlot}, {Chauk{\'e}}, {Labb{\'e}}, {Maseda}, {Muzzin}, {Rix}, {Sobral}, {van de Sande}, {van Dokkum}, {Wild}, \& {Wolf}}]{van_der_Wel_2016ApJS..223...29V}
{van der Wel}, A., {Noeske}, K., {Bezanson}, R., {et~al.} 2016, \apjs, 223, 29

\bibitem[{{Walcher} {et~al.}(2011){Walcher}, {Groves}, {Budav{\'a}ri}, \& {Dale}}]{Walcher_2011Ap&SS.331....1W}
{Walcher}, J., {Groves}, B., {Budav{\'a}ri}, T., \& {Dale}, D. 2011, \apss, 331, 1

\bibitem[{{Weaver} {et~al.}(2022){Weaver}, {Kauffmann}, {Ilbert}, {McCracken}, {Moneti}, {Toft}, {Brammer}, {Shuntov}, {Davidzon}, {Hsieh}, {Laigle}, {Anastasiou}, {Jespersen}, {Vinther}, {Capak}, {Casey}, {McPartland}, {Milvang-Jensen}, {Mobasher}, {Sanders}, {Zalesky}, {Arnouts}, {Aussel}, {Dunlop}, {Faisst}, {Franx}, {Furtak}, {Fynbo}, {Gould}, {Greve}, {Gwyn}, {Kartaltepe}, {Kashino}, {Koekemoer}, {Kokorev}, {Le F{\`e}vre}, {Lilly}, {Masters}, {Magdis}, {Mehta}, {Peng}, {Riechers}, {Salvato}, {Sawicki}, {Scarlata}, {Scoville}, {Shirley}, {Silverman}, {Sneppen}, {Smolc̆i{\'c}}, {Steinhardt}, {Stern}, {Tanaka}, {Taniguchi}, {Teplitz}, {Vaccari}, {Wang}, \& {Zamorani}}]{Weaver_2022ApJS..258...11W}
{Weaver}, J.~R., {Kauffmann}, O.~B., {Ilbert}, O., {et~al.} 2022, \apjs, 258, 11

\bibitem[{{Whitaker} {et~al.}(2017){Whitaker}, {Bezanson}, {van Dokkum}, {Franx}, {van der Wel}, {Brammer}, {F{\"o}rster-Schreiber}, {Giavalisco}, {Labb{\'e}}, {Momcheva}, {Nelson}, \& {Skelton}}]{Whitaker_2017ApJ...838...19W}
{Whitaker}, K.~E., {Bezanson}, R., {van Dokkum}, P.~G., {et~al.} 2017, \apj, 838, 19

\bibitem[{{Whitaker} {et~al.}(2014){Whitaker}, {Franx}, {Leja}, {van Dokkum}, {Henry}, {Skelton}, {Fumagalli}, {Momcheva}, {Brammer}, {Labb{\'e}}, {Nelson}, \& {Rigby}}]{Whitaker_2014ApJ...795..104W}
{Whitaker}, K.~E., {Franx}, M., {Leja}, J., {et~al.} 2014, \apj, 795, 104

\bibitem[{{Whitaker} {et~al.}(2012){Whitaker}, {van Dokkum}, {Brammer}, \& {Franx}}]{Whitaker_2012ApJ...754L..29W}
{Whitaker}, K.~E., {van Dokkum}, P.~G., {Brammer}, G., \& {Franx}, M. 2012, \apjl, 754, L29

\bibitem[{{Whitaker} {et~al.}(2013){Whitaker}, {van Dokkum}, {Brammer}, {Momcheva}, {Skelton}, {Franx}, {Kriek}, {Labb{\'e}}, {Fumagalli}, {Lundgren}, {Nelson}, {Patel}, \& {Rix}}]{Whitaker_2013ApJ...770L..39W}
{Whitaker}, K.~E., {van Dokkum}, P.~G., {Brammer}, G., {et~al.} 2013, \apjl, 770, L39

\bibitem[{{Wilkinson} {et~al.}(2017){Wilkinson}, {Maraston}, {Goddard}, {Thomas}, \& {Parikh}}]{Wilkinson_2017MNRAS.472.4297W}
{Wilkinson}, D.~M., {Maraston}, C., {Goddard}, D., {Thomas}, D., \& {Parikh}, T. 2017, \mnras, 472, 4297

\bibitem[{{Williams} {et~al.}(2009){Williams}, {Quadri}, {Franx}, {van Dokkum}, \& {Labb{\'e}}}]{Williams_2009ApJ...691.1879W}
{Williams}, R.~J., {Quadri}, R.~F., {Franx}, M., {van Dokkum}, P., \& {Labb{\'e}}, I. 2009, \apj, 691, 1879

\bibitem[{{Witt} \& {Gordon}(2000)}]{Witt_2000ApJ...528..799W}
{Witt}, A.~N. \& {Gordon}, K.~D. 2000, \apj, 528, 799

\bibitem[{{Woo} {et~al.}(2024){Woo}, {Walters}, {Archinuk}, {Faber}, {Ellison}, {Teimoorinia}, \& {Iyer}}]{Woo_2024MNRAS.530.4260W}
{Woo}, J., {Walters}, D., {Archinuk}, F., {et~al.} 2024, \mnras, 530, 4260

\bibitem[{{Worthey} {et~al.}(1994){Worthey}, {Faber}, {Gonzalez}, \& {Burstein}}]{Worthey_1994ApJS...94..687W}
{Worthey}, G., {Faber}, S.~M., {Gonzalez}, J.~J., \& {Burstein}, D. 1994, \apjs, 94, 687

\bibitem[{{Wu} {et~al.}(2021){Wu}, {Nelson}, {van der Wel}, {Pillepich}, {Zibetti}, {Bezanson}, {DEugenio}, {Gallazzi}, {Pacifici}, {Straatman}, {Bari{\v{s}}i{\'c}}, {Bell}, {Maseda}, {Muzzin}, {Sobral}, \& {Whitaker}}]{Wu_2021AJ....162..201W}
{Wu}, P.-F., {Nelson}, D., {van der Wel}, A., {et~al.} 2021, \aj, 162, 201

\bibitem[{{Wu} {et~al.}(2018{\natexlab{a}}){Wu}, {van der Wel}, {Bezanson}, {Gallazzi}, {Pacifici}, {Straatman}, {Bari{\v{s}}i{\'c}}, {Bell}, {Chauke}, {van Houdt}, {Franx}, {Muzzin}, {Sobral}, \& {Wild}}]{Wu_2018ApJ...868...37W}
{Wu}, P.-F., {van der Wel}, A., {Bezanson}, R., {et~al.} 2018{\natexlab{a}}, \apj, 868, 37

\bibitem[{{Wu} {et~al.}(2018{\natexlab{b}}){Wu}, {van der Wel}, {Gallazzi}, {Bezanson}, {Pacifici}, {Straatman}, {Franx}, {Bari{\v{s}}i{\'c}}, {Bell}, {Brammer}, {Calhau}, {Chauke}, {van Houdt}, {Maseda}, {Muzzin}, {Rix}, {Sobral}, {Spilker}, {van de Sande}, {van Dokkum}, \& {Wild}}]{Wu_2018ApJ...855...85W}
{Wu}, P.-F., {van der Wel}, A., {Gallazzi}, A., {et~al.} 2018{\natexlab{b}}, \apj, 855, 85

\bibitem[{{Wuyts} {et~al.}(2011){Wuyts}, {F{\"o}rster Schreiber}, {Lutz}, {Nordon}, {Berta}, {Altieri}, {Andreani}, {Aussel}, {Bongiovanni}, {Cepa}, {Cimatti}, {Daddi}, {Elbaz}, {Genzel}, {Koekemoer}, {Magnelli}, {Maiolino}, {McGrath}, {P{\'e}rez Garc{\'\i}a}, {Poglitsch}, {Popesso}, {Pozzi}, {Sanchez-Portal}, {Sturm}, {Tacconi}, \& {Valtchanov}}]{Wuyts_2011ApJ...738..106W}
{Wuyts}, S., {F{\"o}rster Schreiber}, N.~M., {Lutz}, D., {et~al.} 2011, \apj, 738, 106

\bibitem[{{Wuyts} {et~al.}(2007){Wuyts}, {Labb{\'e}}, {Franx}, {Rudnick}, {van Dokkum}, {Fazio}, {F{\"o}rster Schreiber}, {Huang}, {Moorwood}, {Rix}, {R{\"o}ttgering}, \& {van der Werf}}]{Wuyts_2007ApJ...655...51W}
{Wuyts}, S., {Labb{\'e}}, I., {Franx}, M., {et~al.} 2007, \apj, 655, 51

\bibitem[{{York} {et~al.}(2000){York}, {Adelman}, {Anderson}, {Anderson}, {Annis}, {Bahcall}, {Bakken}, {Barkhouser}, {Bastian}, {Berman}, {Boroski}, {Bracker}, {Briegel}, {Briggs}, {Brinkmann}, {Brunner}, {Burles}, {Carey}, {Carr}, {Castander}, {Chen}, {Colestock}, {Connolly}, {Crocker}, {Csabai}, {Czarapata}, {Davis}, {Doi}, {Dombeck}, {Eisenstein}, {Ellman}, {Elms}, {Evans}, {Fan}, {Federwitz}, {Fiscelli}, {Friedman}, {Frieman}, {Fukugita}, {Gillespie}, {Gunn}, {Gurbani}, {de Haas}, {Haldeman}, {Harris}, {Hayes}, {Heckman}, {Hennessy}, {Hindsley}, {Holm}, {Holmgren}, {Huang}, {Hull}, {Husby}, {Ichikawa}, {Ichikawa}, {Ivezi{\'c}}, {Kent}, {Kim}, {Kinney}, {Klaene}, {Kleinman}, {Kleinman}, {Knapp}, {Korienek}, {Kron}, {Kunszt}, {Lamb}, {Lee}, {Leger}, {Limmongkol}, {Lindenmeyer}, {Long}, {Loomis}, {Loveday}, {Lucinio}, {Lupton}, {MacKinnon}, {Mannery}, {Mantsch}, {Margon}, {McGehee}, {McKay}, {Meiksin}, {Merelli}, {Monet}, {Munn}, {Narayanan}, {Nash}, {Neilsen}, {Neswold}, {Newberg}, {Nichol}, {Nicinski},
  {Nonino}, {Okada}, {Okamura}, {Ostriker}, {Owen}, {Pauls}, {Peoples}, {Peterson}, {Petravick}, {Pier}, {Pope}, {Pordes}, {Prosapio}, {Rechenmacher}, {Quinn}, {Richards}, {Richmond}, {Rivetta}, {Rockosi}, {Ruthmansdorfer}, {Sandford}, {Schlegel}, {Schneider}, {Sekiguchi}, {Sergey}, {Shimasaku}, {Siegmund}, {Smee}, {Smith}, {Snedden}, {Stone}, {Stoughton}, {Strauss}, {Stubbs}, {SubbaRao}, {Szalay}, {Szapudi}, {Szokoly}, {Thakar}, {Tremonti}, {Tucker}, {Uomoto}, {Vanden Berk}, {Vogeley}, {Waddell}, {Wang}, {Watanabe}, {Weinberg}, {Yanny}, {Yasuda}, \& {SDSS Collaboration}}]{York_2000AJ....120.1579Y}
{York}, D.~G., {Adelman}, J., {Anderson}, Jr., J.~E., {et~al.} 2000, \aj, 120, 1579

\bibitem[{{Zibetti} \& {Gallazzi}(2022)}]{Zibetti_2022MNRAS.512.1415Z}
{Zibetti}, S. \& {Gallazzi}, A.~R. 2022, \mnras, 512, 1415

\bibitem[{{Zibetti} {et~al.}(2017){Zibetti}, {Gallazzi}, {Ascasibar}, {Charlot}, {Galbany}, {Garc{\'\i}a Benito}, {Kehrig}, {de Lorenzo-C{\'a}ceres}, {Lyubenova}, {Marino}, {M{\'a}rquez}, {S{\'a}nchez}, {van de Ven}, {Walcher}, \& {Wisotzki}}]{Zibetti_2017MNRAS.468.1902Z}
{Zibetti}, S., {Gallazzi}, A.~R., {Ascasibar}, Y., {et~al.} 2017, \mnras, 468, 1902

\end{thebibliography}

\appendix
%

\onecolumn
\section{A comparison between the observed and modeled Lick indices} \label{ap:A}

\begin{figure*}[!h]
    \centering
    \includegraphics[width=14cm]{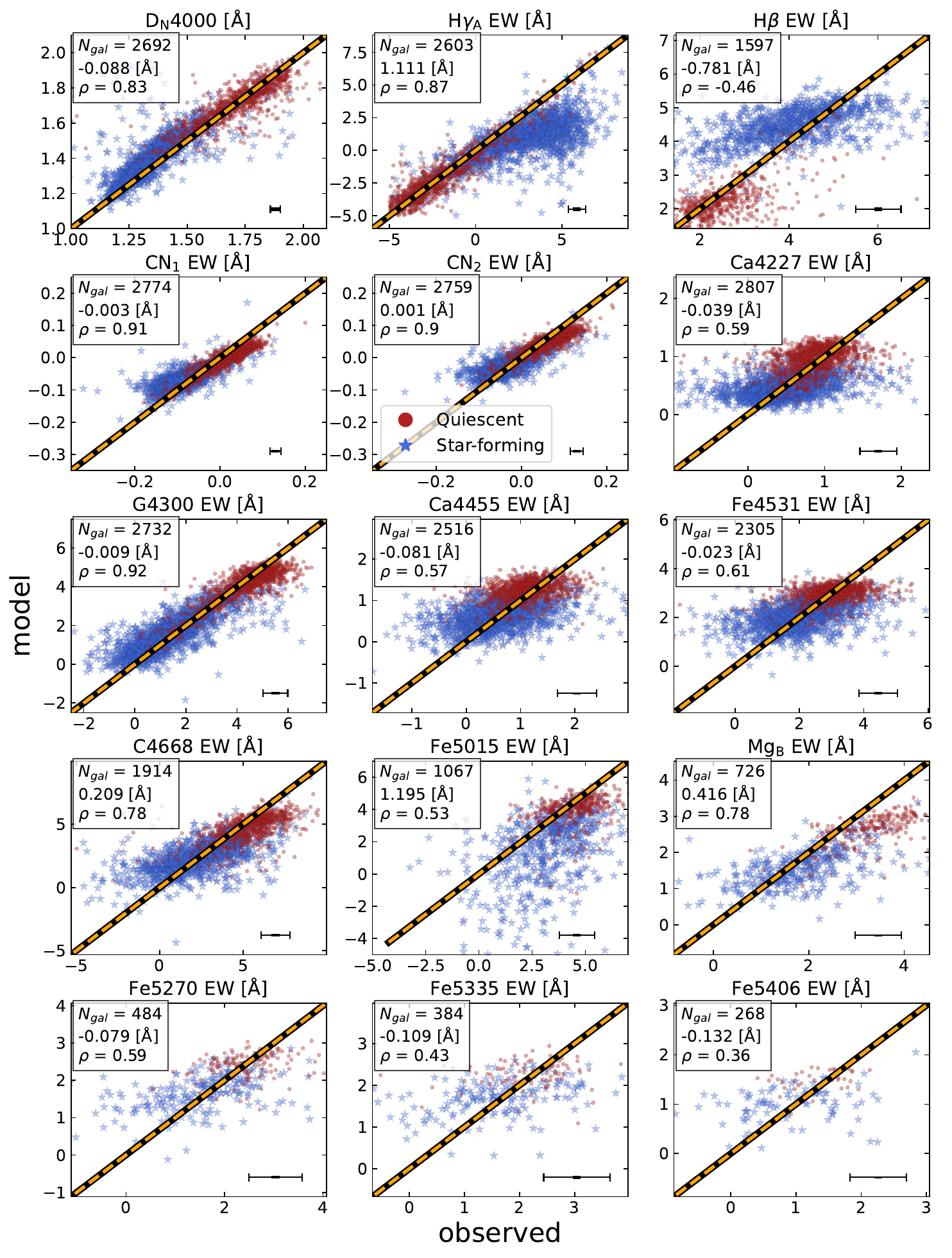}
    \caption{Modeled absorption features from {\tt Prospector} fits compared to observations. Galaxies are color-coded by their $UVJ$ diagram classification as star-forming (blue stars) and quiescent (red points). The absorption-only models are compared to the observed values (corrected for emission). The dashed orange line shows the one-to-one relation. The statistics of the median offsets and the Spearman's rank correlation coefficient ($\rho$) are shown in the top-left corner of each panel. The median uncertainties of each axis are shown in the bottom-right corner of each panel.}
    \label{fig:a1}
\end{figure*}

Here, we present a comparison of 14 additional spectral absorption features, corrected for emission. The results of this comparison are shown in Fig.~\ref{fig:a1}. In each panel, the observed values of a spectral absorption feature are plotted on the \textit{x}-axis. The model values are plotted on the \textit{y}-axis, and were measured from the best-fit SEDs, recovered with {\tt Prospector} by fitting the LEGA-C spectrophotometric SEDs. The galaxies are color-coded by their $UVJ$ diagram classification as star forming and quiescent. 

Similarly to the results shown in Fig.~\ref{fig:hd_fe4383_obs_vs_prd}, the global bimodality of star-forming and quiescent galaxies is reproduced with the model spectra. Overall, we find very strong correlations for most of the Lick indices here. We also measure either small or rather moderate deviations from the one-to-one relation. The absorption features with an average high uncertainty, are also those showing larger scatter. On the other hand, galaxies with an average low uncertainty follow closely the one-to-one relation. A noticeable deviation from the unity slope is seen for the \hg~and \hb~absorption lines of the star-forming galaxies. A possible explanation for this deviation is that light was either over- or under-subtracted from these two features during the stellar emission correction in the observed spectra. In addition, some of the model absorption features such as \hb, Ca4227, Ca4455, and Fe4531, seem to suffer by a limited dynamical range compared to their respective observed relation. This limited limited dynamical range may be related to the use of the continuity SFH, which smooths out any bursts of star formation that would have allowed a larger range of values. Of course, a limitation exists in the metallicity range of the adopted models (MILES and MIST), which extend only up to $\log (Z_\star/\mathrm{Z}_\odot) = 0.2$, whereas the observed spectra may indicate the need for higher metallicities.

A systematic underestimation can also been noted for the \ion{Mg}{B} absorption feature. The underestimation of \ion{Mg}{B} could be a signature of the mismatch in the abundance of the different $\alpha$-elements compared to iron [$\alpha$/Fe], between the SSP models and the observations. Recently, \citet{Pernet_2024A&A...687L..14P} showed that when the assumed stellar population models neglect the individual elemental variations, results in an erroneous modeling of the strength of the absorption lines (including \ion{Mg}{B}). Allowing for the relative change in elemental abundance can better capture the complexity of the chemical evolution in galaxies, and thus properly recovering the strength of the spectral absorption features \citep{Pernet_2024A&A...687L..14P}. 

Overall, the results presented here are a vast improvement over the predicted Lick index measurements in \citet{Nersesian_2024A&A...681A..94N}, where spectroscopy was not complementing the photometric SEDs in the {\tt Prospector} fits. This again implies that reliable age or metallicity determinations can be achieved when photometry is informed by spectroscopy.

\section{Comparison with other literature measurements } \label{ap:B}

\begin{figure*}[h!]
    \centering
    \includegraphics[width=\textwidth]{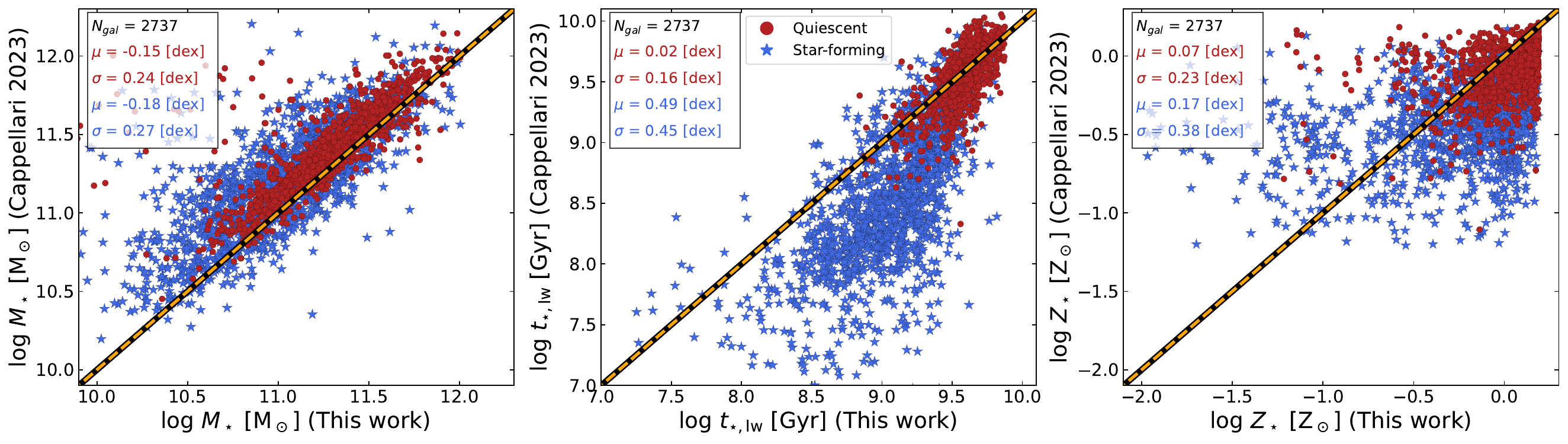}
    \caption{Comparison of main galaxy properties with \citet{Cappellari_2023MNRAS.526.3273C}. Galaxies are color-coded by their $UVJ$ diagram classification as star forming (blue stars) and quiescent (red points). The dashed orange line shows the one-to-one relation. The statistics of the mean offset ($\mu$) and variance ($\sigma$) of each galaxy population, are shown in the top-left corner of each panel.}
    \label{fig:comp_with_cappellari}
\end{figure*}

In this section, we perform a direct comparison of the best-fit properties of our primary sample with those from the recent study by \citet{Cappellari_2023MNRAS.526.3273C}. \citet{Cappellari_2023MNRAS.526.3273C} applied the {\tt PPXF} method to a subsample of 3200 LEGA-C galaxies, using spectroscopy from the LEGA-C survey (DR3) and 28-bands photometry from two different sources, the UltraVISTA catalog by \citet{Muzzin_2013ApJ...777...18M} and the COSMOS2020 catalog by \citet{Weaver_2022ApJS..258...11W}. \citet{Cappellari_2023MNRAS.526.3273C} used three different stellar population synthesis (SPS) models in {\tt pPXF}, including the {\tt FSPS} models generated with the MILES stellar library and MIST isochrones. Since there are many similarities in both our studies, in terms of input data and theoretical models, we want to investigate the consistency of the main derived properties, and whether spectroscopy is capable to constrain the stellar ages and metallicities, regardless of the fitting methods. We note here that \citet{Cappellari_2023MNRAS.526.3273C} used a \citet{Salpeter_1955ApJ...121..161S} IMF. Therefore, we rescaled the stellar masses to a \citet{Chabrier_2003PASP..115..763C} IMF by multiplying them with a constant factor of 0.61.

Figure~\ref{fig:comp_with_cappellari} shows a comparison of three main galaxy properties, namely the stellar mass ($M_\star$), light-weighted ages ($t_\mathrm{\star, lw}$) weighted by the bolometric luminosity, and stellar metallicities ($Z_\star$). We calculated the median offsets for each of the aforementioned properties, and for each galaxy population. Firstly, we find a small systematic offset of 0.16~dex in stellar mass, for the full sample. The mean offset for the quiescent galaxies is slightly reduced at -0.15~dex, whereas the mean offset for the star-forming galaxies is slightly larger at -0.18~dex. These offsets could be due to the differences in the photometric uncertainties, color systematics, and zero point calibrations. Our current stellar mass estimates are consistent with those from the published catalog of LEGA-C galaxies \citep{van_der_Wel_2021ApJS..256...44V}, yielding a mean offset of 0.02~dex and a scatter around the mean of 0.2~dex. This is no surprise, since the released stellar masses by \citet{van_der_Wel_2021ApJS..256...44V} were calculated by fitting the UltraVISTA photometry alone with {\tt Prospector}.

Furthermore, we find a reasonably good agreement and small systematic offsets for the quiescent galaxies in terms of light-weighted ages (0.02~dex) and metallicities (0.07~dex). On the other hand, the ages of the star-forming galaxies that we measure are on average systematically older than those measured by \citet{Cappellari_2023MNRAS.526.3273C}, on the order of $\sim 0.5$~dex and with a large scatter of 0.45~dex. The offset in stellar metallicity still remains reasonable, with a mean offset of 0.17~dex, yet a large scatter of 0.38~dex. The differences in stellar ages could potentially originate from the fact that \citet{Cappellari_2023MNRAS.526.3273C} used more photometric bands in the UV part of the SED (even including the GALEX bands), giving more weight in the blue part of the spectrum, and thus resulting in younger ages for the star-forming galaxies. Another reason could be that \citet{Cappellari_2023MNRAS.526.3273C} uses a different definition for the stellar ages, that is the average $\log$~(age), while we take the logarithm of the average age. Averaging the $\log$~(age) generally yields lower values compared to taking the logarithm of the average age. 

The metallicity estimates also seem to be quite uncertain either for very low metallicity systems or very high metallicity galaxies. Overall, for the full sample, we find a mean offset of 0.29~dex for the stellar ages and 0.13~dex for the stellar metallicities. It is quite surprising that even when including high-resolution spectroscopy, and fitting with the same SPS models, we find such a significant difference both in age and metallicity. \citet{Woo_2024MNRAS.530.4260W} compared the performance of various spectrum-fitting codes, including {\tt pPXF}, by fitting synthetic optical spectra from the IllustrisTNG100-1 simulation \citep{Nelson_2019MNRAS.490.3234N}. The authors found that among the tested spectrum fitting codes, {\tt pPXF} was better at recovering the stellar population properties with typical scatter of 0.11~dex and and offset bias of 0.08~dex. Nevertheless, \citet{Woo_2024MNRAS.530.4260W} showed that the various spectrum-fitting codes yield different systematic systematic biases in the recovery of the stellar properties (especially stellar metallicity).

\section{Example of joint posterior distributions for a star-forming galaxy} \label{ap:C}

\begin{figure*}[h!]
    \centering
    \includegraphics[width=18.5cm]{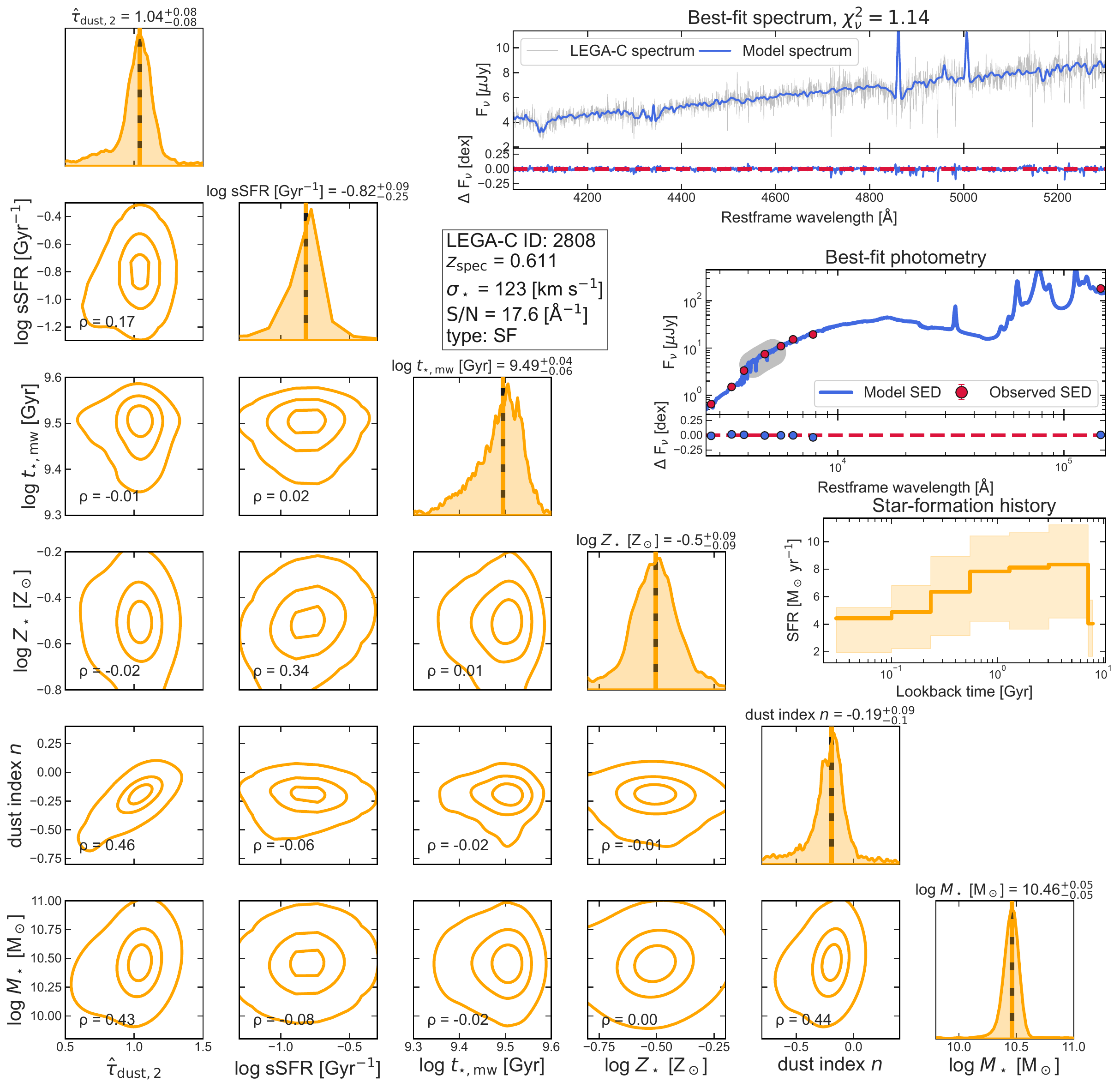}
    \caption{Same as Fig.~\ref{fig:posteriors_q} but for an example of a star-forming galaxy at $z_\mathrm{spec} = 0.611$.}
    \label{fig:posteriors_sf}
\end{figure*}

\begin{figure*}[h!]
    \centering
    \includegraphics[width=\textwidth]{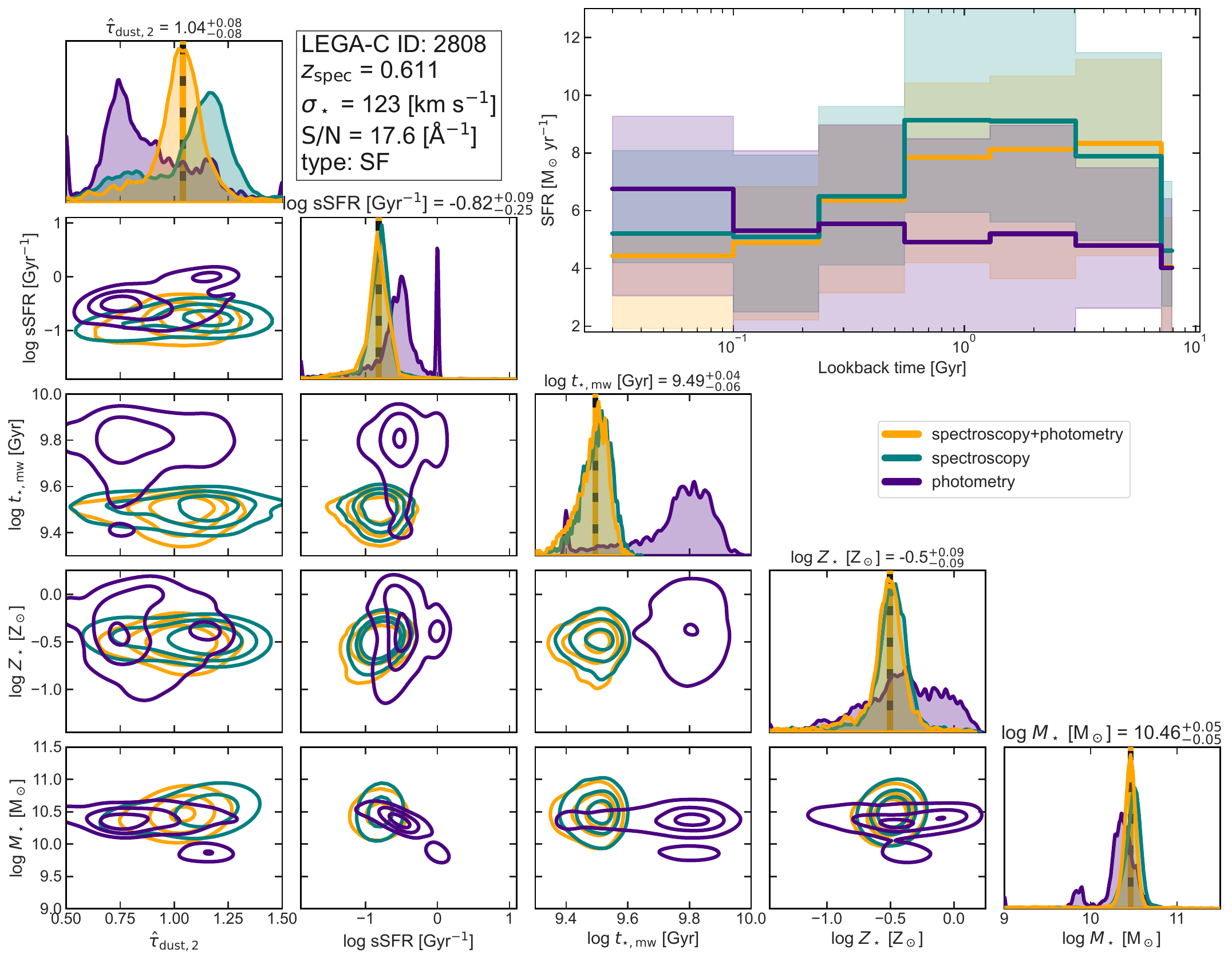}
    \caption{Same as Fig.~\ref{fig:posteriors_q_} but for an example of a star-forming galaxy at $z_\mathrm{spec} = 0.611$.}
    \label{fig:posteriors_sf_}
\end{figure*}

In this section, we present the resulting joint posterior distributions of the main physical quantities (Fig.~\ref{fig:posteriors_sf}), and a comparison of the posteriors retrieved from three different SED fitting runs (Fig.~\ref{fig:posteriors_sf_}), of a handpicked star-forming galaxy with the LEGA-C ID 2808. See main text for the discussion.

\end{document}